\documentclass[aps,pra,twocolumn,superscriptaddress,showpacs,amsmath,amssymb,amsfonts,10pt]{revtex4-2}
\usepackage{graphicx}
\usepackage{dcolumn}
\usepackage{stmaryrd}
\usepackage{latexsym}
\usepackage{amssymb}
\usepackage{amsfonts}
\usepackage{amsmath}
\usepackage{mathtools}
\usepackage{verbatim}
\usepackage{fancybox}
\usepackage{color}
\usepackage{bigdelim}
\usepackage[unicode=true,pdfusetitle,bookmarks=true,bookmarksnumbered=true,bookmarksopen=false, breaklinks=false,pdfborder={0 0 0},pdfborderstyle={},backref=false,colorlinks=false, pdftitle={Directional quantum scattering transducer in cooperative Rydberg metasurfaces}]{hyperref}
\usepackage{float}
\usepackage[capitalize]{cleveref}
\usepackage{epstopdf}
\usepackage{braket}
\usepackage{placeins}
\usepackage{appendix}
\usepackage[dvipsnames]{xcolor}
\usepackage{tikz}
\usepackage{tikz-feynman}
\usepackage{tikz-feynhand}
\usetikzlibrary{shapes.misc}
\tikzset{cross/.style={cross out, draw=black, minimum size=2*(#1-\pgflinewidth), inner sep=0pt, outer sep=0pt},
cross/.default={3pt}}
\usepackage[normalem]{ulem}
\usepackage[straightquotes]{newtxtt}
\usepackage{lmodern}

\usepackage{cleveref}
\usepackage{physics}
\renewcommand{\pv}{\mathbf{p}}
\usetikzlibrary{arrows.meta, decorations.pathmorphing} % Load the necessary libraries

\usepackage{csquotes}

\newcommand{\rv}{\mathbf{r}}
\newcommand{\kv}{\mathbf{k}}

\newcommand{\qv}{\mathbf{q}}

\newcommand{\Ev}{\mathbf{E}}

\newcommand{\zerov}{\mathbf{0}}

\newcommand{\kvp}{\mathbf{k_\parallel}}

\newcommand{\nnl}{\nonumber \\}
\renewcommand{\pv}{\mathbf{p}}
\usepackage{braket}
 \DeclareMathOperator{\sgn}{sgn}
\renewcommand{\Re}{\operatorname{Re}}
\renewcommand{\Im}{\operatorname{Im}}

\definecolor{ColorOne}  {RGB}{214,  40,  57}
\definecolor{ColorTwo}  {RGB}{255, 202,   0}
\definecolor{ColorThree}{RGB}{  0, 114, 178}
\definecolor{ColorFour} {RGB}{155,  32, 229} 
\colorlet   {ColorFive} {ColorOne!50!ColorTwo}

\usepackage{tikz-3dplot}

\begin{document}

\title{Directional quantum scattering transducer in cooperative Rydberg metasurfaces}

\author{Jonas von Milczewski}
\email[Corresponding author, E-Mail:]{jonasvonmilczewski@fas.harvard.edu}
\affiliation{Department of Physics, Harvard University, Cambridge, Massachusetts 02138, USA}

\author{Kelly Werker Smith}
\affiliation{Department of Physics, Harvard University, Cambridge, Massachusetts 02138, USA}
\affiliation{Harvard Quantum Initiative, Harvard University, Cambridge, Massachusetts 02138, USA}

\author{Susanne F. Yelin}
\email[E-Mail:]{syelin@g.harvard.edu}
\affiliation{Department of Physics, Harvard University, Cambridge, Massachusetts 02138, USA}
\date{\today}

\begin{abstract}
We present a single-photon  transduction scheme using four-wave-mixing and quantum scattering in planar, cooperative Rydberg arrays that is both efficient and highly directional and may allow for terahertz-to-optical transduction. In the four-wave-mixing scheme, two lasers drive the system, coherently trapping the system in a dark  ground-state manifold and coupling a signal transition, that may be in the terahertz,  to an idler transition that may be in the optical regime.  The photon-mediated  dipole–dipole interaction between different emitters generates collective super- and subradiant dipolar surface modes, both on the signal and the idler transition. As the array is cooperative with respect to the signal transition, an incident signal photon can efficiently couple into the array and is admixed into dipolar idler modes by way of the drive. Under specific criticality conditions, this admixture is into a superradiant idler mode which primarily decays  into a specific, highly directional optical photon that propagates within the array plane. Outside of the array, this photon may then be coupled into existing quantum devices for further processing.  
Using a scattering-operator formalism we derive  resonance and criticality conditions that govern this two-step process and obtain analytic transduction efficiencies. For lattices of infinite extension, we predict transduction efficiencies into specific spatial directions of up to 50\%, while the overall, undirected transduction efficiency can be higher. An analysis for finite arrays of $N^2$ emitters,  shows that the output is collimated into lobes that narrow as $1/\sqrt{N}$.  Our scheme combines the broadband acceptance of free-space four-wave mixing with the efficiency, directionality and tunability of cooperative metasurfaces, offering a route towards quantum coherent THz detection and processing for astronomical spectroscopy, quantum-networked sparse-aperture imaging and other quantum-sensing applications.
\end{abstract}

\maketitle

\section{Introduction}
\label{sect:intro}

In recent years two-dimensional ordered arrays of quantum emitters have rapidly become a central toolbox for nanophotonics and quantum optics.  When the lattice spacing is comparable to, or smaller than, the relevant optical wavelength, dipole–dipole interactions endow the array with collective super- and sub-radiant surface modes, cooperative Lamb shifts, and highly structured band dispersions \cite{Asenjo-Garcia2017,Shahmoon2017,Bettles2016,Reitz2022,Ruostekoski2023}, which alter the interaction of atoms with light strongly. These cooperative effects may have a myriad of possible applications such as near-unit reflectance \cite{Bettles2016,Shahmoon2017}, photon storage and retrieval \cite{Facchinetti2016,Asenjo-Garcia2017,Manzoni2018,Ballantine2021a,RubiesBigorda2022}, nonlocal entanglement \cite{Guimond2019}, strongly correlated \cite{Pedersen2023} and topological \cite{Perczel2017,Bettles2017} photonic states, quantum computing \cite{Shah2024}, dynamically reconfigurable cavities \cite{CastellsGraells2025} as well as super- and subradiant dynamics \cite{Masson2022,Sierra2022,Robicheaux2021,RubiesBigorda2022b,RubiesBigorda2023,RubiesBigorda2023b}. Similarly, stacking layers of emitter arrays, bulk properties such as near-zero \cite{Ruks2025} and negative refractivity \cite{Ruks2025a} may arise, and bundling several emitters in one array site magnetic dipoles may be realized \cite{Ballantine2021}. Many of these ideas and in particular their experimental realizations \cite{Rui2020,Srakaew2023}, employ atoms with Rydberg transitions. Their strong polarizabilities offer long-range interactions and nonlinear optics \cite{Peyronel2012,Baur2014,Firstenberg2016}, which may enable additional control over an ordered emitter array \cite{Bekenstein2020,Srakaew2023} or low-loss photon-photon interactions \cite{MorenoCardoner2021,Zhang2022}.

A further, distinctive virtue of Rydberg atoms is their ability to  bridge disparate frequency domains as a single atom may simultaneously support connected electric-dipole transitions from the optical to the microwave and terahertz domain. This spectral reach has been exploited to mediate quantum frequency conversion \cite{Huang1992,Hafezi2012,Lambert2019,Lauk2020}: millimetre-wave resonator photons have been coherently up-converted to the optical domain in cryogenic cavity systems \cite{Suleymanzade2020,Kumar2023} and related theory proposals analyze transduction of microwave resonator photons to the optical domain \cite{Gard2017,Hafezi2012,Covey2019,Petrosyan2019}.

The use of cavities and resonators limits the transduction bandwidth. Despite this, efficient, broadband transduction has been demonstrated in free space experiments using six-wave mixing both in cold \cite{Han2018,Vogt2019,Tu2022} and hot \cite{Borwka2023,Li2024} ensembles of Rydberg atoms, allowing for pulsed and continuous-wave signal transduction down to the single-photon limit \cite{Tu2022,Li2024}. The preservation of phase information for such free-space conversion seems to be possible for signals containing many photons \cite{Tu2022}. However, at the single photon level, the phase of the outgoing field strongly depends on where in the interaction region the signal is transduced. 

In contrast, planar cooperative arrays of Rydberg atoms offer local Rydberg control and in-situ tunability of geometric parameters such as lattice constant and array size, potentially enabling efficient, directional, broadband and mode-selective transduction at the single-photon level. Past work on transduction with planar Rydberg arrays has investigated amplification schemes to strengthen weak signals \cite{Nill2024}. However, such amplifiers cannot perfectly replicate a quantum state (the "no-cloning theorem") \cite{repeater_review}; noise is added during the amplification process that can impede the preservation of the coherence properties of the input state and may preclude the possibility of subsequent quantum information processing.

In this paper, we propose a cooperatively-enhanced quantum transduction scheme based on two-dimensional arrays of Rydberg atoms, along with a quantum scattering formalism to analyze this scheme.  The scheme transduces individual photons, that are incident on the array from one frequency range to specific modes in another frequency. Our approach is designed to transfer photons from frequencies where suitable quantum devices are scarce to more accessible ranges, such as optical or centimeter/millimeter waves, which hold well-developed quantum technologies \cite{Lambert2019,Lauk2020}. By transducing into specific photon modes, the transduced photons may then be fed into existing quantum devices such as optical waveguides or detectors for further processing.

One potential application of this technology lies in the field of astronomy, particularly in the Terahertz (THz) regime. Spectral fingerprints of ions, atoms and molecules relevant to the origin of planetary systems and the star-formation history of galaxies lie in this band  \cite{Kulesa2011,Blake2009}. 
Signals in this range are often weak due to atmospheric interference, in particular from water vapor, and the frequencies fall between the traditional ranges of radio and infrared astronomy, where existing detectors are more sensitive. This scarcity of low-noise, single-photon resolving detectors creates a challenge for direct, coherent detection of THz signals. By converting these signals to a more manageable frequency range with the potential to preserve coherence information, our scheme provides a means for investigating this largely unexplored regime using quantum devices. For example, single photon resolution may allow for successful detection even with higher noise levels \cite{Matsuo2016}, while greater single-photon sensitivity would enhance foreground removal in cosmic-microwave-background surveys
\cite{Bennett2003,Finkbeiner2003} and strengthen searches
for dark-matter axions, which demand low dark
count rates at THz frequencies \cite{Adams2022,Berger2022}.

\subsection{Overview of the approach}

Our approach employs an ordered array of Rydberg atoms (see \cref{fig:setup}(\textbf{a})), similar to existing experimental implementations \cite{Rui2020,Bluvstein2022,Srakaew2023,Shao2024}. Each atom has a signal transition, which may be in the THz range, and an idler transition that the incoming photon is transduced to (see \cref{fig:setup}(\textbf{b})). Due to their large electric dipole moment, atoms with Rydberg transitions in the GHz-THz range interact strongly with incoming fields in that frequency range. Furthermore, in such an ordered emitter array there are dipolar surface modes corresponding to the atomic Rydberg transitions (for reviews see \cite{Reitz2022,Ruostekoski2023}).

\tikzset{
  atom/.style      = {ball color=cyan!60!black, circle, inner sep=2.2pt},
  laseraxis/.style = {red!80!black, very thick, -stealth},
  laserfoot/.style = {red!60,      fill=red!30,  opacity=.25}
}
\tdplotsetmaincoords{60}{30}  
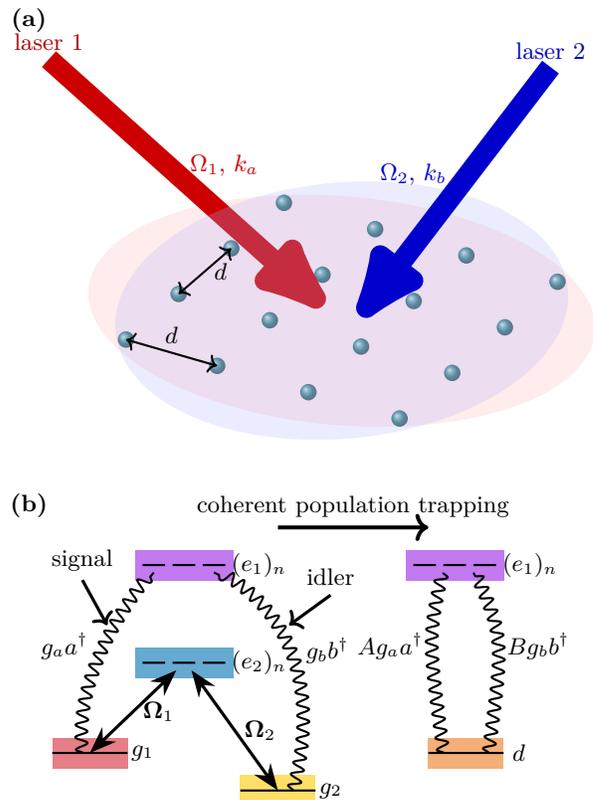
\begin{figure}
    \centering

\tikzset{
  atom/.style   = {ball color=cyan!60!black, circle, inner sep=2.2pt},
  laxisR/.style = {red!80!black,  line width=2.8mm, -{Latex[round]}},
  lfootR/.style = {red!70,        fill=red!30,  opacity=.25},
  laxisB/.style = {blue!80!black, line width=2.8mm, -{Latex[round]}},
  lfootB/.style = {blue!70,       fill=blue!30, opacity=.25}
}
\begin{tikzpicture}
\begin{scope}[local bounding box=A,tdplot_main_coords,scale=1.4]

\foreach \i in {0,...,3}
  \foreach \j in {0,...,3}
    \node[atom] at (\i,\j,0) {};

\coordinate (C) at (1.5,1.5,0);  
\coordinate (C1) at (1.3,1.5,0);
\coordinate (C2) at (1.7,1.5,0);

\coordinate (Sred) at (-2,2,1.5);

\node[red!70!black,above] at (Sred) {laser 1};
\begin{scope}[canvas is xy plane at z=0]             
\fill[lfootR] (1.5,1.5) ellipse (2.5 and 2.1);
\draw[laxisR] (Sred) --
      node[pos=.55,above]{$\ \ \ \ \ \ \Omega_1,\,k_a$} (C1);
\end{scope}

\coordinate (Sblue) at (3.5,2,3);

\begin{scope}[
canvas is xy plane at z=0,
shift={(1.5,1.5)},                 
rotate=10]                         
\fill[lfootB] (0,0) ellipse (2.1 and 2.5);
\end{scope}
      
\node[blue!70!black,above] at (Sblue) {laser 2};
\draw[laxisB] (Sblue) --
      node[pos=.55,above]{$\Omega_2,\,k_b\ \ \ \ \ \ \ \ $} (C2);

\begin{scope}[canvas is xy plane at z=0, thick]
  \draw[<->] (0,0) -- node[above] {$d$} (1,0);   
  \draw[<->] (0,1) -- node[right] {$d$} (0,2);   
\end{scope}

\end{scope}
\begin{scope}[xshift=3.02cm, yshift=-8cm,local bounding box=B]
\begin{scope}[shift={(-2.5,2)}]

\fill[ColorOne!60] (-1.5,0.3) rectangle (-0.5,0.7);

\fill[ColorTwo!60] (1, -0.2) rectangle (2, 0.2);

\draw[thick] (-1.5,0.5) -- (-0.5,0.5);
\node at (-.3,0.5) {$g_1$};  

\draw[thick] (1,0) -- (2,0);
\node at (2.7-.5,0) {$g_2$};

\fill[ColorThree!60] (0.2-.6,1.8-.3) rectangle (1.5-.6,2.2-.3);

\fill[ColorFour!60] (0.2-.6,2.8) rectangle (1.5-.6,3.2);

\draw[thick] (0.3-.6,2-.3) -- (0.6-.6,2-.3);
\draw[thick] (0.7-.6,2-.3) -- (1.0-.6,2-.3);
\draw[thick] (1.1-.6,2-.3) -- (1.4-.6,2-.3);
\node at (1.25,1.7) {$(e_2)_n$};

\draw[thick] (0.3-.6,3) -- (0.6-.6,3);
\draw[thick] (0.7-.6,3) -- (1.0-.6,3);
\draw[thick] (1.1-.6,3) -- (1.4-.6,3);
\node at (1.25,3) {$(e_1)_n$};

\draw[{Stealth[scale=1.2]}-{Stealth[scale=1.2]}, line width=1.2pt] (-1.0,0.5) -- (0.85-.6-.1,2-.3-.1) node[midway, right] {\(\mathbf{\Omega}_1\)};
\draw[{Stealth[scale=1.2]}-{Stealth[scale=1.2]}, line width=1.2pt] (1.5,0) -- (0.85-.6+.1,2-.3-.1) node[midway, right] {\(\mathbf{\Omega}_2\)};

\draw[decorate, decoration={snake, amplitude=.8mm, segment length=1.6mm},thick] 
    (-1.1,0.5) .. controls (-1.2,1.5) and (-0.5,2.5) .. (0.85-.6-.4,3-.1) node[midway, left] {$g_a a^\dagger \ $};

\draw[decorate, decoration={snake, amplitude=.8mm, segment length=1.6mm},thick] 
    (1.7,0) .. controls (2.2-.4+.3,1.5) and (1.5,2.5) .. (0.85-.6+.4,3-.1) node[midway, right] {$ \ g_b b^\dagger$};

\fill[ColorFive!60] (3.5,0.3) rectangle (4.5,0.7);

\draw[thick] (3.5,0.5) -- (4.5,0.5); 
\node at (4.7,.5) {$d$};

\fill[ColorFour!60] (3.2,2.8) rectangle (4.5,3.2);
\draw[thick] (3.3,3) -- (3.6,3);
\draw[thick] (3.7,3) -- (4.0,3);
\draw[thick] (4.1,3) -- (4.4,3);
\node at (4.85,3) {$(e_1)_n$};

\draw[decorate, decoration={snake, amplitude=.8mm, segment length=1.6mm},thick]
    (3.7,0.5) .. controls (3.5,2.0) .. (3.7,3-.1) node[midway, left] {$A  g_a a^\dagger$};

\draw[decorate, decoration={snake, amplitude=.8mm, segment length=1.6mm},thick]
    (4.3,0.5) .. controls (4.5,2.0) .. (4.1,3-.1) node[midway, right] {$B g_b b^\dagger$};
\draw[->, line width=1.2pt] (-1.1,2.8) -- (-0.77, 2.2) node[pos=0,above] {signal};
\draw[->, line width=1.2pt] (2.2,2.6) -- (1.65, 2.2) node[pos=0,above] {idler};

\draw[thick, ->, line width=1.5pt] (1.5,3.5) -- (3.5,3.5) node[midway, above] {coherent population trapping};
\end{scope}
\end{scope}
\node[align=right] at (-1.29,4.25) {$\textbf{(a)}$};
\node[align=right] at (-1.29,-2.2) {$\textbf{(b)}$};
\end{tikzpicture}
    \caption{\textbf{Schematic view of the experimental setup and level scheme of an individual atom site.} (\textbf{a}) Rydberg atoms are arranged in a square lattice with spacing $d$ in the $x$-$y$ plane at $z=0$. The array is illuminated (ellipses) by two driving lasers, laser 1 (red) and laser 2 (blue), with Rabi polarizations $\mathbf{\Omega_1}$, $\mathbf{\Omega_2}$ and wavevectors $\kv_{L_1}$, $\kv_{L_2}$, respectively. (\textbf{b}) The level structure of each atom site features a double $\Lambda$-system which shares the ground states $g_1$ and $g_2$. In the lower $\Lambda$-system the lasers 1 and 2 couple the ground states to a  manifold of excited states $(e_2)_n$. In the upper $\Lambda$-system the ground states  couple to another manifold of excited states  $(e_1)_n$. 
    The $g_1 \leftrightarrow e_1$ and $g_2 \leftrightarrow e_1$ transitions are referred to as the signal and idler transition, which respectively couple to quantized electromagnetic fields $a$ and $b$ with vacuum coupling strength $g_a$ and $g_b$.
    Due to the strong drive on the lower $\Lambda$-system, the ground state manifold is coherently trapped in a dark state superposition $d$ of the two ground states, which now couples to the excited state manifold $(e_1)_n$ through the signal field $a $ and the idler field $b$ with effective coupling strengths $Ag_a$ and $Bg_b$, as described in the main text.}
    \label{fig:setup}
\end{figure}

These modes originate from both the ordered structure into which the atoms are placed and also the dipole-dipole interaction that is mediated by photons, which may make the atom-photon interaction non-local. Due to the periodic structure, these modes can be highly selective in their interaction with light \cite{Bettles2016,Shahmoon2017,Asenjo-Garcia2017} (similar to constructive and destructive interference patterns of classical dipole arrays). The relative importance of these modes and the selective radiance is controlled by the ratio between the lattice constant of the array and the wavelength of the given transition. 

For small ratios, the array is said to be cooperative \cite{Asenjo-Garcia2017} with respect to that transition and collective effects may strongly increase or suppress the scattering cross section for an incoming THz signal photon. With increasing ratio, the relevance of these many-body excitations diminishes step-wise as cooperativity thresholds are passed where the degree of cooperativity decreases. For a vanishing degree of cooperativity, the system behaves as an uncorrelated array of individual atoms and the dipolar modes lose their selectivity with respect to their interaction with light. As signals in the GHz-THz range have wavelengths that are larger than typical lattice spacings in experimental realizations of Rydberg arrays, in this work we will study instances where the array is cooperative with respect to the signal transition. Similarly, as transduced wavelengths in the optical regime may be smaller than experimentally realizable lattice spacings, we will typically study instances in which the array is only weakly cooperative with respect to the idler transition.

To enable the transduction of incoming signal photons, the atomic array is driven by two coherent lasers on two auxiliary transitions which implements a four-wave mixing process \cite{Lukin2000} (see \cref{fig:setup}(\textbf{b})). By means of the  drive, a subspace, which we refer to as a ground-state subspace, is coherently population trapped into a corresponding dark state and, as a result of this, excitations on the signal transition are mixed into excitations on the idler transition. Importantly, such a level scheme where both the idler and the signal transition share a common atomic level is crucial for a quantum-coherent transduction process, because by virtue of the level scheme an incoming signal photon cannot be transduced into several signal and idler photons. Notably, this level scheme thus differs from existing implementations used for intensity squeezing \cite{Niu2023,Sim2025}, where the combined number of signal and idler photons is not conserved. 

Such multiplication of photons could instead be achieved by switching the idler transition and the transition driven by laser 2 in \cref{fig:setup}(b) or by implementing an avalanche detection scheme \cite{Nill2024}, however this will lead to the loss of coherence information.   

On both the idler and the signal transition there are associated dipolar surface modes which are characterized by their in-plane momentum. Approaching specific in-plane momenta these surface modes become superradiant and develop a large decay rate that scales with system size and diverges for an array of infinite extension (for a cooperative array these modes are near the light cone \cite{Shahmoon2017,Asenjo-Garcia2017}). To obtain efficient absorption of an incoming photon, the photon needs to resonantly couple to a \emph{cooperative} dipolar surface mode on the signal transition. That is, it needs to match the energy of the surface mode that has the same in-plane momentum as the incoming photon. In addition, because the dipolar mode is cooperative, it is extremely selective in the photon modes it can couple to. This selectivity increases overlap with the incoming photon. If the photon is coupled into  a dipolar surface mode on the signal transition, the four-wave mixing process then mixes this mode into dipolar surface modes on both the signal and the idler transition. If this mixing is into a superradiant surface mode on the idler transition, then the decay of the atomic excitation is predominantly into a few photonic modes with specific wavevectors parallel to the array plane and a frequency near the idler transition frequency.

Thus, an incoming photon with frequency near the signal transition frequency may be efficiently transduced into a few, photonic evanescent modes with frequency near the idler transition frequency if, dependent on its wavevector and polarization, it fulfills both a resonance and a criticality condition. The resonance condition stems from resonantly coupling the incoming photon to a surface mode on the signal transition, while the criticality condition demands that the idler surface mode be superradiant. These transduced photonic modes may then be used for further processing using existing quantum devices.

In the following, we describe the array system from first principles, including the relevant level scheme of the individual atoms and the coherent population trapping using coherent lasers along with the subsequent four-wave mixing and drive grating \cite{Lukin2000} (\cref{sect:system}). To determine which photonic modes an incoming plane-wave photon is processed/scattered into, we use a scattering operator formalism that maps an in-state to its out-state (\cref{sect:methods}). The scattering operator formalism uncovers the analytic structure of the dipolar surface modes which arise due to the signal and idler transitions at every atomic site. These modes yield instability conditions that enable predominant scattering of incoming signal photons into specific idler photon modes and we relate these conditions to existing literature on dipolar excitations (\cref{sect:methods}).

Our approach from first principles may equivalently be described using a purely dipolar approach, in which the photonic degrees of freedom have been eliminated \cite{Asenjo-Garcia2017}. This, however, comes with the caveat that while the effect of photons on the array is explicit, the effect of the array on incoming photons is not transparent. Both approaches hold similar levels of complexity. Similarly, we find that in the single-photon regime, equivalent scattering behavior may be obtained in a semi-classical approach using Maxwell's equations for an infinite array of dipoles with polarizabilities obtained from steady state Maxwell-Bloch equations (\cref{sect:methods}).  Analyzing the behavior of transduction efficiencies near the critical scattering instabilities we illuminate how the scattering  of the incoming photon predominantly goes into a few critical photon modes and how these instabilities depend on the direction of the wavevector and the polarization of the incoming photon (\cref{sect:results}). These critical modes are confined to the array plane and propagate into specific in-plane directions. 

Matching both the resonance condition on the signal transition and the criticality condition on the idler transition, both at normal and angled incidence we find transduction efficiencies  into a \emph{single} specific, critical in-plane mode of up to $1/2$. The overall transduction efficiency into \emph{any} idler mode can be higher than $1/2$ and may be closer to $1$. To assess the effects of finite rather than infinite arrays, we complement these efficiencies with scattering calculations for finite arrays. We find that when matching the resonance and the criticality condition, there still is residual scattering into non-critical modes and the in-plane scattering is not perfectly confined but rather forms a scattering lobe. Increasing the number of emitters $N^2$ in the array, the confinement of the lobe to the array plane improves as $1/\sqrt{N}$.  Finally, we discuss how, by placing detectors and optical fibers at the edge of the array, devices such as the one proposed here may be used in sensing applications and we comment on our findings (\cref{sect:discussion}). 

\section{System}\label{sect:system}
We consider an infinite, periodic array of identical, point-like atoms, which can be polarized (see \cref{fig:setup}). This array is contained within the $x$-$y$ plane at $z=0$. Each atom features an effective level structure of a double-$\Lambda$-system, consisting of two, rotationally symmetric $\ell=0$ states $g_1$, $g_2$ and two   manifolds $(e_{1})_{n}$ and $(e_{2})_{n}$ of  oriented excited states with angular momentum $\ell=1$, where $n=x,y,z$ denotes the spatial polarization directions  of the excited state. 

The upper $\Lambda$-system consists of $g_1$, $g_2$ and the $e_1$ manifold and features the signal transition $g_1 \leftrightarrow e_1$ and the idler transition $g_2 \leftrightarrow e_1$. Due to the weak fields on the signal and idler transition,    the $g_1 \leftrightarrow e_1$ (signal) and $g_2 \leftrightarrow e_1$  (idler) transitions couple to a quantized electromagnetic field, where for simplicity  we assume that they couple to two different photon fields in free space, whose creation operators are denoted by $a^{\dagger}$ and $b^{\dagger}$.

The lower $\Lambda$-system, consisting of $g_1, g_2$ and $e_2$ is driven by two strong lasers with wavevectors $\kv_{L_1}$, $\kv_{L_2}$ and Rabi frequencies/polarizations $(\mathbf{\Omega}_{1})_{n}, (\mathbf{\Omega}_{2})_{n}$ on the $g_1\leftrightarrow e_2$ and $g_2\leftrightarrow e_2$ transitions, respectively.  This coherently traps any superposition between $g_1$ and $g_2$ in the corresponding dark state as perturbations on the signal and idler fields, $a^\dagger$ and $b^\dagger$, are extremely weak in comparison \cite{Lukin2000}. We thus project the ground state manifold $g_1, g_2$ onto the corresponding dark state such that under the rotating-wave approximation and in a rotating frame (setting $\hbar= c= \epsilon_0=1$) we obtain an effective Hamiltonian
\begin{align}
    \hat{H}= \hat{H}_0 + \hat{V},\label{Ham}
\end{align}
with
\begin{align}
    \hat{H}_0&=  \sum_{\kv\mu} \left(\omega_\kv- \omega_{e_1}+ \omega_{g_1}\right) a^{\dagger}_{\kv\mu} a^{\phantom{\dagger}}_{\kv\mu}\nnl
    &+  \sum_{\kv\mu} \left(\omega_\kv- \omega_{e_1}+ \omega_{g_2}\right)  b^{\dagger}_{\kv\mu} b^{\phantom{\dagger}}_{\kv\mu},
    \end{align}
    and
    \begin{align}
    \hat{V}&=   \frac{A^* }{\sqrt{L^3}} \sum_{\kv\mu, i,n}  g_{\kv\mu,n,a} a_{\kv\mu} e^{i (\kv+\kv_{L_2}) \rv_i } \ket{e_{1,i,n}}\bra{\text{dark}_i}\nnl%+ (h.c.)\nnl
    &+   \frac{B^*}{\sqrt{L^3}} \sum_{\kv\mu, i,n}   g_{\kv\mu,n,b} b_{\kv\mu} e^{i (\kv+\kv_{L_1} ) \rv_i }  \ket{e_{1,i,n}}\bra{\text{dark}_i} \nnl
    &+ (h.c.), 
\end{align}
where we refer to $\hat{H}_0$ as the non-interacting Hamiltonian and to $\hat{V}$ as the interacting part.
Here $L^3$ denotes the system volume, and $\kv,\mu$ denote a photon's wavevector and polarization, while $\omega_{\kv}= |\kv|$ denotes its energy. The energies of the ground and excited states are given by $\omega_{g_1}, \omega_{g_2}$ and $\omega_{e_1}, \omega_{e_2}$, while $\rv_i$ denotes the position of atom site $i$. The coupling constant is given by $g_{\kv\mu,n,\sigma}=\wp_{\sigma} \vec{\mathbf{e}}_n \cdot \vec{\mathbf{\epsilon}}_{\kv\mu} \sqrt{\frac{\omega_\kv}{2  }}$, where $\wp_{\sigma}, \sigma \in \{a,b\}$ denote the dipole moments of the $e_1 \leftrightarrow g_1$ and $e_1 \leftrightarrow g_2$ transitions, respectively, $\vec{\mathbf{\epsilon}}_{\kv\mu}$ is a unit polarization vector and $\vec{\mathbf{e}}_n$ is a unit vector in direction $n=x,y,z$. 

In order for the ground state manifold to be coherently trapped in a dark state superposition, in writing \cref{Ham} we have assumed that the driving lasers share the same polarization such that 
\begin{align}
    \sum_{n} (\mathbf{\Omega}_{1})_{n}   \ket{e_{2,i,n}}&= \Omega_1 \ket{e_2,i },\nnl
     \sum_{n} (\mathbf{\Omega}_{2})_{n}   \ket{e_{2,i,n}}&= \Omega_2 \ket{e_2,i },
\end{align}
and at every lattice site $i$ the dark state $ \ket{\text{dark}_i}$ of the lower $\Lambda$-system is given by
\begin{align}
   \ket{\text{dark}_i} &= \frac{\Omega_2 e^{i \kv_{L_2} \rv_i }\ket{g_{1,i}}-\Omega_1 e^{i \kv_{L_1} \rv_i }\ket{g_{2,i}}}{\sqrt{|\Omega_1|^2 + |\Omega_2|^2}}\nnl
    &= A^{*} e^{i \kv_{L_2} \rv_i }\ket{g_{1,i}}+ B^{*} e^{i \kv_{L_1} \rv_i }\ket{g_{2,i}}. 
\end{align}
As a result, rather than individually coupling to the signal or the idler transition, the signal ($a^\dagger$) and idler ($b^\dagger$) photons both couple to a transition $\text{dark} \leftrightarrow e_1$ between the dark state and the $e_1$ excited state. This enables the four-wave mixing process between the different transitions. The $a^\dagger$ and $b^\dagger$ photons mediate the dipole-dipole interaction between different atomic sites. One can treat the photons explicitly, or one can equivalently integrate out the two photonic fields to arrive at an effective Hamiltonian of interacting atomic emitters \cite{Asenjo-Garcia2017}. Photons may then be coupled into or out of the system using input-output relations. As we study the fate of photons which are incident on the array system, interact with it, and then scatter off it, we here choose to treat the photonic fields explicitly. This explicit treatment makes the effect on incoming photons more transparent. As we will see in the following sections, the wavevectors of the driving lasers $\kv_{L_1}, \kv_{L_2}$ induce a drive grating \cite{Hemmer1995} which displaces the in-plane momentum of the transduced photon mode by the difference of the in-plane components of the wavevectors $\kv_{L_1}$ and $\kv_{L_2}$.

Throughout this paper we will repeatedly make use of the following index conventions: $i,j$ will denote atomic sites, $n,m,o$ denote spatial orientations in $x,y$ or $z$-direction, $\kv, \kv', \kv''$ denote three dimensional wavevectors, while their in- and out-of-plane components are denoted by $\kvp$ and $\kv_\perp$. Within subscripts, $\sigma$ and $\sigma'=a,b$ refer to the corresponding transitions, where $\kv_{L_\sigma}$ denotes $\kv_{L_a}= \kv_{L_1}$ and $\kv_{L_b}= \kv_{L_2}$ and $\omega_\sigma$ denotes $\omega_a=\omega_{g_1}$ and $\omega_b= \omega_{g_2}$, while outside of subscripts  $\sigma,\sigma'$ denote $A$ and $B$. Polarization directions are denoted by $\mu$ and $\mu'$, i.e. in $\vec{\mathbf{\epsilon}}_{\kv\mu}$ they each denote the two directions orthogonal to $\kv$. Finally, $\pv$ will denote reciprocal lattice vectors which, for a square lattice of spacing $d$, are contained in $\pv \in (2\pi/d) \mathbb{Z}_2$.

\section{Methods}\label{sect:methods}
Having specified the emitter array along with the atomic structure of the individual emitters, in the following we study the physical process in which a signal photon is incident on the emitter array. To evaluate the scattering of incoming signal photons of type $a$ along with the transduction of said photons into idler photons of type $b$, we consider the scattering of photons from the lattice structure using a scattering operator $\hat{S}$ \cite{TaylorBook}. Given an incoming photon state at $t\to -\infty$, long before the scattering process, the scattering operator $\hat{S}$ can be used to determine which outgoing states at $t\to \infty $ this photon state will be scattered to.  We go over the scattering approach and its application to transduction in \cref{methods_Smatrix} and \cref{SApproachAndTransduction}. In \cref{DipolarVSPhotonModes}, we connect our approach to previous works \cite{Asenjo-Garcia2017, Shahmoon2017}.

It turns out that in the single-excitation limit considered here, the quantum-mechanical $\hat{S}$-operator treatment can be equivalently related to a solution of Maxwell's equations for classical dipoles with (cross)-polarizabilities obtained from single-atom steady-state Maxwell-Bloch equations. This alternative approach will be introduced in \cref{Maxwell_methods}. Both approaches hold a similar level of complexity as they contain the same Green's functions. These approaches are only equivalent in the linear/single-excitation limit, as the semi-classical treatment does not account for optical saturation/excitation blockade. While the scattering operator approach is explicit about the quantum state of a scattered photon and allows the study of the analytical properties of the scattering process, the  semi-classical approach is not as transparent about the state of the scattered photon. However, the semi-classical approach can easily be solved numerically for finite, rather than infinite, emitter arrays. This is particularly relevant for the question of experimental feasibility, as the mode picture, developed throughout this paper, only becomes exact in the limit of infinite emitters, while for finite emitter arrays it is only approximate.  In both approaches, for simplicity, we will disregard decay channels that are not into the two electric field operator modes $a$ and $b$. These decay channels may be added manually, by adding appropriate decay widths in the dipolar self-energies  or the cross-polarizabilities used in \cref{Maxwell_methods}.

\subsection{The scattering operator}\label{methods_Smatrix}
In our scattering approach, we assume the driven lattice structure is prepared in its zero-excitation ground state $\ket{0}= \bigotimes_i \ket{\text{dark}_i}$  where every atom site $i$ is prepared in its dark state. We then consider the scattering of an incoming signal photon of type $a$ with wavevector $\kv$ and polarization $\mu$ off the lattice structure which is prepared in $\ket{0}$. To this end, we define the in/out-states (which are eigenstates of the non-interacting Hamiltonian $\hat{H}_0$)
\begin{align}
\ket{\kv\mu a} &= a^{\dagger}_{\kv\mu} \ket{0},\nnl
\ket{\kv\mu b} &= b^{\dagger}_{\kv\mu} \ket{0},
\end{align}
which gives corresponding asymptotes $e^{-i \hat{H}_0 t}\ket{\kv\mu a}$ and $e^{-i \hat{H}_0 t}\ket{\kv\mu b}$ under time-evolution with the non-interacting Hamiltonian $\hat{H}_0$. 

Let us consider the scattering of an incoming photon of type $a$ with wavevector $\kv$ and polarization $\mu$ and let us furthermore presume that the wavefunction of this photon may be written as $\ket{\psi(t)}= e^{-i \hat{H}t }\ket{\psi(0)}$. Then, at $t \to -\infty$ this photon's  time evolution is effectively governed by the non-interacting Hamiltonian and it behaves like an asymptotic state:  $\ket{\psi(t)}\xrightarrow{t\to - \infty} e^{-i \hat{H}_0 t}\ket{\kv\mu a} $, such that $\ket{\psi(0)}=\lim_{t\to \infty} e^{- i \hat{H} t} e^{i \hat{H}_0 t}\ket{\kv\mu a}$.  Similarly, at $t'\to \infty$ the scattered state  $\ket{\psi(t)}$ evolves like a superposition of asymptotic states:
\begin{align}
    &\ket{\psi(t')}= e^{-i \hat{H}t' }\ket{\psi(0)}\nnl &\xrightarrow{t'\to  \infty}  \sum_{\kv' \mu'} c_{\kv' \mu' a }   e^{-i \hat{H}_0 t'}\ket{\kv'\mu' a} + c_{\kv' \mu' b }   e^{-i \hat{H}_0 t'}\ket{\kv'\mu' b}.
\end{align}
Thus, defining the $\hat{S}$ operator as 
\begin{align}
\hat{S} 
&= \lim_{t\to\infty}e^{i \hat{H}_0 t}  e^{- 2 i  \hat{H} t} e^{i \hat{H}_0 t}, \label{Sdef}
\end{align} one may compute
$c_{\kv' \mu' \sigma }=\bra{\kv'\mu' \sigma} \hat{S} \ket{\kv\mu a} $, $\sigma \in \{a,b\}$. Hence, an incoming state that at $t\to -\infty$ behaves like the asymptote of $\ket{\kv \mu a}$, is scattered onto a state that at $t\to\infty$ behaves like the asymptote of $\hat{S}  \ket{\kv\mu a}$ \cite{TaylorBook}:
\begin{align}
     \ket{\kv\mu a} \xrightarrow[]{\text{scattering}} \hat{S}  \ket{\kv\mu a}.
\end{align}

From \cref{Sdef}, one may show \cite{TaylorBook}  that given eigenstates $\chi$ and $\phi$ of the free Hamiltonian $\hat{H}_0$ with  energies $E_{\chi/\phi}$, the scattering matrix elements of $\hat{S}$ can be computed as 
\begin{align}
    \bra{\chi} \hat{S} \ket{\phi} = \bra{\chi}  \ket{\phi}+ \frac{1}{2}  \bra{\chi} \left[ \hat{V}\frac{1}{\rho - \hat{H}}+\frac{1}{\rho - \hat{H}} \hat{V}\right]\ket{\phi}, \label{Smatrix}
\end{align} where $\rho= (E_{\phi}+ E_{\chi})/2 +i 0^+$. To solve for the scattering matrix elements $\bra{\kv'\mu', \sigma} \hat{S}  \ket{\kv\mu a} $, we  solve for the matrix elements of the retarded Green's function $(\rho -\hat{H})^{-1}$ using the resolvent identity
\begin{align}
    \frac{1}{\rho-\hat{H}}= \frac{1}{\rho-\hat{H}_0}+ \frac{1}{\rho-\hat{H}} \hat{V} \frac{1}{\rho-\hat{H}_0}\label{resolventidentity}. 
\end{align}

As we have taken the rotating-wave approximation, the Hamiltonian is excitation-preserving and thus for an in-state with a single excitation one obtains a closed, self-consistent system of equations that can be solved explicitly. For example, consider a matrix element of the form $\bra{\kv'\mu'b} \frac{1}{\rho-\hat{H}} \ket{\kv\mu a}$: applying the resolvent identity \cref{resolventidentity}, one finds that this matrix element is related to a superposition of matrix elements $\bra{\kv'\mu'b} \frac{1}{\rho-\hat{H}} \ket{i,n}$, where $\ket{i,n}= \ket{e_{1,i,n}} \bigotimes_{j\neq i } \ket{\text{dark}_{j}}$. Applying the identity once again, one obtains that these matrix elements in turn are given by superpositions of $\bra{\kv'\mu'b} \frac{1}{\rho-\hat{H}} \ket{\kv''\mu'' a}$ and $\bra{\kv'\mu'b} \frac{1}{\rho-\hat{H}} \ket{\kv''\mu'' b}$. Solving  for the matrix elements of $\frac{1}{\rho-\hat{H}}$ and using these solutions within \cref{Smatrix}  one obtains the matrix elements of the $\hat{S}$ operator, where for $\bra{\kv'\mu'b} \hat{S} \ket{\kv\mu a}$ one may use $\rho= \omega_\kv -  \omega_{e_1} + \omega_{g_1} + i 0^+ $.  Explicit expressions of these matrix elements for a square lattice of lattice constant $d$ can be found in \cref{app_explicitcalculation} and the action of the $\hat{S}$ operator on the in-state $\ket{(\kvp,k_\perp) \mu a}$ is given by:
\begin{align}
&\hat{S} \ket{(\kvp,k_\perp) \mu a}=\ket{(\kvp,k_\perp) \mu a}\nnl 
&-\frac{ i }{ d^2}\sum_{_{\substack{ \sigma=a,b \\ s=+,-}}  }\sum_{\mu' \pv nm}    \Bigg[ \Theta\left(\omega_\kv+ \omega_{g_1}- \omega_{\sigma}-|\kvp +\pv+\kv_{L,\sigma a}  |\right) \nnl 
&\ \ \ \ \ \ \ \ \ \ \ \ \times \frac{A^* \sigma g_{\kv \mu, n,a } g_{\kv'(\pv, s,\sigma) \mu',m, \sigma} (\omega_{\kv}+ \omega_{g_1}- \omega_{\sigma})}{\sqrt{ -(\kvp +\pv+ \kv_{L, \sigma a} )^2 + \left(\omega_{\kv}+ \omega_{g_1}- \omega_{\sigma}\right)^2} } \nnl
&\ \ \ \ \ \ \ \ \ \ \ \ \times \left( \rho \mathbb{I}- \rho P(\kvp,a)\right)^{-1}_{n,m} \ket{\kv'(\pv, s,\sigma) \mu' \sigma}   \Bigg]. \label{Soperatormain}
\end{align}Here, 
\begin{align}
    \kv'(\pv, s,\sigma)_\parallel&= \kvp +\pv+ \kv_{L, \sigma a},\nnl
     \kv'(\pv, s,\sigma)_\perp&=s \sqrt{ -(\kvp + \pv+ \kv_{L, \sigma a})^2 + \left(\omega_\kv+ \omega_{g_1}- \omega_{\sigma}\right)^2}\label{k_scattered}
\end{align}
is the wavevector of an outgoing photon, $\kv_{L,\sigma\sigma'}= (\kv_{L_\sigma}- \kv_{L_{\sigma'}})_\parallel$ gives the in-plane component of the drive grating \cite{Hemmer1995},  the $\pv$ denote reciprocal lattice vectors $\pv \in (2\pi/d) \mathbb{Z}_2$, $s\in\{+,-\}$, and
\begin{align}
     &P(\kvp,a)_{m,n}= \sum_\sigma D(\kvp +\kv_{L, \sigma a},\sigma)_{m,n} \nnl
     &=\sum_{\substack{\sigma\pv\\ k_\perp \mu }}  \frac{|\sigma|^2}{L \rho d^2}\frac{g_{(\kvp+\pv+ \kv_{L,\sigma a}, k_\perp)\mu,n,\sigma} g_{(\kvp+\pv+ \kv_{L,\sigma a}, k_\perp)\mu,m,\sigma} }{\rho- (\omega_{(\kvp+\pv+ \kv_{L,\sigma a}, k_\perp)}- \omega_{e_1}+ \omega_{\sigma})} \label{in-planeFourier}.
\end{align} 
Note that due to the Heaviside $\Theta$ appearing in \cref{Soperatormain}, the perpendicular component of $\kv'(\pv, s,\sigma)$ (\cref{k_scattered}) is always real, $\kv'(\pv, s,\sigma)_\perp \in \mathbb{R}$. Furthermore, we point out that $D$ (\cref{in-planeFourier}) has a component $\propto 1/ \rho$, such that $\rho=0+i0^+$ (resonance) does not imply a vanishing $\rho D$.

As can be seen in \cref{Soperatormain,k_scattered}, an incoming $a$ photon with wavevector $(\kvp, k_\perp)$ and polarization $\mu$ is scattered into both $a$ and $b$ photons. The in-plane momentum of the scattered photons $\kv'(\pv, s,\sigma)_\parallel$ is given by the in-plane momentum of the incoming photon $\kvp$ displaced by a reciprocal lattice vector $\pv$ and by the grating difference $\kv_{L, \sigma a }$ (which vanishes when scattering into $a$ photons, $\kv_{L,aa}=\zerov$). The out-of-plane momentum $\kv'(\pv, s,\sigma)_\perp$ is obtained from conservation of energy, taking into account the energy difference between the two ground states when scattering into $b$ photons.

\subsection{The scattering approach and transduction}\label{SApproachAndTransduction}

By treating the photonic fields explicitly, the scattering approach is well suited to analytically analyze the transduction properties of infinite emitter arrays. In the following sections, we explain how the scattering approach can be used to analyze \emph{transduction efficiency} and \emph{mode selectivity}.

\subsubsection{Transduction efficiency}\label{methods_efficiency}
We define \emph{transduction efficiency} as the probability for an incoming signal photon to be scattered into one of the desired outgoing photonic modes. In particular, from \cref{Soperatormain}, we can determine the probability for an incoming photon $\ket{(\kvp, k_\perp) \mu a}$ to be scattered into any $b$ mode as well as the probability for an incoming photon to be scattered into specific $b$ modes. 

Projecting the scattering operator $\hat{S}$ into its $a$ and $b$ subspaces (\textit{i.e.} setting either $\sigma=a$ or $\sigma=b$ in \cref{Soperatormain}, respectively), the operator  can be decomposed as $\hat{S}=\hat{S}_a+ \hat{S}_b$ and the overall transduction efficiency from $a \to b$ is given by (for details see \cref{App_a_to_b_efficiency}):
\begin{align}
   \Big| \hat{S}_b &\ket{(\kvp,k_\perp) \mu a}\Big|^2 =\frac{ |A|^2    \omega_{\kv}}{ k_\perp d^2} \sum_{ nmn'm'}  \Big[ g_{\kv \mu, n,a }g_{\kv \mu, n',a } \nnl
   &\times 2 i \Im \rho D(\kvp+ \kv_{L,ba},b)_{m, m'} \nnl
   & \times\left( \rho \mathbb{I}- \rho P(\kvp,a)\right)^{-1}_{n,m} \left( \rho \mathbb{I}- \rho P(\kvp,a)\right)^{*,-1}_{n',m'}\Big] \label{conv_efficiency_atob}
\end{align} and the probability to remain in any $a$ mode is given by:
\begin{align}
     \left| \hat{S}_a\ket{(\kvp,k_\perp) \mu a}\right|^2 =1-\left| \hat{S}_b\ket{(\kvp,k_\perp) \mu a}\right|^2. 
\end{align}

Similarly, to determine the probability to be scattered into specific $b$-modes, one can define a projection operator $\hat{P}_{\text{specific}}$ onto specific $b$-modes and compute the transduction efficiency into these specific modes as: 
\begin{align}
    \left|\hat{P}_{\text{specific}} \hat{S}_b \ket{(\kvp,k_\perp) \mu a}\right|^2. 
\end{align}
    \label{specific_efficiency}
Since we will typically parametrize the momentum of the outgoing modes  $\kv'(\pv, s,b)$ by the reciprocal lattice vectors $\pv$, computation of these specific transduction probabilities will amount to a restriction of the sum over $\pv$ in \cref{Soperatormain}; for example, to $|\pv|$ smaller than a certain threshold value. This will translate to a modified computation of $\Im \rho D(\kvp+ \kv_{L,ba},b)_{m, m'}$ within \cref{conv_efficiency_atob}. For details see \cref{App_a_to_b_efficiency}.

\subsubsection{Scattering instabilities and mode selectivity}\label{methods_instabilities}
We define \emph{mode selectivity} as the ability of the transduction scheme to individually target specific outgoing photonic modes, characterized by wavevector and polarization. The scattering approach yields \emph{scattering instabilities} of the emitter array which can be used to analyze the mode selection properties of the array. In the following, we analyze the analytical structure of the action of the $\hat{S}$ operator in \cref{Soperatormain} to illuminate how mode selectivity arises in the scattering process on a technical level and in \cref{DipolarVSPhotonModes}  we will relate this to the underlying dipolar surface modes.

The $\hat{S}$ operator (\cref{DipolarVSPhotonModes}) scatters incoming $a$ photons with wavevector $(\kvp, k_\perp)$ into $a$ and $b$ photons with wavevectors $\kv'(\pv, s,\sigma)$, $\sigma$=$a,b$, $s$=$+,-$ whose in-plane component   $\kv'(\pv, s,\sigma)_\parallel$ is shifted from $\kvp$ by a reciprocal lattice vector $\pv\in (2 \pi/d) \mathbb{Z}_2$ and the grating of the driving lasers $\kv_{L, \sigma a}$. This wavevector is energy conserving, i.e. $\omega_{(\kvp, k_\perp)}+ \omega_{g_1}=  \omega_{\kv'(\pv, s,\sigma)}+ \omega_\sigma$, and furthermore only modes with $\kv'(\pv, s,\sigma)_\perp \in \mathbb{R}$ are allowed for out-states, as only modes with $\kv'(\pv, s,\sigma)_\perp \in \mathbb{R}$ are solutions to Maxwell's equations in free space and can thus radiate away from the array. For this reason, we refer to such modes as \emph{free space}, \emph{radiant}, or \emph{light-like} modes. For small values of $ \omega_{\kv'(\pv, s,\sigma)} d $,  this severely restricts the set of outgoing modes. If only $\pv=0$ yields an outgoing mode, then the array is fully cooperative with respect to the resonance. For larger values of $ \omega_{\kv'(\pv, s,\sigma)} d $, the number of allowed modes that can be scattered into increases, \textit{i.e.}  the degree of cooperativity decreases, and eventually  there is a quasi-continuum of allowed modes that can be scattered into. 
Within the allowed modes to scatter into, from  \cref{Soperatormain,k_scattered} one can see that there exist incoming wavevectors $(\kvp, k_\perp)$ for which one can find suitable wavevectors $\pv \in (2\pi/d) \mathbb{Z}_2$   such that $|\kv'(\pv, s,\sigma)_\perp |\to 0^+$, rendering the fraction in the expression (\cref{Soperatormain}) unbounded. As the scattering process is unitary, this suggests that a similar divergence occurs within $P(\kvp,a)$, preventing some terms in \cref{Soperatormain} from diverging and causing other terms without an unbounded fraction to vanish. This concentrates the probability of scattering into particular modes, giving rise to mode selectivity.

The matrix-valued function $P(\kvp, a)$ in \cref{in-planeFourier} is related to the dyadic Green's function \cite{Novotny2012} within the rotating-wave approximation. Carrying out an in-plane Fourier transform to real space of the $D$ function appearing in \cref{in-planeFourier}, one finds that $ \rho D(\rv, \sigma)= (\rho+ \omega_{e_1}- \omega_\sigma)^2 |\sigma|^2 \wp_\sigma^2  K^{+} (\rv,\rho+ \omega_{e_1}- \omega_\sigma)$.  The $K^{+}$ is obtained from the classical dyadic Green's function $G(\rv, k)= K^+ (\rv, k)- K^-(\rv, k)$ upon carrying out the rotating-wave approximation \cite{Novotny2012,J_rgensen_2022} (for further details see \cref{App_Dyadic_Green}). The imaginary parts of both functions coincide, $\Im G= \Im K^+$ \cite{J_rgensen_2022}, and a finite, closed expression exists (see \cref{App_Dyadic_Green}). The real parts of their in-plane Fourier transforms differ and need to be regularized manually by setting $\Re G(\rv=\zerov,\omega)=\Re K^+(\rv=\zerov, \omega)=0 $. As mentioned in \cref{sect:system}, one may choose to integrate out the photon fields. In that case, the dyadic Green's function $G$ gives the interaction between different dipole emitters \cite{Asenjo-Garcia2017}. 

As suggested above, a divergence similar to the unbounded fraction in \cref{Soperatormain} occurs within $P(\kvp,a)$. The eigenvectors and eigenvalues of $P(\kvp,a)$ can be related to the polarizations, energies and decay rates of dipolar excitations,  while $\left( \rho \mathbb{I}- \rho P(\kvp,a)\right)^{-1}$ is proportional to the propagator of dipolar excitations. Like its counter-part that is obtained when the rotating-wave approximation is not taken,  the dipolar self-energy  $\rho P(\kvp,a)$ has instabilities for which its entries diverge and as a result so do its eigenenergies. In \cref{App_Dyadic_Green}, we show that these instabilities are governed by the same instability conditions, whether the rotating-wave approximation is undertaken or not.  These instabilities occur whenever there exists a suitable reciprocal wavevector $\pv \in (2\pi/d) \mathbb{Z}_2$ such that $(\rho+ \omega_{e_1}- \omega_\sigma)^2-(\kvp+\kv_{L,\sigma\sigma'} + \pv)^2$ becomes arbitrarily close to $0$. As a result, the self-energy $\rho P(\kvp, a)$ has a contribution with an out-of-plane wavevector component that is approaching the critical point $k_\perp= \pm \sqrt{(\rho+ \omega_{e_1}- \omega_\sigma)^2-(\kvp+\kv_{L,\sigma\sigma'} + \pv)^2} \to 0$ between free space modes $k_\perp\in \mathbb{R}$, which may propagate in vacuum, and evanescent modes $k_\perp\in i \mathbb{R}$, which may not propagate in vacuum (see \cref{App_Dyadic_Green}).  While the fraction in \cref{Soperatormain} may only have such an instability result from a free space mode approaching this point, instabilities in  $P(\kvp,a)$ may come from both free space or evanescent modes nearing the critical point. 

These  instability conditions for $P(\kvp,a)$ are derived from \cref{in-planeFourier} by carrying out the $k_\perp$-integral using contour integration. For details see \cref{App_Dyadic_Green}. In the following, we sketch a quick explanation of the origin of the instability condition. The denominator in \cref{in-planeFourier} has a pole at $k_\perp= \pm \sqrt{(\rho+ \omega_{e_1}- \omega_\sigma)^2-(\kvp+\kv_{L,\sigma\sigma'} + \pv)^2} $ which may be located close to the real or the imaginary axis. Evaluating the residue of this pole yields a contribution to the imaginary (real) part of $\rho P(\kvp, a)$ if $k_\perp$ is close to the real (imaginary) axis. In either case, it results in a contribution that has a component $\propto 1/k_\perp $, which leads to a  divergence in $\rho P(\kvp, a)$   when   $k_\perp=\pm \sqrt{(\rho+ \omega_{e_1}- \omega_\sigma)^2-(\kvp+\kv_{L,\sigma\sigma'} + \pv)^2 }\to 0 $, corresponding to coupling/decay to photons propagating parallel to the array. 

In summary, whenever there is a divergence in $\rho P(\kvp, a)$, only terms within \cref{Soperatormain} that have a balancing divergence coming from an unbounded fraction can remain non-vanishing. All other terms vanish and thus the number of modes that are scattered into is greatly reduced to those which fulfill the criticality condition. In \cref{DipolarVSPhotonModes} we relate the origin of the divergences in $\rho P(\kvp, a)$ to the underlying dipolar modes.

\subsection{Connections between dipolar modes and photon modes}\label{DipolarVSPhotonModes}

The action of the $\hat{S}$ operator on the in-state $\ket{(\kvp,k_\perp) \mu a}$ in \cref{Soperatormain} is closely related to the collective dipole modes studied in Refs. \cite{Shahmoon2017,Asenjo-Garcia2017}. Due to the resonant four-wave mixing, an incoming photon couples to the dipole modes associated with the $e_1 \leftrightarrow g_1$ transition as well as the $e_1 \leftrightarrow g_2$ transition. These dipole modes may then propagate through the system, before eventually coupling to a photonic out-state which propagates away from the lattice. This structure can be seen in \cref{Soperatormain}: the coupling vertices between the in-/out-photons and the dipole modes are  proportional to $g_{\kv \mu, n,a }$ and $g_{\kv'(\pv, s,\sigma) \mu',m, \sigma}$. 
Similarly, the inverse matrix $\left( \rho \mathbb{I}- \rho P(\kvp,a)\right)^{-1}$ represents the Green's function of a dipole excitation  with wavevector $\kvp$ while for $\sigma=a,b$ the $\rho D(\kvp +\kv_{L, \sigma a},\sigma)$ are its self-energy corrections due to the two dipole transitions.

To make the connection between the dipolar modes --studied in previous works \cite{Shahmoon2017,Asenjo-Garcia2017}-- and the photon modes --studied here-- more explicit, we note that for a given transition $\sigma=a$ or $\sigma =b $, a dipolar excitation with  in-plane momentum  $\kv'_{\parallel}$  in the first Brillouin zone is made up of photonic modes with  $\kv'(\pv, s,\sigma)_\parallel \in  \left\{\kv'_{\parallel}+ \pv | \pv \in (2 \pi /d) \mathbb{Z}_2 \right\}$. If  all such modes have that $\kv'(\pv, s,\sigma)_\perp \in i \mathbb{R} $, then the corresponding dipolar mode lies outside the light cone. If there exist modes with $\kv'(\pv, s,\sigma)_\perp \in \mathbb{R} $, then the dipolar mode lies within the light cone. The lattice structure is cooperative with respect to one of the transitions, $\sigma$, if for some or all  dipolar excitations within the light cone  --parametrized by their  in-plane momentum  $\kv'_{\parallel}$-- there exist only two  photonic wavevectors that fulfill 
\begin{align}
&\kv'(\pv, s,\sigma)_\parallel \in  \left\{\kv'_{\parallel}+ \pv | \pv \in (2 \pi /d) \mathbb{Z}_2 \right\},\nnl
&\kv'(\pv, s,\sigma)_\perp \in \mathbb{R},\label{condition_cooperative}
\end{align}
 namely $\kv'(\pv=\zerov, +,\sigma)$ and $\kv'(\pv=\zerov,-,\sigma)$. With a decreasing degree of cooperativity, more photonic wavevectors with $\pv \neq \zerov, \ \pv \in  (2 \pi /d) \mathbb{Z}_2 $ that fulfill the conditions in \cref{condition_cooperative} become available. The importance of cooperativity in our scheme may be understood from the conditions in \cref{condition_cooperative}. Cooperativity limits the photonic modes that a dipolar modes can couple to, which increases overlap with individual photonic modes.
 
 As shown in Ref. \cite{Asenjo-Garcia2017}, near the light cone, light-like dipolar modes with a transversal polarization develop an unbounded decay rate. This can also be seen from  the expression for the imaginary part of $\rho D(\kvp, \sigma)$ in \cref{App_Dyadic_Green}, which is proportional to the decay rate. The decay is into specific photonic modes with $k_\perp= \pm \sqrt{(\rho+ \omega_{e_1}- \omega_\sigma)^2-(\kvp+\kv_{L,\sigma\sigma'} + \pv)^2} \to 0$, $k_\perp \in \mathbb{R}$. From the perspective of dipolar modes, the in-state of an $a$ photon couples to a specific light-like dipolar mode on the $a$ transition which contains many different  photonic  $ a$ modes, both evanescent and free-space. Through the coherent population trapping it furthermore couples to a dipolar mode on the $b$ transition with $\kvp + \kv_{L,ba}$ and energy $\omega_\kv + \omega_{g_1}- \omega_{g_2}$ (provided this energy is positive). If that dipolar excitation is light-like, then it will be able to scatter into the out-states as well. As we consider processes where $\omega_\kv + \omega_{g_1}- \omega_{g_2}$ is several times larger than $\omega_\kv$, in the context of our work the dipolar excitation on the $b$ transition is always light-like.

In \cref{sect:results} we will show that, when the system is near a critical point where $k_\perp= \pm \sqrt{(\rho+ \omega_{e_1}- \omega_\sigma)^2-(\kvp+\kv_{L,\sigma\sigma'} + \pv)^2} \to 0$ for select $b$ modes (i.e. these dipolar modes are near the light cone), scattering into these critical modes is strongly enhanced while scattering into all other $b$ modes is suppressed. Approaching such a critical point, the inverse propagator in \cref{Soperatormain} diverges and for specific values of $\mu'$ and $\pv$ so does the fraction in \cref{Soperatormain}. Effectively, this asymptotically leads to total internal reflection along the surface into well defined surface modes and corresponding electromagnetic modes. In \cref{sect:discussion}, we will elaborate on how these  populated modes may then be used for efficient quantum sensing  using four-wave mixing. While previous treatments of emitter arrays have made Markov-type approximations such as $\omega_{\kv}\approx \omega_{e_1}- \omega_{g_1}$ by replacing $\omega_{\kv}$ with $\omega_{e_1}- \omega_{g_1}$ in the dyadic Green's function \cite{Asenjo-Garcia2017} or throughout the scattered field expression \cite{Shahmoon2017}, we do not undertake such approximations as they are not necessary in the context of our work and may obscure the aforementioned instabilities.

\subsection{Classical Maxwell equations in steady-state}\label{Maxwell_methods}

 The action of the scattering operator $\hat{S}$ on an in-state $\ket{(\kvp, k_\perp)\mu a}$ (\cref{Soperatormain}) may be viewed as a quantized analog of a solution of classical Maxwell equations under the assumption that every 4-level emitter is in the single-emitter steady state  under the strong coherent drive. This steady state yields single-emitter (cross)-polarizabilities $\alpha_{\sigma \sigma'}$ which can be used to study the collective emitter behavior.  
 
 Identifying weak electric fields $\Ev_a$ and  $\Ev_b$ as the  $\omega_{(\kvp,k_\perp)}=k_a$ and $\omega_{\kv'}= k_b$  frequency components of the electric field, the corresponding steady-state (cross)-polarizations at the emitter sites $\rv_i$ are given by \cite{Lukin2000}:
\begin{align}
    \mathbf{P}_\sigma(\rv_i) &= \sum_{\sigma '}\alpha_{\sigma \sigma' } e^{i \kv_{L,\sigma\sigma'} \rv_i} \Ev_{\sigma'}(\rv_i), \\
    \alpha_{\sigma \sigma'}&=i \frac{\wp_\sigma \wp_{\sigma'} \sigma (\sigma')^{*} }{\frac{\gamma_a +\gamma_b}{2}- i \delta}, \label{polarizability}
\end{align}
where $\gamma_\sigma =- 2 k^2_\sigma \wp_\sigma^2 |\sigma|^2 \Im G(\rv=0, k_\sigma)$ \footnote{For the two treatments to be equivalently relatable, we use here the field frequency $k_\sigma$ instead of the atomic frequency $\omega_{e_1}- \omega_a$. Alternatively, carrying out a Markov approximation $(\omega_{e_1}- \omega_a)^3\approx k_\sigma^3 $ within the approach introduced in \cref{methods_Smatrix}, and using  atomic frequencies within $\gamma_\sigma$ here, these approaches may also be related equivalently.} denotes the vacuum decay rates and $\delta= k_a - (\omega_{e_1}- \omega_{a})= k_b - (\omega_{e_1}- \omega_{b}) $ denotes the detuning from the atomic transitions where the Lamb shift is absorbed into the atomic transition frequencies $\omega_{e_1}- \omega_{\sigma}$. We operate on 4-photon resonance. These equations are solved by:
\begin{align}
    \Ev_\sigma(\rv)= \Ev^{0}_\sigma (\rv) - k^{2}_{\sigma} \sum_{i} G(\rv-\rv_i, k_\sigma) \mathbf{P}_{\sigma} (\rv_i),\label{Maxwell_to_solve}
\end{align}
and in \cref{App_Maxwell} we show that upon taking the rotating-wave approximation, the solution for an incoming plane wave of type $a$,  $\Ev^{0}_{a}(\rv= (\rv_\parallel, z))=  \Ev_0 e^{i \kvp \rv_\parallel} e^{i k_\perp z}$, $\Ev^{0}_{b}=\zerov$,  can be equivalently related to  the action of the $\hat{S} $ operator on the in-state $\ket{(\kvp,k_\perp) \mu a}$ in \cref{Soperatormain}. Importantly, while \cref{Maxwell_to_solve} may be solved analytically (for details see \cref{App_Maxwell}) for an infinite emitter array, it may also be solved numerically for a finite emitter array of arbitrary constellation. As a result, this approach is well-suited to study the scattering properties of finite emitter arrays which will be done in \cref{maxwell_calcs}.

\section{Results}\label{sect:results}

In the following section, we analyze the transduction efficiency and scattering properties of the setup introduced in \cref{sect:system}. As such a system may be used to transduce an incoming signal from a frequency range such as the GHz-THz spectrum to, for example, optical wavelengths, we will typically study instances for which the array is cooperative, or near cooperative, with respect to the incoming signal, while for the transduced idler the degree of cooperativity is typically much lower.

 In order to achieve efficient, spatially directed transduction, the transduction process needs to achieve two goals. First, to enable efficient transduction, an incoming signal photon needs to be absorbed by the emitter array with a high probability. That is, the emitter array needs to have a sizable scattering cross section with an incoming signal photon. After coupling into the system, the four-wave mixing process then mixes the two types of dipolar modes and eventually some of these decay into idler photons. In \cref{sect_conv_efficiency}, we thus study how the overall transduction efficiency into idler photons depends on the dark state mixing ratio and the cooperativity of the array. Second, to achieve spatially directed transduction after coupling into the array, the decay into idler photons should predominantly occur into very few modes. In \cref{mode_selectivity}, we therefore study how mode selectivity arises through scattering into critical modes. Finally, in \cref{maxwell_calcs}, we study how the mode picture developed in  \cref{sect:methods} fares in finite rather than infinite emitter arrays.

 For simplicity, we will only present results for a vanishing grating \cite{Lukin2000} ($\kv_{L,ba}=\zerov$) and for incoming signal photons that are resonant with the signal transition ($\rho=i 0^+\leftrightarrow\omega_\kv= \omega_{e_1}- \omega_{g_1}$). The grating shifts the in-plane momentum of the outgoing transduced photon and can therefore be used to tune to critical transduced modes. Similarly, while the bare detuning $\delta$ enters the effective detuning, the effective detuning has additional contributions stemming from the dipolar modes that can be tuned by changing the lattice constant $d$.   Finally, throughout this section, we will show results in which the rotating-wave-approximation is not carried out within $\rho D$. As discussed in \cref{methods_instabilities}, this does not affect the criticality conditions and may merely induce a shift of the effective detuning. In \cref{RWA_no_RWA_comp}, we compare our results to results for which the rotating-wave-approximation was used and we find no qualitative and only faint quantitative differences between the methods.

\subsection{Overall transduction efficiency from $a$ photons into $b$ photons}
\label{sect_conv_efficiency}

\begin{figure}[t]
    \centering
    \includegraphics[width=\linewidth]{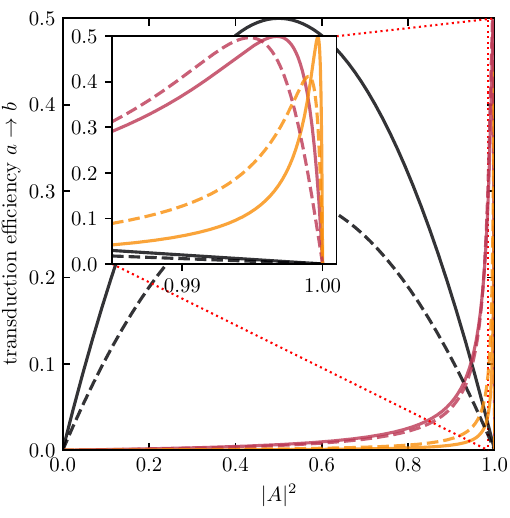}

    \caption{\textbf{Transduction efficiency at normal incidence for varying dark state mixing ratios $|A|$.} For a vanishing grating ($\kv_{L,ba}= \zerov$), the transduction efficiency $\left| S_b\ket{(\kvp,k_\perp) \mu a}\right|^2$ is shown for resonant photons at normal incidence as a function of the squared dark state mixing ratio $|A|^2$. Curves for $\omega_{\kv}= 0.2 \times (2\pi/d)$ (solid) and $\omega_{\kv}= 0.495 \times (2\pi/d)$ (dashed) are shown for $\omega_{g_1}- \omega_{g_2} =0$ (black), $\frac{5}{2} \times (2 \pi/d) $ (red) and $5 \times(2 \pi/d) $ (orange). The efficiency is peaked around a value of $|A|$ which approaches $|A|\to 1$ for increasing transduction energies. At normal incidence an efficiency of up to $\frac{1}{2}$ can be attained.}
    \label{trans_efficiency_Adependence}
\end{figure}

\begin{figure*}[ht!]
    \centering
 \begin{tikzpicture}
    \node[anchor=south west,inner sep=0] (image) at (0,0) {\includegraphics[width=\linewidth]{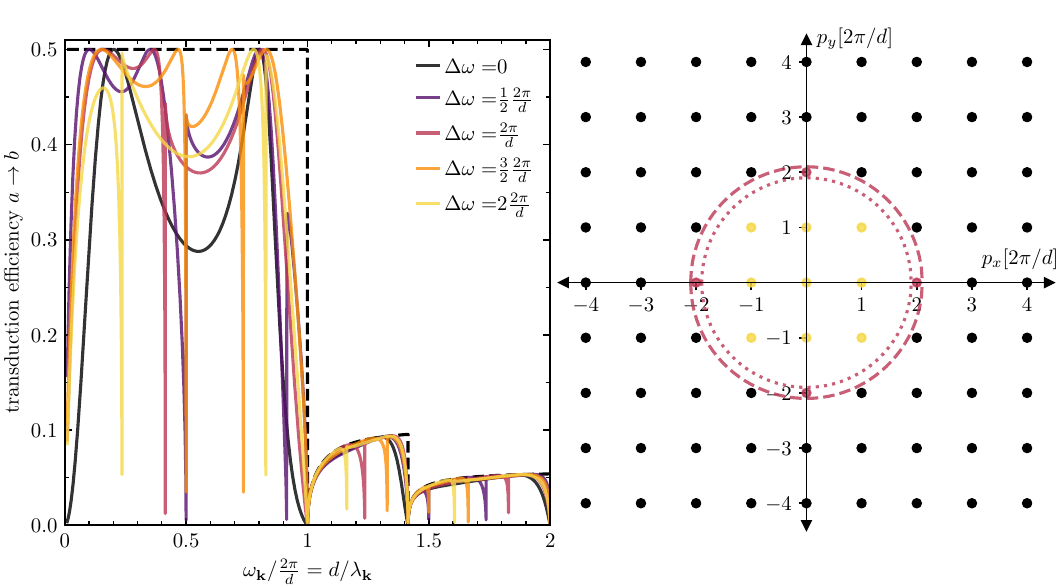}};
    \begin{scope}[x={(image.south east)},y={(image.north west)}]
            \node[align=right] at (0.54,0.96) {$\textbf{(b)}$};
        \node[align=right] at (0.01,0.96) {$\textbf{(a)}$};
        \fill[ fill=blue!80,  opacity=1] (0.06,.94) rectangle  (.292,.98);
        \fill[ fill=blue!80,  opacity=.7] (.292,.94) rectangle  (.386,.98);
        \fill[ fill=blue!80,  opacity=.5] (.386,.94) rectangle  (.522,.98);
        \node[align=right] at (.175,.995) {fully cooperative};
        \node[align=right] at (.405,.995) {decreasing cooperativity $\longrightarrow$};
\end{scope}
\end{tikzpicture}  
       \caption{\textbf{Transduction efficiency at normal incidence for  varying incoming photon energies.} (\textbf{a}) The transduction efficiency  of a resonant ($\rho=0+ i 0^+$) $a$ photon  into $b$ modes at the optimal value of $|A|$ is shown for different transduction energies $\Delta \omega= \omega_{g_1}- \omega_{g_2}$ as a function of the energy of the incoming photon $\omega_{\kv}$. The efficiency (solid) is shown for an $a$ photon at  normal incidence  ($\kvp=\zerov, k_\perp= \omega_{\kv}$) and a vanishing grating ($\kv_{L,ba}=\zerov$) for  transduction energies $\Delta \omega$ of $0$ (black), $\frac{1}{2}\frac{2\pi}{d}$ (purple), $\frac{2 \pi}{d}$ (red), $\frac{3}{2} \frac{2 \pi}{d}$ (orange) and $2 \frac{2\pi}{d}$ (yellow). At normal incidence, for $\omega_\kv< 2 \pi/d$ the signal transition is fully cooperative and the  efficiency can reach values up to $\frac{1}{2}$. For lesser degrees of cooperativity the maximum efficiency is lowered. The efficiency shows sharp spikes towards $0$ at the instabilities discussed in \cref{sect:methods} when 
       $\omega_{\kv'}= \omega_{\kv} +\Delta \omega =\left| \kv'(\pv, s, b)_\parallel \right|= \left|\kvp + \frac{2 \pi}{d} (m,n)+ \kv_{L,ba} \right|$ for some  $m,n\in \mathbb{Z}$, i.e. the perpendicular wavevector component of the outgoing photon is vanishingly small. For example, for $\Delta \omega= \frac{3}{2}\frac{2 \pi}{d}$, the shortest lattice vectors that can fulfill this condition are   $[\pm 2,0] (2\pi/d)$ and $[0,\pm 2] (2\pi/d)$, which leads to the spiked feature at $\omega_\kv= \frac{1}{2} \frac{2 \pi}{d}$. The next longer lattice vectors are   $[\pm 2,\pm 1] (2\pi/d)$ and $[\pm 1,\pm 2] (2\pi/d)$, which lead to an instability at $\omega_\kv= \left(\sqrt{5} - \frac{3}{2}\right) \frac{2 \pi}{d}\approx 0.74 \frac{2 \pi}{d} $. Additionally, the transduction energy is shown for when the real part of $\rho \mathbb{I}- \rho P(\kvp, a)$ is set to zero (black, dashed). In that case, the efficiency does not depend on the transduction energy $\Delta\omega$ and the spiked features are no longer present. For any $\omega_\kv\leq \frac{2 \pi}{d}$  the maximum efficiency is attained, while for $\omega_\kv> \frac{2 \pi}{d}$, the efficiency depends on $\omega_\kv$ and has discontinuities whenever new modes turn critical (i.e. they turn from evanescent to light-like) on the $a$ transition (as opposed to modes turning critical on the $b$ transition which leads to the spiked features). (\textbf{b}) Reciprocal lattice vectors for $\omega_{\kv'}= 2 (2 \pi/d) \pm 0^{+}$ ($-$ dotted, $+$ dashed) in units of $2 \pi/d $ . Bound, evanescent modes are shown as black dots, while free space modes are shown in yellow. Modes inside the circle (light cone) have a real perpendicular momentum component $\kv'_\perp= \pm \sqrt{(\omega_{\kv'})^2 - |\pv|^2}\in \mathbb{R}$ and are free space modes, while modes outside have an imaginary perpendicular component, making them evanescent. If a mode lies on the light cone, it has a vanishing perpendicular component and is therefore critical.   Going from $\omega_{\kv'}= 2 \times (2 \pi/d) - 0^{+}$ (dotted circle) to $\omega_{\kv'}= 2 \times (2 \pi/d) + 0^{+}$ (dashed circle), the modes with $[\pm 2,0] (2\pi/d)$ and $[0,\pm 2] (2\pi/d)$ (red dots) turn into free space modes and may therefore be present in $\hat{S}_b\ket{(\kvp,k_\perp) \mu a}$. For $\omega_{\kv'}= 2 (2 \pi/d)$ these modes are critical and lie along the light cone, leading to the instabilities and the spiked features in (\textbf{a}). Similarly, on the signal transition for $\omega_\kv< 2 \pi/d$, only $\pv=\zerov$ is radiant, while at $\omega_\kv= 2 \pi/d$, $\omega_\kv= \sqrt{2} 2 \pi/d$ and  $\omega_\kv= 4 \pi/d$, modes such as $\pv= [1,0] \frac{2\pi}{d} $, $\pv= [1,1] \frac{2\pi}{d} $  and $\pv= [2,0] \frac{2\pi}{d} $, turn radiative. For $\kvp \neq \zerov$ the same picture applies with the center of the circle displaced from the origin by $\kvp$. }
    \label{transduction_bestA}
\end{figure*}

To begin, we consider the probability for an incoming photon in an $a$ mode to be scattered into any $b$ mode as defined in \cref{conv_efficiency_atob}. We determine how that probability depends on the dark state mixing ratio $|A|$ (where $|B|= \sqrt{1-|A|^2}$) and the degree of cooperativity. Here, we assume that the incoming signal photon is normally incident upon the array ($\kvp=\zerov$, $k_\perp= \omega_{\kv}$) and has an arbitrary in-plane polarization $\mu$.

\subsubsection{Dark state mixing ratio}
 First, in \cref{trans_efficiency_Adependence}, we show how the transduction efficiency  $\left| \hat{S}_b\ket{(\kvp,k_\perp) \mu a}\right|^2$ depends on the dark state mixing ratio $|A|$ for different transduction energies $\Delta\omega= \omega_{g_1}- \omega_{g_2}$ and $\wp_a= \wp_b$. On resonance, the absolute strength of the dipole moments $\wp_a, \wp_b$ does not play a role and instead only their relative strength $\wp_a/\wp_b$ matters.    At $|A|=0$ $(|B|=1)$ and $|A|=1$ $(|B|=0)$ no transduction can occur because one of the driving lasers has  a vanishing Rabi frequency. In between these limits, the efficiency is peaked and with increasing transduction energy the location of this peak approaches $|A|=1$ asymptotically. For an  incoming photon energy of $\omega_\kv= 0.2\times \frac{2\pi}{d}$ the maximum efficiency of $\frac{1}{2}$ for normal incidence can be achieved, while for sub-optimal values of $\omega_{\kv}$ the maximum efficiency with respect to $|A|$ is lower.

\begin{figure*}[ht!]
    \centering
 \begin{tikzpicture}
    \node[anchor=south west,inner sep=0] (image) at (0,0) {\includegraphics[width=\textwidth]{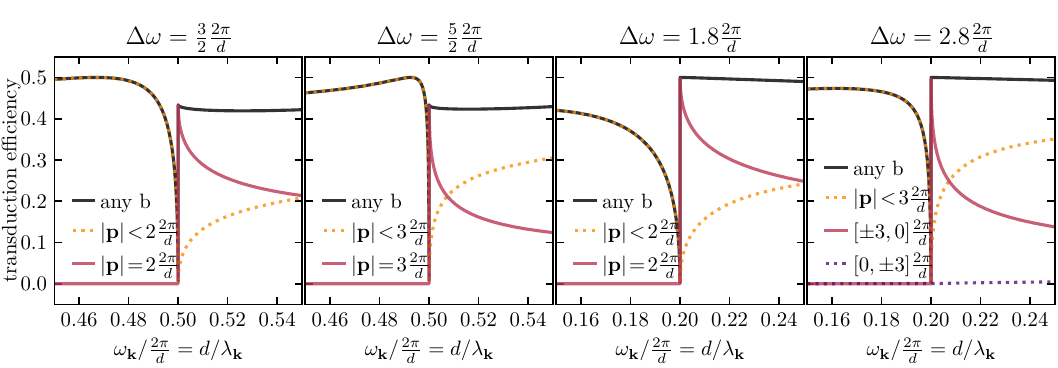}};
    \begin{scope}[x={(image.south east)},y={(image.north west)}]
            \node[align=right] at (0.53+.005,0.93) {$\textbf{(c)}$};
        \node[align=right] at (0.02,0.93) {$\textbf{(a)}$};
        \node[align=right] at (0.29+.007,0.93) {$\textbf{(b)}$};
         \node[align=right] at (.77+.005,0.93) {$\textbf{(d)}$};
\end{scope}
\end{tikzpicture}
    \caption{\textbf{Transduction efficiency near critical modes.} For resonant, $y$-polarized $a $ photons at normal incidence ($\kvp=\zerov$) the transduction efficiency into any $b$ mode (\cref{conv_efficiency_atob}, solid, black) along with the transduction efficiency into specific $b$ modes (\cref{specific_efficiency}, orange, red, purple) is shown. These efficiencies are shown as function of the incoming photon energy $\omega_{\kv}$ for different transduction energies $\Delta \omega= \omega_{g_1}- \omega_{g_2}= \frac{3}{2} (2 \pi/d)$ (\textbf{a}), $\frac{5}{2} ( 2 \pi/d)$ (\textbf{b}), $1.8 ( 2 \pi/d)$ (\textbf{c}) and $2.8  (2 \pi/d) $ (\textbf{d}), such that the energy and momenta of the outgoing transduced modes are given by $\omega_{\kv'}= \omega_\kv+ \Delta \omega$ and  $\kv'(\pv, s,b), \pv \in (2 \pi/d) \mathbb{Z}_2$. Thus at $\omega_{\kv}= \frac{1}{2} ( 2\pi/d)$ (\textbf{a},\textbf{b}) and $0.2( 2 \pi/d)$ (\textbf{c},\textbf{d}), modes with $|\pv|=2  (2 \pi/d)$  (\textbf{a},\textbf{c}) and $|\pv|=3 (2 \pi/d)$ (\textbf{b,d}) turn critical. As a result, approaching the instability  from below, the overall $a\to b$ efficiency (solid, black) drops to zero and rapidly increases again to values near $\frac{1}{2}$ beyond the instability. Below the instability, all of the transduction is into the lower modes inside the light cone (orange, dotted) which have $|\pv|<2 \times (2\pi/d) $(\textbf{a},\textbf{c}) and $|\pv|<3\times (2\pi/d)  $(\textbf{b},\textbf{d}). Slightly, beyond the instability, transduction is predominantly into the new modes (solid, red). At normal incidence, if the incoming photon is $y$-polarized, then the polarization of the outgoing photon must have a polarization component in the $y$-direction. Hence, no weight is placed into modes such as the $\pv= (0,\pm 3) (2\pi/d)$ mode (dotted, purple) in (\textbf{d}), as free-space photons are transversely-polarized.}
    \label{efficiency_near_resonance}
\end{figure*}

\subsubsection{Degree of cooperativity}

Next, we analyze how the degree of cooperativity affects the transduction efficiency at the optimal values of the dark state mixing ratio $|A|$ found in \cref{trans_efficiency_Adependence}. Specifically, in \cref{transduction_bestA}(a), the transduction efficiency for the optimal value of the dark state mixing ratio $|A|$,
\begin{align}
 \max_{|A| \in [0,1]}   \left| \hat{S}_b\ket{(\kvp,k_\perp) \mu a}\right|^2,\label{optimal_transduction}
\end{align}
is shown as a function of the initial energy $\omega_\kv$ for different transduction energies $\Delta \omega= \omega_{g_1}- \omega_{g_2}$ at normal incidence. 

For a fully   cooperative array with $\omega_\kv< 2\pi/d$, the transduction efficiency can reach values of up to $\frac{1}{2}$. However, with a decreasing degree of cooperativity the maximum efficiency decreases rapidly once a further free space  mode in $a$ becomes available (i.e. an evanescent  $a $ mode turns into a free space $a$ mode) when $\omega_{\kv}> \left|\kvp + \frac{2 \pi}{d} (m,n)\right|$ for some $ m,n\in \mathbb{Z}$ (see also \cref{transduction_bestA}(b)). Note that  for normal incidence we have $\kvp=\zerov$. Sharp drops in efficiency occur whenever a critical point in the $b$ modes is approached at which the energy of the outgoing photon fulfills the relation $\omega_{\kv'}= \omega_{\kv} + \omega_{g_1}- \omega_{g_2} \approx |\kv'(\pv, s, b)_\parallel|= \left|\kvp + \pv+ \kv_{L,ba} \right|$ for some  $\pv\in (2\pi/d)\mathbb{Z}_2$, i.e. the perpendicular component of the wavevector $\kv'(\pv, s, b)_\perp$ of the outgoing photon is vanishingly small, as discussed in \cref{sect:methods} and depicted in \cref{transduction_bestA}(b). At normal incidence for a vanishing drive grating this simplifies to  $\omega_{\kv'} \approx \frac{2\pi}{d} \sqrt{m^2+n^2}$ for some $m,n \in \mathbb{Z}$. Furthermore, when one sets the real part of $(\rho \mathbb{I}- \rho P(\kvp,a))$ in \cref{conv_efficiency_atob} to zero, the maximum conversion efficiency for an incoming $a$ photon  at normal incidence with energy $ \omega_\kv< 2\pi/ d $ is attained for all transduction energies. This highlights the role of the interplay between the detuning $\delta =\Re \rho= \omega_{\kv} + \omega_{g_1}- \omega_{e_1} $ and the cooperative resonance shifts given by the real parts of $\rho D(\kvp,a)$ and $\rho D(\kvp + \kv_{L,ba}, b)$ within $\rho P(\kvp, a)$, as also highlighted in Ref. \cite{Shahmoon2017}. That is, maximally efficient transduction at normal incidence is possible when the incoming light is cooperative and resonant with a dipolar surface mode, i.e. the real part  of $ \rho \mathbb{I}-  \rho D(\kvp + \kv_{L,ba}, b) -  \rho D(\kvp,a)$ vanishes. This may be achieved by having these three components vanish individually (e.g. $\rho D(\kvp,a)$ vanishes for $\omega_\kv\approx 0.2, 0.8 \times (2\pi/d) $ \cite{Shahmoon2017}) or by compensating non-vanishing resonance shifts with appropriate detunings \cite{Shahmoon2017}.

Due to the coherent population trapping, the effective dipole strengths are given by $A^* \wp_a$ and $B^* \wp_b$ and the results shown in \cref{trans_efficiency_Adependence}, which consider varying mixing ratios $|A|$ for $\wp_a=\wp_b$, would indeed change for different $\wp_a/\wp_b$. However, when tuning $|A|\in [0,1]$ (and as a result $|B|= \sqrt{1- |A|^2}$) to the value that yields maximum transduction efficiency, the dependence on $\wp_a/\wp_b$ diminishes and as a result all other results shown in this paper, including \cref{transduction_bestA}, do not depend on $\wp_a/\wp_b$.

\subsection{Mode selectivity}
\label{mode_selectivity}
In this subsection, we demonstrate how scattering instabilities can be used to select specific outgoing photon modes in the cases of normally incident and non-normally incident signal photons.

\subsubsection{Normal incidence}

In \cref{efficiency_near_resonance}, the behavior of the maximum overall transduction efficiency (\cref{optimal_transduction}) is shown near different instabilities as a function of the energy of the \emph{normally incident} incoming photon, polarized in the $y$-direction. For transduction energies of 
$\Delta \omega/\frac{2 \pi}{d}= 1.5 $ (\textbf{a}), $2.5$ (\textbf{b}), $1.8 $ (\textbf{c}) and $2.8 $ (\textbf{d}) there are instabilities at incoming energies $\omega_{\kv}= 0.5 \times( 2\pi/d)$ (\textbf{a},\textbf{b}) and $0.2 \times( 2 \pi/d)$ (\textbf{c},\textbf{d}), where new modes 
$ \kv'(\pv, s,b)=\left(\pv ,s \sqrt{   \left(\omega_\kv+ \Delta \omega\right)^2-  |\pv|^2}\right)$ turn critical. Thus, for outgoing $b$ photons with energies  $\omega_{\kv'}=\omega_{\kv}+ \Delta \omega = 2 \times( 2 \pi/d )$ (\textbf{a},\textbf{c}), the new modes have in-plane momenta  $\pv=(0, \pm 2) (2\pi/d)$ and    $\pv=( \pm 2,0) (2\pi/d)$, while for $\omega_{\kv'}= 3 \times( 2 \pi/d )$ (\textbf{b},\textbf{d}), the new modes have   $\pv=(0, \pm 3) (2\pi/d)$ and    $\pv=( \pm 3,0) (2\pi/d)$. 

Dissecting the transduction process, \cref{efficiency_near_resonance}(\textbf{a}) shows that, approaching the instability at $\omega_\kv= 0.5 \times (2 \pi/d)$ from below, the overall transduction efficiency sharply drops from values near $\frac{1}{2}$ to $0$ and all transduction is into modes with $|\pv|< 2\times (2 \pi/d)$ (obtained by using a projection $\hat{P}_{\text{specific}}$ to modes $\kv'(\pv, s,b)$ with $|\pv|< 2\times (2 \pi/d)$  within \cref{specific_efficiency}).
Beyond the instability, the overall transduction efficiency sharply increases again and, initially, the majority of the transduction is into the new modes ($|\pv|= 2\times (2\pi/d)$) as discussed in \cref{methods_instabilities}. Further away from the transition, the preferential scattering into the new modes is less pronounced and the weight distributes  more evenly across all available modes. 

Similar behavior can be seen in \cref{efficiency_near_resonance}(\textbf{b}) where the new modes are given instead by $\pv=(0, \pm 3) (2\pi/d)$ and    $\pv=( \pm 3,0) (2\pi/d)$. Notably, for incoming photon energies of $0.2 \times( 2 \pi/d)$ (\textbf{c},\textbf{d}), the transduction efficiency is much closer to 0.5 beyond the instability, highlighting the adversarial effect the cooperative shift (i.e. the real part of $[\rho \mathbb{I}- \rho P(\kvp,a)]$) may have when it does not conspire with the detuning $\delta =\Re \rho= \omega_{\kv}+ \omega_{g_1}- \omega_{e_1}$. Furthermore, for an incoming photon polarized in the $y$-direction,  \cref{efficiency_near_resonance}(\textbf{d}) shows that  the transduction into new modes is solely into the modes with $\pv= [\pm 3,0] (2\pi/d)$, which are transverse to the incoming photon's polarization, and not into modes with $\pv= [0,\pm 3] (2\pi/d)$.  This occurs because, for $(\kvp)_{x}= (\kv_{L,ba})_x=0$ and/or $(\kvp)_{y}= (\kv_{L,ba})_y=0$, the matrix $(\rho \mathbb{I}- \rho P(\kvp, a))$ is diagonal. Thus, if the incoming photon is polarized in the $y$-direction, i.e. the $\vec{\epsilon}_{\kv \mu}$ within $g_{\kv \mu, a}$ points in $y$-direction, then from \cref{Soperatormain} one can see that the outgoing photon must have a non-vanishing polarization component in the $y$-direction. However,  scattering into $\pv= [0,\pm 3] (2\pi/d)$ with a non-vanishing polarization component in the $y$-direction would not produce a transversally polarized photon and is therefore forbidden. 

\begin{figure*}[ht!]
\centering
 \begin{tikzpicture}
    \node[anchor=south west,inner sep=0] (image) at (0,0) {\includegraphics[width=\textwidth]{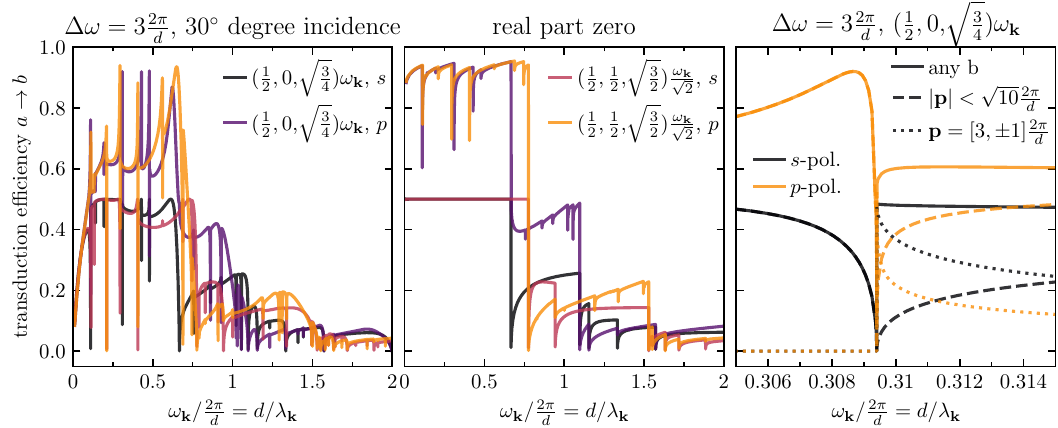}};
    \begin{scope}[x={(image.south east)},y={(image.north west)}]
            \node[align=right] at (0.704,0.95) {$\textbf{(c)}$};
        \node[align=right] at (0.02,0.95) {$\textbf{(a)}$};
        \node[align=right] at (0.395,0.95) {$\textbf{(b)}$};
\end{scope}
\end{tikzpicture}
    \caption{\textbf{Transduction efficiency for $30^{\circ}$ degree incidence.} (\textbf{a}) Transduction efficiency of a resonant ($\rho=i0^+$) $a$ photon into $b$ modes at the optimal value of $|A|$ as function of the incoming photon energy $\omega_{\kv}$ for different wavevectors and polarizations. The transduction energy is given by $\Delta\omega=\omega_{g_1}-\omega_{g_2}= 3  (\frac{2 \pi}{ d})$. The incoming photon has a wavevector either along one of the lattice directions $(\kvp, k_\perp)= (1/2,0,\sqrt{3/4} )\omega_\kv$ (black, purple) or diagonal to it $(\kvp, k_\perp)= (1/2\sqrt{2},1/2\sqrt{2},\sqrt{3/4} )\omega_\kv$ (red, orange) with polarizations either parallel to the plane of incidence ("p", purple, orange) or parallel to the array ("s", black, red). The efficiency shows sharp spikes towards $0$ near the instabilities discussed in the main text. For polarizations parallel to the array, the efficiency is bounded by $1/2$, while for polarizations parallel to the plane of incidence higher efficiencies may be achieved. (\textbf{b}) Transduction efficiencies for the same parameters as in (\textbf{a})  with the real part of $(\rho\mathbb{I}- \rho P(\kvp,a))$ set to $0$. The efficiencies give an upper bound to the efficiencies shown in (\textbf{a}). In both (\textbf{a}) and (\textbf{b}), with a decreasing degree of cooperativity, the efficiency decreases rapidly and the critical energies for that are the same in (\textbf{a}) and (\textbf{b}). Note that the critical energy up to which incoming photons are still cooperative depends on $\kvp$ and as a result the sharp drop in efficiency occurs at different energies for the two different wavevectors.
    When the real part is set to zero, the spikes stemming from $b$-modes turning critical are only observed for $p$-polarized modes, the transduction efficiency of $s$-polarized modes on the other hand, does not depend on the transduction energy as also seen  in \cref{transduction_bestA}. 
    (\textbf{c}) Transduction efficiency into different modes near the critical point at $\omega_{\kv}= 0.3094 (2\pi/d)$ with $(\kvp, k_\perp)= (1/2,0,\sqrt{3}/2 )\omega_\kv$ and $\Delta \omega= 3 \times (2 \pi/d)$ (this corresponds to the black and purple curves shown in (\textbf{a})), where the mode  $\kv'(\pv,s,b)=\left(\kvp+ \pv,s \sqrt{(\omega_\kv + \Delta \omega)^2-|\kvp+ \pv|^2 }\right)$ with $\pv=[3, \pm 1] (2\pi/d)$ turns critical. Solid lines denote transduction into any $b$-mode, while dashed lines denote transduction into lower modes inside the light cone and dotted lines denote transduction into the new modes which turn critical at the critical energy. The efficiencies are given both for "$s$"- and "$p$"-polarized photons.  Below the critical point, transduction is entirely into the lower modes, while slightly  above the critical point the new modes are strongly scattered into. While for "$p$"-polarized modes the overall efficiency may be higher than $\frac{1}{2}$, transduction efficiency into the critical modes is lower than $\frac{1}{2}$ for both polarizations.}
    \label{angled_fig}
\end{figure*}
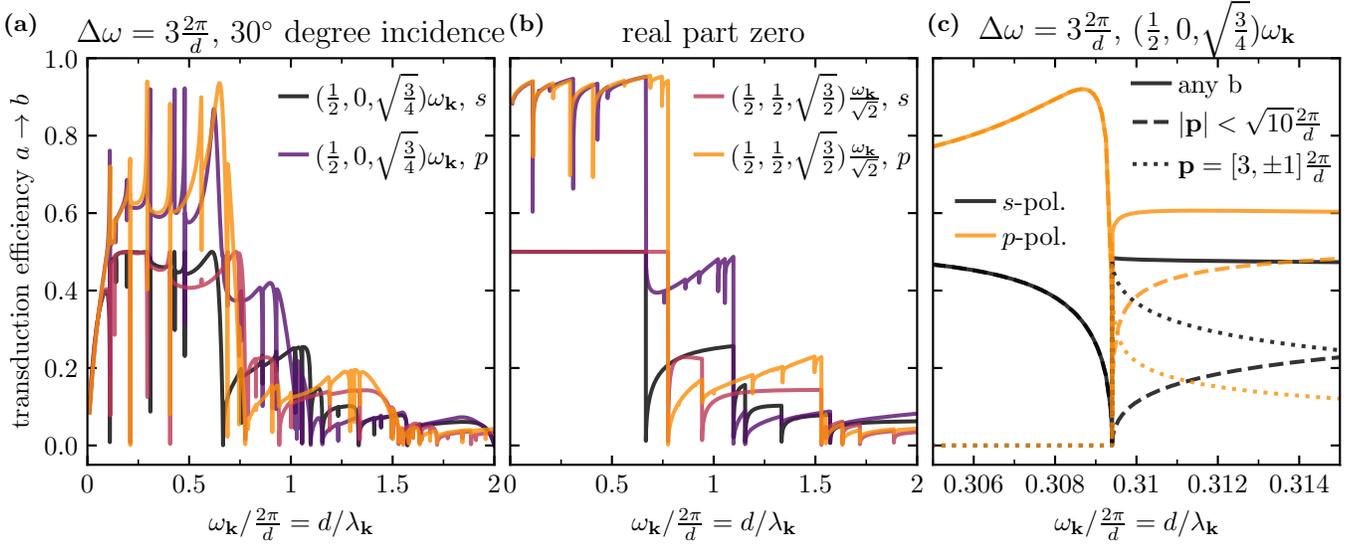

The spiked drops in overall transduction efficiency, shown in \cref{transduction_bestA,efficiency_near_resonance} whenever a new mode turns critical, result from a Lamb shift that diverges as a critical point is approached from below (see for example Ref. \cite{Shahmoon2017}). This shifts the coupling of an incoming photon into idler modes out of resonance. These spiked features are relatively narrow in energy and will be broadened in any experimental setup where incoming signals or driving lasers might not be perfectly monochromatic. While, experimentally, the critical points therefore might not show up as sharp features in the overall transduction efficiency, in \cref{efficiency_near_resonance} it can be seen that the predominant scattering into the new, directed modes is usually broader in energy than the sharp drops in efficiency, allowing for better experimental observability.

\subsubsection{Non-normal incidence}

The results discussed in this section so far  have been for the case of normal incidence, where an efficiency of up to $\frac{1}{2}$ could be achieved. Intuitively, it is reasonable to expect total internal reflection to be facilitated when the incidence is at an angle rather than normal to the array, possibly allowing for transduction efficiencies higher than $\frac{1}{2}$. This raises the question  of whether transduction  into specific, critical modes  can be achieved with efficiencies higher than $\frac{1}{2}$. At non-normal incidence ($\kvp\neq\zerov$), the criticality conditions for an outgoing mode ($ |\kvp + \pv|^2 \approx (\omega_\kv + \Delta \omega)^2 $) are no longer symmetric with respect to the in-plane directions, but rather distinguish specific in-plane directions over others. In addition, at non-normal incidence, the two polarization directions are no longer equivalent (at normal incidence both polarization directions are in-plane directions). Rather, it becomes relevant to distinguish between modes polarized parallel to the array ("$s$") and modes polarized parallel to the plane of incidence ("$p$").

On a technical level, the difference in scattering behavior between $s$- and $p$-polarized incoming modes roots back to the structure of the $(\rho \mathbb{I}- \rho P(\kvp, a))$ matrix: its $x$-$z$ and $y$-$z$ components vanish, i.e. $(\rho \mathbb{I}- \rho P(\kvp, a))_{m,n}=0$ for $m=z, n=x,y$ and $n=z, m=x,y$; and the same goes for its inverse. As a result, $s$-polarized modes do not interact with the $n=m=z$ component of $(\rho \mathbb{I}- \rho P(\kvp, a))^{-1}$. On the other hand, $p$-polarized modes have a non-vanishing polarization component in the $z$-direction, leading to different qualitative behavior. Given that $s$-polarized modes at non-normal incidence interact with the same entries of $(\rho \mathbb{I}- \rho P(\kvp, a))^{-1}$ as modes at normal incidence (regardless of their polarization), it is thus conceivable that, qualitatively, modes at normal incidence behave similarly to  $s$-polarized modes at non-normal incidence.

In \cref{angled_fig}(\textbf{a}) we show overall transduction efficiencies obtained with a $30^{\circ}$ angle of incidence. Results are shown for an incident wavevector along one of the lattice directions ($\kv= (\kvp, k_\perp)= [\sin(30^{\circ})\omega_{\kv},0,\cos(30^{\circ})\omega_{\kv} ]$)
and at a $45^{\circ}$ angle to it ($\kvp= \left[\frac{1}{\sqrt{2}},\frac{1}{\sqrt{2}}\right]\sin(30^{\circ})\omega_{\kv}$, $k_\perp= \cos(30^{\circ})\omega_{\kv}$), each for $s$- and $p$-polarized modes. The transduction energy is held constant at $\Delta \omega= 3 \times (2 \pi/d)$ and the energy of the incident photon $\omega_{\kv}$  is varied.

To begin, let us restrict our analysis to $s$-polarized modes. It can be seen that $s$-polarized modes have an overall transduction efficiency of up to $\frac{1}{2}$. Like at normal incidence, when an instability is approached where $\omega_{\kv'}= \omega_{\kv} +\Delta \omega  \approx|\kv'(\pv,s, b)_\parallel|=|\kvp + \pv|$ for some  $\pv \in (2\pi/d)\mathbb{Z}_2 $, the efficiency spikes down to $0$ and then quickly recovers. Notably, for  different orientations of  $\kvp$ vectors, the instabilities occur at different energies $\omega_{\kv}$, as can be seen in  \cref{angled_fig}(\textbf{a}) and easily verified by the instability condition. 

Similarly, the different orientations of  $\kvp$ vectors have different requirements for cooperativity: For an incident photon with an in-plane wavevector $\kvp= [1,0]\sin(30^{\circ})\omega_{\kv}$, the requirement to be fully cooperative is $\omega_{\kv}< \frac{2}{3} (\frac{2\pi}{d})$ (obtained from the condition that $(\omega_\kv)^2- [\sin(30^{\circ})\omega_\kv- 2\pi/d]^2=0$), while for $\kvp= \left[\frac{1}{\sqrt{2}},\frac{1}{\sqrt{2}}\right]\sin(30^{\circ})\omega_{\kv}$, it is $\omega_{\kv}<\frac{\sqrt{2}}{3} \left(\sqrt{7}-\sqrt{1} \right)(\frac{2\pi}{d})$. As for the case of normal incidence, with decreasing degree of cooperativity, the transduction efficiency diminishes. This can also be seen in \cref{angled_fig}(\textbf{b}) where the same transduction efficiencies are shown with the real part of $(\rho \mathbb{I}- \rho P(\kvp, a))$ set to zero. 

The efficiencies shown in \cref{angled_fig}(\textbf{a}) are upper bounded by the efficiencies shown in \cref{angled_fig}(\textbf{b}), where the real part was set to zero. For cooperative,  $s$-polarized modes the transduction efficiency is constant at $1/2$ when the real part is discarded. As the degree of cooperativity of the incoming $s$-polarized modes is decreased, the transduction efficiency decreases rapidly, which, depending on the orientation of $\kvp$, occurs at different incoming energies. Like at normal incidence, discarding the real part, for $s$-polarized modes there are are no efficiency spikes stemming from $b$ modes turning critical. Furthermore, as also seen in \cref{transduction_bestA}(\textbf{a}), the overall transduction efficiency  is independent of the transduction energy. Apart from the varying conditions for cooperativity and criticality, the scattering of  $s$-polarized modes at non-normal incidence is thus similar to scattering at normal incidence.

For $p$-polarized modes i.e. modes with a nonzero $z$-component of the polarization, \cref{angled_fig}(\textbf{a}) and \cref{angled_fig}(\textbf{b}) show that  overall transduction efficiencies higher than $\frac{1}{2}$ may be achieved. The conditions for cooperativity  and criticality for $p$-polarized modes are the same as for $s$-polarized modes and thus in \cref{angled_fig}(\textbf{a}) the spikes to $0$ near an instability as well as the decreases in degree of cooperativity occur at the same energies $\omega_\kv$. However, for cooperative, $p$-polarized modes near an instability, the efficiency additionally spikes upwards before spiking down to $0$ and recovering. Another difference between polarizations arises when discarding the  real part of $(\rho \mathbb{I}- \rho P(\kvp, a))^{-1}$; for cooperative  $p$-polarized modes, \cref{angled_fig}(\textbf{b}) shows that the overall efficiency does  depend on the energy of the incoming photon and it has spikes to $0$ at the locations of the instabilities.

Finally, in  \cref{angled_fig}(\textbf{c}), the transduction efficiency and its relative weight for different modes are shown for energies $\omega_{\kv}$ near the critical point at $\omega_{\kv}= 0.3094 (2\pi/d)$ with $\kvp= [1,0]\sin(30^{\circ})\omega_{\kv}$, $k_\perp= \cos(30^{\circ})\omega_{\kv}$ and $\omega_{g_1}- \omega_{g_2}= 3 \times (2 \pi/d)$. Here the modes $\kv'(\pv,s,b) $ with $\pv=[3, \pm 1] (2\pi/d)$(but not those with $\pv=[-3, \pm 1] (2\pi/d)$) turn critical and it can be seen that below the critical point the transduction is into the lower modes, while slightly above the critical point the new modes are strongly preferenced. As seen before in  \cref{angled_fig}(\textbf{a}), the transduction efficiency of $s$-polarized modes is bounded by $1/2$, while the $p$-polarized modes may feature higher efficiencies. However, beyond the critical point the probability to transduce into the new modes with $\pv=[3, \pm 1] (2\pi/d)$ is still around $1/2$ regardless of polarization. Beyond the critical point the transduction efficiency into the new modes with $\pv=[3, \pm 1 ] 2\pi/d $ depends strongly on the polarization and falls off faster for $p$- than for $s$- polarized modes. This can be explained by the differing polarization overlap in \cref{Soperatormain} that an incoming mode with polarization $\mu=s,p$ may have with the outgoing modes 
$ \propto  \sum_{\mu'} \vec{\mathbf{\epsilon}}_{(\kvp, k_\perp) \mu} 
 \left( \rho \mathbb{I}- \rho P(\kvp,a)\right)^{-1}   \vec{\mathbf{\epsilon}}_{\kv'(\pv,+,b) \mu'}  $ with $\pv=[3, \pm 1] (2\pi/d)$.

\subsection{Scattering properties of finite emitter arrays}\label{maxwell_calcs}

\begin{figure*}[ht!]
    \centering
     \begin{tikzpicture}
    \node[anchor=south west,inner sep=0] (image) at (0,0) {\includegraphics[width=\textwidth]{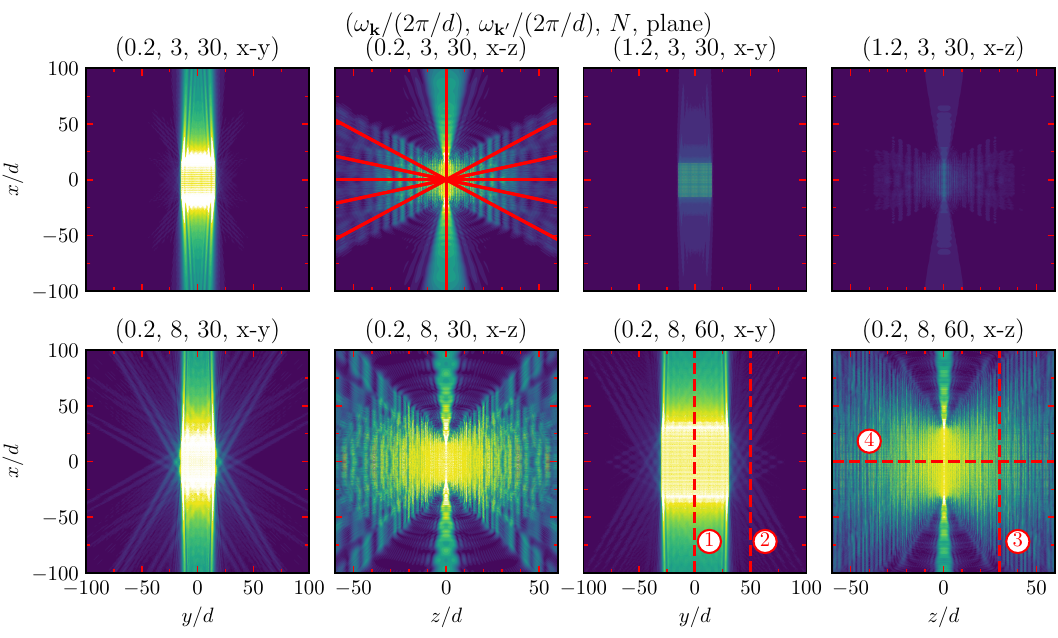}};
    \begin{scope}[x={(image.south east)},y={(image.north west)}]
        \node[align=right] at (0.06,0.92) {$\textbf{(a)}$};
        \node[align=right] at (0.56,0.92) {$\textbf{(b)}$};
        \node[align=right] at (0.06,0.48) {$\textbf{(c)}$};
        \node[align=right] at (0.56,0.48) {$\textbf{(d)}$};
         \draw[dashed, opacity=0.3, line width=1pt] (.539,.1) -- (.539,.94);
          \draw[dashed, opacity=0.3, line width=1pt] (.05,.51) -- (1,.51);
\end{scope}
\end{tikzpicture}
    \caption{\textbf{Scattered $b$-type electric field strengths at normal incidence.} For an incident, resonant, plane wave electric field polarized in $y$-direction $\Ev^{0}_a(\rv)= \Ev_0 e^{i \kvp \rv_\parallel +i k_\perp z}$ with $(\kvp, k_\perp)= (0,0,1)\omega_\kv$ and frequency $\omega_\kv$, the strength of the scattered $b$ electric field $|\Ev_b(\rv)|$ is shown in the $z=0$ plane (left plots) and the $y=0$ plane (right plots). The energies of the incoming $a$ field are  $k_a=\omega_\kv=0.2 (2\pi/d)$ (\textbf{a},\textbf{c},\textbf{d}) and $\omega_\kv=1.2 (2\pi/d)$(\textbf{b}), while the energies of the outgoing $b$ field are $k_b=\omega_{\kv'}= 3(2\pi/d)$ (\textbf{a},\textbf{b}) and $8(2\pi/d)$ (\textbf{c},\textbf{d}). The grid sizes $N^2$ used are $30 \times 30 $ (\textbf{a},\textbf{b},\textbf{c}) and $60 \times 60 $(\textbf{d}) emitters. For each of these combinations of incoming field and field energies, there is a critical mode in the $x$-direction with $k_\perp=0$  available.  The red lines in (\textbf{a}) show the directions of free space modes $(\kvp +\pv,\pm \sqrt{\omega_{\kv'}^2 - |\kvp +\pv|^2 })$ with $\pv \in \mathbb{Z}_2 (2\pi/d)$ and $\kvp=\zerov$. (As the plot is in the $y=0$ plane, only modes with $(\kvp +\pv)_{y}=0$ are shown.)  The numbered, red dashed lines in (\textbf{d}) give the directions of  the Fourier transforms shown in \cref{fig:Fourier_Maxwell}. The electric field strength is given in units of $|\Ev_0|$ and the color scale ranges are $[0,3]$ in subfigures (\textbf{a},\textbf{b},\textbf{c}) and $[0,6]$ in  (\textbf{d}). In all plots it can be seen that in the $z=0$ ($x$-$y$) plane there is strong scattering in the $x$-direction, as expected, while other (off-resonant, see main text) modes are populated weakly (left plots). While the mode in the $x$-direction with $k_\perp=0$ is critical in all these plots, also modes with $|k_\perp|>0 $ are seen to be populated (right plots). Furthermore, the scattering in the $x$-direction is not perfectly confined to the $z=0$ plane, but rather a lobe is formed which, comparing $(\textbf{c})$ and $(\textbf{d})$,  tightens upon increasing the emitter number. Comparing a fully cooperative incoming $a$ field with  $\omega_\kv= 0.2(2\pi/d)$ (\textbf{a}) with a less cooperative field (\textbf{b}), the scattering is qualitatively the same; however the overall transduced $b$-field strength is much weaker. For larger outgoing energies, more modes with $|k_\perp|>0 $ are available, forming closer to a continuum in the right plots of (\textbf{c,d}). Increasing the number of emitter from $30 \times 30$ (\textbf{c}) to $60 \times 60$  (\textbf{d}), a higher proportion of the field is scattered into the critical modes in the $x$-direction. }
    \label{maxwell_normal_incidence}
\end{figure*}

To evaluate the effect of an array with a finite rather than an infinite number of emitters, in the following we present scattering calculations using classical Maxwell equations and the (cross-)polarizabilities $\alpha_{\sigma \sigma'}$ as discussed in \cref{Maxwell_methods}. As mentioned there and shown in \cref{App_Maxwell}, the electric field scattered off such an emitter array for an incoming plane wave closely corresponds  to  the action of the $\hat{S}$ operator on a plane wave photon in-state. 

\subsubsection{Normal incidence}
In \cref{maxwell_normal_incidence}, we show the strength of the scattered $b$ electric field $|\Ev_b|$ for an incoming, $y$-polarized, resonant plane wave  $\Ev^0_a=\Ev_0 e^{i \kvp \rv_\parallel +i k_\perp z}$ propagating \emph{normal} to the array plane with $(\kvp, k_\perp)= (0,0,1)\omega_\kv$. Different combinations of energies of the incoming $a$ field $k_a=\omega_\kv$, energies of the scattered $b$ field $k_b= \omega_{\kv'}$ and array sizes are shown, both in the $x$-$y$ plane at $z=0$ and in the $x$-$z$ plane at $y=0$. The array is centered at $x=y=0$.  

In \cref{maxwell_normal_incidence}(\textbf{a}), we show results for a cooperative energy  $\omega_\kv= 0.2 (2\pi/d)$ which is transduced to an energy  $\omega_{\kv'}= 3 (2\pi/d)$ in a $30 \times  30 $ emitter array. It can be seen that in the array plane ($x$-$y$) the transduced $b$ field is strongly scattered into the $x$-direction, which contains the critical modes with $\pv= [\pm 3,0 ]  (2 \pi/d)$ (the modes with $\pv= [0, \pm 3]  (2 \pi/d)$ are not scattered into due to the lacking polarization overlap as also seen in \cref{efficiency_near_resonance}(\textbf{d})).    Similarly, in the $x$-$z$ plane it can be seen that the field scattered in $x$-direction is not perfectly confined to the array plane (which one would expect for an infinite emitter array), but rather develops a lobe-like spread in the $z$-direction. This "beam spread" effect and its dependence on the number of emitters in the array will be analyzed further in \cref{beam spread}. Additionally, in the $x$-$z$ plane (right plots),  scattering into the direction of the available free space modes $(\kvp +\pv,\pm \sqrt{\omega_{\kv'}^2 - |\kvp +\pv|^2 })$ with  $\pv \in \mathbb{Z}_2 (2\pi/d)$ and $\kvp=\zerov$ is visible. For illustration, the propagation direction of these modes is shown in \cref{maxwell_normal_incidence}(\textbf{a}) using solid red lines (as the plot is in the $x$-$z$ plane, only modes with $(\kvp+ \pv)_y =0$ are shown).

In \cref{maxwell_normal_incidence}(\textbf{b}), the scattered field is instead shown for a greater incoming energy $\omega_\kv= 1.2 (2\pi/d)$ that is not fully cooperative. The same qualitative behavior as the fully cooperative case can be seen, albeit with much weaker field strengths. This is consistent with the results discussed in \cref{sect_conv_efficiency} as the criticality of the $ \pv= [\pm 3,0 ]  (2 \pi/d)$ $b$-modes does not primarily depend on $\omega_\kv$ (but rather on $\omega_{\kv'}$ and $\kvp$) and thus one still sees preferential scattering into the $ \pv= [\pm 3,0 ]  (2 \pi/d)$ modes. However as seen in \cref{transduction_bestA} the maximum transduction efficiency strongly depends on $\omega_\kv$, in particular when $\omega_\kv$ is not fully cooperative, and thus the scattered $b$-field is weaker.

\begin{figure*}[ht!]
    \centering
     \begin{tikzpicture}
    \node[anchor=south west,inner sep=0] (image) at (0,0) {\includegraphics[width=\textwidth]{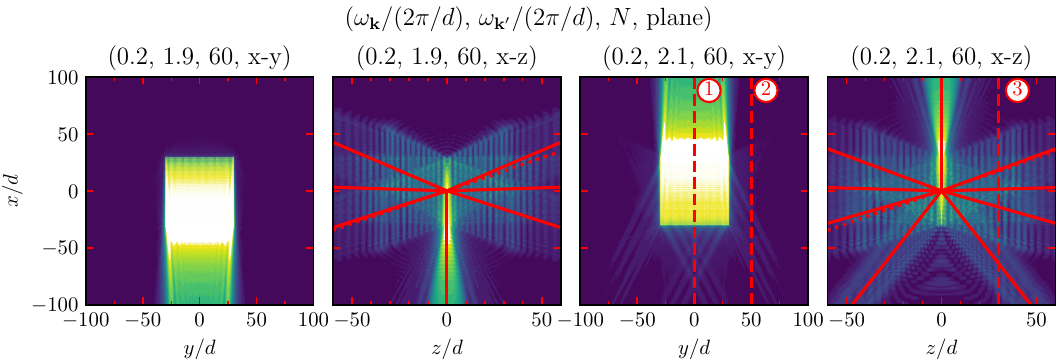}};
    \begin{scope}[x={(image.south east)},y={(image.north west)}]
        \node[align=right] at (0.07,0.85) {$\textbf{(a)}$};
        \node[align=right] at (0.55,0.85) {$\textbf{(b)}$};
      \draw[dashed, opacity=0.3, line width=1pt] (.54,.15) -- (.54,.8);   
\end{scope}
\end{tikzpicture}
\caption{\textbf{Scattered $b$-type electric field strengths for $30^\circ$-degree incidence.} For an incident, resonant, plane wave electric field polarized in the $y$-direction (in-plane) $\Ev^{0}_a(\rv)= \Ev_0 e^{i (\kvp,k_\perp)\cdot( \rv_\parallel, z)}$, the strength of the scattered $b$ electric field $|\Ev_b(\rv)|$ is shown in the $z=0$ plane (left plots) and the $y=0$ plane (right plots). Here,  $(\kvp, k_\perp)= (\sin(30^\circ),0,\cos(30^\circ))\omega_\kv$,  $\Ev^{0}=(0,1,0) |\Ev^{0}|$ and the energy of the incoming $a$ field is  $k_a=\omega_\kv=0.2 (2\pi/d)$, while the   energies of the outgoing $b$ field are $k_b=\omega_{\kv'}= 1.9(2\pi/d)$ (\textbf{a}) and $2.1(2\pi/d)$ (\textbf{b}).  In the $x$-$y$ plane, strong scattering into the negative  $x$-direction  is seen in (\textbf{a}), while in (\textbf{b}) strong scattering into positive $x$-direction is visible. This reflects the critical modes pointed out in the main text: for $\kvp=(0.1,0) (2\pi/d)$ in (\textbf{a}) the mode with $\pv=(-2,0) (2\pi/d)$ is critical and in  (\textbf{b}) the mode with $\pv=(2,0) (2\pi/d)$ is critical. In the $x$-$z$ plane, a corresponding scattering lobe is seen in the $x$-direction which has a spread in $z$-direction with increasing distance from the array. Additionally, scattering into free space modes is visible in the $x$-$z$ plots and in the $x$-$y$ plot faint scattering into off-resonant modes with $\kvp +\pv\approx (-2+0.1, \pm 1) (2\pi/d)$ can be seen. A grid size of  $60 \times 60 $ emitters was used. The dotted red lines  show the propagation direction of the incident plane wave $\Ev^{0}_a $. The numbered, red dashed lines  give the directions of  the Fourier transforms shown in \cref{fig:Fourier_Maxwell_nonnormal}. The solid red lines in the right plots show the directions of free space modes $(\kvp+ \pv, \pm \sqrt{\omega^2_{\kv}- |\kvp+\pv|^2})$ with $\pv\in (2\pi/d) \mathbb{Z}$. As in \cref{maxwell_normal_incidence}, since the right plots are in the $y=0$ plane, only modes with $(\kvp+ \pv)_y=0$ are shown here.  The electric field strength is given in units of $|\Ev_0|$ and the color scale range is $[0,4]$.}
    \label{fig:non_normal_maxwell}
\end{figure*}
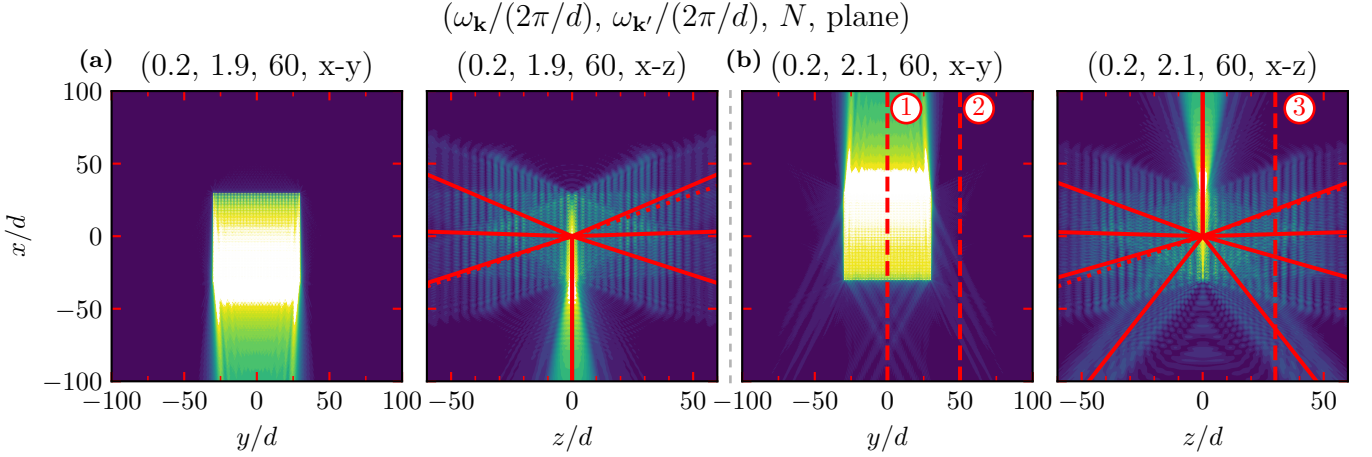

In \cref{maxwell_normal_incidence}(\textbf{c}) and  (\textbf{d}), we show results for $\omega_\kv= 0.2 (2\pi/d)$ and $\omega_{\kv'}= 8 (2\pi/d)$ with $30\times  30$ and $60\times  60$ emitters, respectively. Again, in the array plane ($x$-$y$), the scattering is mainly into the $x$-direction, containing the critical modes $ \pv= [\pm 8,0 ]  (2 \pi/d)$;  however, several other scattering directions are faintly visible. Similarly, in the $x$-$z$ plane one can see scattering into free space modes and because the outgoing energy is larger now, there are more free space modes available. Again, it can be seen that the scattering in the $x$-direction is not fully confined to the array plane, but rather there is a spread in the $z$-direction which gets tighter when the number of emitters is increased in \cref{maxwell_normal_incidence}(\textbf{d}). Note, that in addition to the scattered $b$ electric field $\Ev_b$, there is also a  scattered $a$ electric field $\Ev_a$, which, however, we do not show here. As mentioned above, in the $x$-$y$ plot in \cref{maxwell_normal_incidence}(\textbf{c}), apart from the strong scattering in the $x$-direction, there are also fainter scattering directions visible. These directions are approximately given by modes with $\pv\approx[\pm 7, \pm 4](2\pi /d )$ and $\pv\approx[\pm 4, \pm 7] (2\pi /d )$. However, to conserve energy, the directions must differ from these wavevectors slightly. In the case of infinitely many emitters, only modes with $\pv \in (2\pi /d )\mathbb{Z}_2$ may be scattered into and that would violate energy conservation. For a finite amount of emitters, the restraint on $\pv$ is only approximate and thus these modes may be populated. With increasing number of emitters, however, scattering into these modes becomes asymptotically weaker as can be seen when comparing \cref{maxwell_normal_incidence}(\textbf{c}) and \cref{maxwell_normal_incidence}(\textbf{d}).  In \cref{Fourier_scattering}, to analyze the spectral makeup of the scattered signal, a Fourier analysis of the scattered fields is performed for the fields shown in \cref{maxwell_normal_incidence}(\textbf{d}), confirming the qualitative understanding of scattering into modes.

\subsubsection{Non-normal incidence}
The scattering at normal incidence shown in \cref{maxwell_normal_incidence} was symmetric under exchanges $x\leftrightarrow -x $, $y\leftrightarrow -y $ and $z\leftrightarrow -z $ due to the vanishing $\kvp$ component of the wavevector.  However, at a general angle of incidence, these symmetries are lost (or at least partially lost). In \cref{fig:non_normal_maxwell}, we show scattered field strengths for a resonant, $y$-polarized $a$-field $\Ev^{0}_a(\rv)= \Ev_0 e^{i \kvp \rv_\parallel +i k_\perp z}$  incident at a $30^\circ$ degree angle with $(\kvp, k_\perp)= (\sin(30^\circ),0,\cos(30^\circ))\omega_\kv$, $\omega_\kv=0.2 (2\pi/d)$ and $60\times  60$ emitters, both for outgoing energies $\omega_{\kv'}= 1.9 (2\pi/d)$ and $\omega_{\kv'}= 2.1 (2\pi/d)$. As a result, for $\kvp= [0.1, 0 ](2\pi/d)$ and thus for $\omega_{\kv'}= 1.9 (2\pi/d)$, the $\pv= [-2,0](2\pi/d)$ mode is critical, while for $\omega_{\kv'}= 2.1 (2\pi/d)$ the mode with $\pv= [2,0](2\pi/d)$ is critical (for these modes, $k_\perp=\pm \sqrt{(\omega_{\kv'})- |\kvp + \pv |^2 }=0$). Hence, while for $\omega_{\kv'}= 1.9 (2\pi/d)$ the preferential scattering is into the negative $x$-direction, for $\omega_{\kv'}= 2.1 (2\pi/d)$ it is into the positive $x$-direction. 

\begin{figure*}[ht!]
    \centering
     \begin{tikzpicture}
    \node[anchor=south west,inner sep=0] (image) at (0,0) {\includegraphics[width=\linewidth]{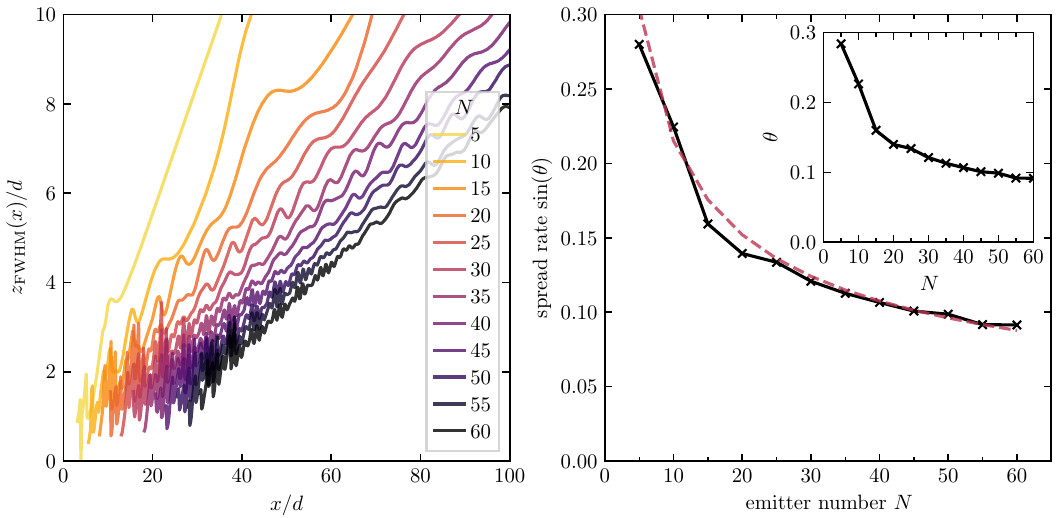}};
    \begin{scope}[x={(image.south east)},y={(image.north west)}]
        \node[align=right] at (0.02,0.98) {$\textbf{(a)}$};
        \node[align=right] at (0.51,0.98) {$\textbf{(b)}$};
\end{scope}
\end{tikzpicture}
    \caption{\textbf{Spread of the critical scattering lobes in the $z$-direction.} \textbf{(a)} For the same parameters as in \cref{maxwell_normal_incidence}(\textbf{a}), the FWHM of the critical scattering lobe in the $x$-direction is shown for different emitter numbers, $N$, between $5$ (yellow) and $60$ (black), corresponding to $5\times 5$ and $60 \times 60 $ emitters, respectively. For every point $\mathbf{x}=(x,0,0)$ along the $x$-axis, the value $z_{\text{FWHM}}(x)$ is determined for which the electric field strength at $\mathbf{x'}=(x,0,z_{\text{FWHM}}(x))$ has decreased to half its peak value at $\mathbf{x}$. With increasing distance from the array,  $z_{\text{FWHM}}(x)$ increases and  for larger emitter numbers it decreases. \textbf{(b)} The spread rate $\sin (\theta)$, the linear slope fit of $z_{\text{FWHM}}(x)$ with $x$, is shown (black crosses) as a function of the emitter number $N$ (corresponding to $N\times N$ emitters). The spread rate decreases with increasing emitter number approximately as $\propto 1/\sqrt{N}$ as can be seen by the fit (dashed red) and as was noted recently \cite{Rusconi_priv_comm}. The inset shows the opening angles $\theta$ determined from the spread rate $\sin (\theta)$.}
    \label{FWHM_analysis}
\end{figure*}

This is shown in \cref{fig:non_normal_maxwell}(\textbf{a});  for $\omega_{\kv'}= 1.9 (2\pi/d)$ in the $z=0$ ($x$-$y$) plane (left plot), the $b$-field is strong in a lobe that starts at the array and points towards the negative $x$-direction, while almost no field is scattered into the positive $x$-direction. Additionally, in the $x$-$z$ plane ($y=0$, right plot in (\textbf{a})),  a strong lobe at $z=0$ in the negative $x$-direction can be seen, but not in the positive $x$-direction.  Similarly, for $\omega_{\kv'}= 2.1 (2\pi/d)$ strong scattering into a confined critical mode with $\pv=[2,0] (2\pi/d)$ can be seen in \cref{fig:non_normal_maxwell}(\textbf{b})  and as a result there is a strong lobe in the positive $x$-direction starting at the array. This emphasizes the strong directionality of the transduction scattering which does not only depend on the direction of the incoming signal field, but also on the criticality conditions of the collective dipolar modes on the idler transition.

As seen similarly in \cref{maxwell_normal_incidence}(\textbf{c},\textbf{d}),  for $\omega_{\kv'}= 2.1 (2\pi/d)$, in the $x$-$y$ plane, there is weak scattering into off-resonant modes with approximately $\kvp +\pv \approx(-2+0.1, \pm 1) (2\pi/d)$. In the $x$-$z$ plane, in addition to scattering into the critical mode, weaker scattering into other free space modes can be seen, which gets weaker with increasing emitter number. Note that because $(\kvp)_y=0 $ here, once again $(\rho \mathbb{I}- \rho P(\kvp,a))$ is diagonal and as a result, the outgoing field must have a field component in the $y$-direction. Thus, modes with a wavevector purely in the $y$-direction would be forbidden to scatter into. However, there are no $\pv\in (2\pi/d) \mathbb{Z}_2$ that fulfill $(\kvp+ \pv)_x =0 $ and additionally these modes are not critical. If a drive grating $\kv_{L,ba}$ is included, this may become a relevant point to consider. To further illuminate the  directionality of the scattering, in \cref{Fourier_scattering}  $x$-direction Fourier transforms of the data shown in \cref{fig:non_normal_maxwell}(\textbf{b}) are performed, confirming the mode picture used to describe the scattering.

\subsubsection{Beam spread}
\label{beam spread}
When transducing an incoming photon into a critical mode in an infinite array, the critical mode propagates with momentum that is parallel to the array plane and is therefore perfectly confined within this plane. For a finite  array however, as mentioned in the discussions regarding the results shown in \cref{maxwell_normal_incidence,fig:non_normal_maxwell}, when scattering into a critical mode a lobe forms which is not perfectly confined to the array plane. With increasing distance from the array, this lobe spreads in the $z$-direction, which tightens with increasing emitter number.

For experimental implementations, a strong degree of confinement, i.e. a narrow scattering lobe, is desirable as it both improves coherence properties and allows for more efficient coupling into an optical fiber or detection within a photon counter. To quantify the increasing degree of confinement, in \cref{FWHM_analysis} we investigate the behavior of the full width half max (FWHM) of the critical scattering lobe as a function of the emitter number. We use the parameters used in \cref{maxwell_normal_incidence}(\textbf{a}), a $y$-polarized field at normal incidence with energy $\omega_\kv= 0.2 (2 \pi/d)$ and outgoing energy $\omega_{\kv'}= 3 (2\pi /d)$, such that the modes with $\pv=[\pm 3,0] (2\pi/d)$ in the positive/negative $x$-direction are critical and we vary the emitter number between 5$\times$5 and 60$\times$60 emitters. Considering a point along the $x$-axis with $y=z=0$, the electric field strength is strongest in the array plane at $z=0$ and, with increasing distance $|z|$ from the array, the field strength decays (with oscillations). To determine the FWHM, for every point along the $x$-axis ($y=0$) we determine the value of $z_\text{FWHM} (x)$ at which the field strength has decayed to half the peak strength it had at $z=0$. This is shown in \cref{FWHM_analysis}(\textbf{a}) for varying emitter numbers. 

As can be seen in \cref{FWHM_analysis}(\textbf{a}), moving along the $x$-axis away from the array, with increasing distance the FWHM increases: the lobe gets broader in $z$-direction. This increase is almost linear and shows oscillations around an increasing mean. Closer to the array (at low values of $x$) these oscillations are stronger and for this reason we only determine the FWHM outside of the array (the size of which varies with emitter number). With increasing emitter number, the oscillations around the mean become weaker, the FWHM decreases and, most importantly, the rate at which the FWHM increases with $x$ becomes smaller. In \cref{FWHM_analysis}(\textbf{b}), we show the spread rate as a function of the emitter number. The spread rate is determined as the slope of a linear fit of $z_\text{FWHM} (x)$ and corresponds to the sine of the opening angle $\theta$ of the lobe, $\sin(\theta)$. The inset shows the corresponding opening angle $\theta$. It can be seen that the spread rate decreases with increasing emitter number and for $N\times N$ emitters the spread rate is approximately proportional to $\propto 1/\sqrt{N}$ as has also been noted recently \cite{Rusconi_priv_comm}.

\section{Discussion and Outlook}\label{sect:discussion}

In this work, we have presented a transduction scheme using driven four-wave mixing in a cooperative, two-dimensional emitter array, which allows for efficient and mode-selective one-to-one transduction of single THz (signal) photons into well-defined optical (idler) photons. To this end, we have developed an $S$-matrix scattering formalism that treats the photonic field explicitly and which may be equivalently related to a semi-classical treatment using Maxwell's equations, such that both finite and infinite emitter arrays may be studied. In the envisioned use scenario ---emitter arrays which are cooperative with respect to the signal photons but uncooperative with respect to idler photons--- high overall transduction efficiencies can be achieved (we have observed efficiencies of up to $90$\% in this work), which may however spread across many photonic idler modes, such that the outgoing signal propagates into many directions. As a result, the outgoing signal at any spatial point is a superposition of many plane waves, and therefore does not have a well-defined wavevector.

However, satisfying a resonance condition on the signal transition and a criticality condition on the idler transition, the transduction process may become highly-directional, in that it allows transduction from specific incoming photon modes into specific outgoing idler photon modes with efficiency of up to $50\%$. Notably, in this case the transduction is almost exclusively into the specific, critical modes, which are highly directional and have a well-defined wavevector. The remainder of the scattered signal consists of signal photon modes at the original signal frequency. Importantly, to match the incoming photon, the resonance and criticality conditions may be modified in-situ by altering parameters such as the lattice spacing, the drive grating or the incidence angle of the signal. As a result, frequency detunings of the incoming signal photon from the signal transition can be compensated for and a broad sensitivity window can be swept. 

In this work, we did not account for non-radiative losses, which would likely weaken the transduction efficiency, though it is unclear to what extent. When the array is cooperative with respect to the signal transition, the corresponding cooperative decay width scales as $(\lambda/d)^2$, which may dominate non-radiative decay widths. Similarly, on the idler transition this transduction scheme builds on coupling to superradiant modes whose cooperative decay widths diverge with emitter number. Thus it remains an open question to what extent non-radiative losses have a detrimental effect on this transduction scheme and whether these effects can be partially cured by considering larger emitter arrays. Beyond that, our work does not touch upon the time scales imposed by the coherent population trapping in the lower $\Lambda$-system, which needs to be sufficiently fast to provide the trapping into the dark state. While this time scale has been studied for single atoms, it may also experience cooperative corrections. Similarly, apart from the coherence time imposed by the population trapping, the transduction process itself is not instantaneous and an excitation will remain in the array for a given amount of time before being emitted. During this time, other photons arriving will not find the array in the dark ground state and as a result nonlinear effects may arise. While this may lead to interesting interaction effects, it may also limit the detection frequency. In principle, our approach may also be extended to the nonlinear regime to treat the simultaneous incidence of two or more photons, albeit with increasing analytical complexity as the number of photons increases.

In the transduction process, the dipolar modes of the signal and the idler transition both play a role. The cooperativity of the dipolar signal modes along with the resonance condition ensure efficient coupling of an incoming photon into the array. Meanwhile, the criticality condition on the idler transition ensures a channeled photon emission. While we have shown that for the frequencies and array extensions considered in \cref{maxwell_calcs} appreciable efficiency and directionality of the transduction process may be achieved (which improves as $1/\sqrt{N}$), it may also be beneficial to understand  how signal and idler transition frequencies impose requirements on the emitter number to reach a specified level of efficiency and directionality. 

To be useful for quantum information processing applications, the transduction process must be both efficient at the single photon level and preserve phase coherence of input superposition states \cite{VanDevender07}.  Future work could investigate how cooperativity affects the preservation of phase coherence and explore applications beyond simply transducing an incoming signal to a frequency for which efficient detectors are available, such as performing further processing on the transduced signal. While for some applications having several critical idler modes to couple into may be undesirable, it may also be a feature: At normal incidence, the polarization of the incoming signal may determine which modes may be coupled into and at which strength. Interfering the emission into different directions at a single site may thus be used to determine polarization information. Beyond a single array,  combining several such sensors into a network could allow for high-resolution sparse-aperture imaging or field interferometry of weak signals \cite{Gottesman2012,Khabiboulline2019}. For example, several Rydberg array devices could be used for a phase interferometry scheme where coupled optical fibers at each array collect the transduced photonic state. Such interferometric schemes could then extract information about the spatial propagation of the incoming signal photons.

While dark, subradiant dipolar modes are typically hard to couple into due to their weak coupling to the electric field, this scheme may also be used to couple into modes that would otherwise be dark, as the transduction process  mixes dipolar modes between the two transitions. After coupling into a dark state, the mixing process may then be turned off either through control of the driving lasers or, using Rydberg interactions, by switching the internal state of a nearby Rydberg atom.  

\section{Acknowledgements}
 We thank Aziza Suleymanzade, Stefan Ostermann, Oriol Rubies-Bigorda and Cosimo Rusconi for inspiring discussions. We thank the NSF through QuSEC (OMA-2326787), PHY-2207972, the CUA PFC (PHY-2317134) and AFOSR through FA9550-24-1-0311. K.W.S. acknowledges support from the Natural Sciences and Engineering Research Council of Canada (NSERC) through a PGS D fellowship.

\appendix

\begin{widetext}
    
\section{Calculation of matrix elements}\label{app_explicitcalculation}
To compute the matrix elements of the resolvent $\frac{1}{\rho -\hat{H}}$ and the $\hat{S}$ operator explicitly we note that the action of the interacting part of the Hamiltonian on the single-excitation states is given by 

\begin{align}
        \hat{V}\ket{\kv\mu a }&=  \frac{1}{\sqrt{V}}\sum_{i,n} g_{\kv \mu, n,a }A^*  e^{i \kv_{L_2} \rv_i }e^{i \kv \rv_i }\ket{i,n},\\
    \hat{V}\ket{\kv\mu b }&=  \frac{1}{\sqrt{V}}\sum_{i,n} g_{\kv \mu, n,b }B^*  e^{i \kv_{L_1} \rv_i }e^{i \kv \rv_i }\ket{i,n},\\
       \hat{V}\ket{i,n}&=  \frac{1}{\sqrt{V}}\sum_{\kv'\mu'} g_{\kv'\mu',n, a} A e^{-i \kv' \rv_i } e^{-i \kv_{L_2} \rv_i }\ket{\kv'\mu' a}\\
       &+ \frac{1}{\sqrt{V}}\sum_{\kv'\mu'} g_{\kv'\mu',n, b} B e^{-i \kv' \rv_i } e^{-i \kv_{L_1} \rv_i }\ket{\kv'\mu' b},
\end{align}
such that
\begin{align}
    \bra{\kv'\mu'b} \frac{1}{\rho-\hat{H}} \ket{\kv\mu a}&=  \frac{1}{\sqrt{V}}\sum_{i,n} \underbrace{\frac{g_{\kv\mu,n,a} A^*}{\rho- (\omega_{\kv}- \omega_{e_1}+ \omega_{g_1})} }_{\equiv H_{\kv\mu,n,a}}  \bra{\kv'\mu'b} \frac{1}{\rho-\hat{H}} \ket{i,n} e^{i \kv \rv_i} e^{i \kv_{L_2} \rv_i },\\
        \bra{\kv'\mu'b} \frac{1}{\rho-\hat{H}} \ket{\kv\mu b}&=\frac{\delta_{\kv' \mu', \kv \mu }}{\rho- (\omega_\kv- \omega_{e_1}+ \omega_{g_2})} \\
        &+\frac{1}{\sqrt{V}}\sum_{i,n} \underbrace{\frac{g_{\kv\mu,n,b} B^*}{\rho- (\omega_{\kv}- \omega_{e_1}+ \omega_{g_2})} }_{\equiv H_{\kv\mu,n,b}}  \bra{\kv'\mu'b} \frac{1}{\rho-\hat{H}} \ket{i,n} e^{i \kv \rv_i} e^{i \kv_{L_1} \rv_i },\\
      \bra{\kv'\mu'b} \frac{1}{\rho-\hat{H}} \ket{i,n}&=  \frac{1}{\sqrt{V}} \sum_{\kv''\mu''} \underbrace{\frac{g_{\kv''\mu'',n, a} A }{\rho}}_{\equiv L_{\kv''\mu'',n,a}}  \bra{\kv'\mu'b} \frac{1}{\rho-\hat{H}} \ket{\kv''\mu''a} e^{-i \kv'' \rv_i} e^{-i \kv_{L_2} \rv_i }\\
      &+  \frac{1}{\sqrt{V}} \sum_{\kv''\mu''} \underbrace{\frac{g_{\kv''\mu'',n, b} B }{\rho}}_{\equiv L_{\kv''\mu'',n,b}}  \bra{\kv'\mu'b} \frac{1}{\rho-\hat{H}} \ket{\kv''\mu''b} e^{-i \kv'' \rv_i} e^{-i \kv_{L_1} \rv_i }.
\end{align}
Note that here $V=L^3$ corresponds to the volume of the system and is not to be confused with the interacting part $\hat{V}$ of the Hamiltonian. Solving this set of equations one obtains 
\begin{align}
    \bra{\kv'\mu'b} \frac{1}{\rho-\hat{H}} \ket{i,n}&= \frac{1}{\sqrt{V}} \sum_{m}  \left( \mathbb{I}- P[\kv_\parallel^{'},b]\right)^{-1}_{n,m}   \frac{L_{\kv^{'}\mu',m,b}}{\rho- (\omega_{\kv'}- \omega_{e_1}+ \omega_{g_2})} e^{-i \kv^{'}_\parallel \rv_i} e^{-i \kv_{L_{1}} \rv_i }, 
\end{align}
where 
\begin{align}
    P(\kvp,\sigma')&\equiv \sum_\sigma D(\kvp +\kv_{L_\sigma}- \kv_{L_{\sigma'}},\sigma)_{m,n},\\
     D(\kvp ,\sigma)_{m,n}&\equiv \sum_{\pv} \frac{1}{L d^2} \sum_{k_\perp \mu} L_{\kvp  + \pv, k_\perp\mu,m,\sigma}  H_{\kvp  + \pv, k_\perp\mu,n,\sigma} , \label{D_def}
\end{align} 
and we have made use of the identity 
\begin{align}
    \sum_i e^{i \kv \rv_i}= \left(\frac{2 \pi}{d}\right)^2 \sum_{\pv} \delta(\kvp- \pv),
\end{align}
with $\kvp$  the component of $\kv=(\kvp, k_\perp)$ that is parallel to the lattice plane and $\pv \in (2 \pi /d) \mathbb{Z}_2$ a vector within the reciprocal space of the lattice. From the definition in \cref{D_def} on can see, that $D(\kvp+ \pv ,\sigma)=D(\kvp ,\sigma)$, i.e. it is invariant under translation by reciprocal lattice vectors.  Using
\begin{align}
    \rho = \frac{\omega_{\kv} + \omega_{\kv'} - 2 \omega_{e_1} + \omega_{g_1} + \omega_{g_2}}{2}+ i 0^{+}= \omega_{\kv}+ \omega_{g_1}- \omega_{e_1}+ i 0^{+}
\end{align} one thus finds that 
\begin{align}
 \bra{\kv' \mu' b} \hat{S} \ket{\kv \mu a}&=-\frac{ i }{ d^2}\sum_{\pv nm}   \Bigg(\delta_{\kvp, \kv_{\parallel}^{'}- \pv+ \delta \kv_\parallel} \delta_{k^{'}_z,\pm \sqrt{ -(\kv_\parallel + \pv- \delta \kv_\parallel)^2 + \left(\omega_{\kv}+ \omega_{g_1}- \omega_{g_2}\right)^2} } \\
& \times \frac{A^* B g_{\kv \mu, n,a } g_{\kv'\mu',m, b} (\omega_{\kv}+ \omega_{g_1}- \omega_{g_2})}{\sqrt{ -(\kv_\parallel + \pv- \delta \kv_\parallel)^2 + \left(\omega_{\kv}+ \omega_{g_1}- \omega_{g_2}\right)^2}}  \left( \rho \mathbb{I}- \rho P(\kv_\parallel,a)\right)^{-1}_{n,m}
\\
 &\times
  \Theta\left[\omega_{\kv}+ \omega_{g_1}- \omega_{g_2}\geq \left|\kv_\parallel + \pv-\delta \kv_\parallel \right| \right] \Bigg).
\end{align}
Similarly, using 
\begin{align}
    \rho=\frac{\omega_\kv+ \omega_{\kv'}}{2}  + \omega_{g_1} - \omega_{e_1}+ i 0^+ = \omega_\kv+ \omega_{g_1} - \omega_{e_1}+ i 0^+ 
\end{align}
one finds 
\begin{align}
     \bra{\kv' \mu' a} \hat{S} \ket{\kv \mu a}=\delta_{\kv'\mu',\kv\mu}    - \frac{  i}{ d^2}\sum_{\pv nm } &\Bigg( \delta_{\kv_\parallel^{'},\kv_\parallel+ \pv}\delta_{k_{\perp}^{'}, \pm \sqrt{\omega_{\kv}^2- (\kvp+ \pv)^2}} \Theta\left[  \omega_{\kv} \geq |\kv_\parallel + \pv| \right]\\
     &\times |A|^2  \left( \rho \mathbb{I}- \rho P(\kv_\parallel,a)\right)^{-1}_{n,m}   \frac{ g_{\kv \mu, n,a}  g_{\kv'\mu',m,a}\omega_{\kv}}{\sqrt{\omega_{\kv}^2- (\kvp+ \pv)^2}} \Bigg).
\end{align}
Finally, for $\rho=(\omega_\kv- \omega_{e_1}+ \omega_{g_1})/2 + i0^+ $ one obtains
\begin{align}
    \bra{j,o} \hat{S} \ket{\kv \mu a}&=\frac{1}{2}  \frac{1}{\sqrt{V}}  \sum_m g_{\kv\mu, m,a} A^{*} e^{i \kv \rv_j } e^{i \kv_{L_2} \rv_j}\left(1+\frac{\rho}{\rho - (\omega_\kv - \omega_{e_1}+ \omega_{g_1})}\right) \left( \rho  \mathbb{I}- \rho P(\kvp, a)\right)^{-1}_{m,o} ,
\end{align}
which vanishes unless $\rho=i 0^+ \leftrightarrow
\omega_\kv= \omega_{e_1}- \omega_{g_1}$ in which case the term in parentheses will provide a factor of 2. Even then, for an infinite volume system, i.e. $V\to \infty$, this term will not contribute to scattering amplitudes.

As a result, the action of the $\hat{S}$ operator is given by  
\begin{align}
\hat{S}\ket{(\kvp,k_\perp) \mu a}&=\ket{(\kvp,k_\perp) \mu a}\nnl
&-\frac{ i }{ d^2}\sum_{\mu' \pv nm}    \sum_{s=+,-}\Bigg[ \frac{|A|^2 g_{\kv \mu, n,a } g_{\kv'(\pv, s,a) \mu',m, a} \omega_{\kv}}{\sqrt{ -(\kvp +\pv)^2 + \omega_{\kv}^2} } \left( \rho \mathbb{I}- \rho P(\kvp,a)\right)^{-1}_{n,m} \nnl
&\times \Theta\left(\omega_\kv \geq|\kvp +\pv |\right)\ket{\kv'(\pv, s,a) \mu' a} \Bigg]\nnl
&-\frac{ i }{ d^2}\sum_{\mu' \pv nm}    \sum_{s=+,-}\Bigg[ \frac{A^* B g_{\kv \mu, n,a } g_{\kv'(\pv, s,b) \mu',m, b} (\omega_{\kv}+ \omega_{g_1}- \omega_{g_2})}{\sqrt{ -(\kvp +\pv+\kv_{L,ba})^2 + \left(\omega_{\kv}+ \omega_{g_1}- \omega_{g_2}\right)^2} } \left( \rho \mathbb{I}- \rho P(\kvp,a)\right)^{-1}_{n,m} \nnl
&\times \Theta\left(\omega_\kv+ \omega_{g_1}- \omega_{g_2}\geq|\kvp +\pv+ \kv_{L,ba}|\right) \ket{\kv'(\pv, s,b) \mu' b}   \Bigg],\label{Soperatoraction}
\end{align}
where
\begin{align}
    \kv'(\pv, s,a) &=\left(\kvp +\pv, s \sqrt{ -(\kvp + \pv)^2 + \omega_\kv^2}\right),\\
    \kv'(\pv, s,b) &=\left(\kvp +\pv+\kv_{L,ba}, s \sqrt{ -(\kvp + \pv+ \kv_{L,ba})^2 + \left(\omega_\kv+ \omega_{g_1}- \omega_{g_2}\right)^2}\right).
\end{align}

When using these equations, special care needs to be taken when using the Kronecker-$\delta$ for the momentum in $z$-direction. Where applicable, the Kronecker-$\delta$ should be regarded as short-hand for: $\delta_{x,y}=\frac{2\pi}{L} \delta(x-y)$ to avoid spurious divergences. Similarly $\frac{1}{L}\sum_{k_z} $ should be regarded as  $\frac{1}{2 \pi} \int dk_z$ in these cases.

\subsection{Probability to scatter into $a$ and $b$ modes}\label{App_a_to_b_efficiency}

In the following we compute the probability for an incoming photon in $\ket{(\kvp,k_\perp) \mu a}$ to be scattered into a $b$ mode. To this end, we split the $\hat{S}$ operator into its projections onto $a$ and $b$ modes $\hat{S}=\hat{S}_a+\hat{S}_b$ and compute 
\begin{align}
   &\bra{(\kvp,k_\perp^{'}) \mu a}\hat{S}^{\dagger}_b  \hat{S}^{\phantom{\dagger}}_b\ket{(\kvp,k_\perp^{\phantom{'}}) \mu a}= \frac{ |A|^2  |B|^2  \omega_{(\kvp, k_\perp)}}{ k_\perp d^4} \sum_{\mu' \pv nmn'm'}    \sum_{s=+,-} \Bigg[\left( \rho \mathbb{I}- \rho P(\kvp,a)\right)^{-1}_{n,m} \left( \rho \mathbb{I}- \rho P(\kvp,a)\right)^{*,-1}_{n',m'}\nnl 
   &\times \frac{g_{(\kvp, k_\perp) \mu, n,a }g_{(\kvp, k^{'}_\perp) \mu, n',a }  g_{\kv'(\pv, s,b) \mu',m, b} g_{\kv'(\pv, s,b) \mu',m', b} (\omega_{(\kvp, k_\perp)}+ \omega_{g_1}- \omega_{g_2}) }{\sqrt{ -(\kvp +\pv+\kv_{L,ba})^2 + \left(\omega_{(\kvp, k_\perp)}+ \omega_{g_1}- \omega_{g_2}\right)^2}  }  \nnl
&\times \Theta\left(\omega_{(\kvp, k_\perp)}+ \omega_{g_1}- \omega_{g_2}\geq|\kvp +\pv+ \kv_{L,ba}|\right)   \frac{2\pi}{L}  \delta\left(k_\perp \pm k_\perp^{'}\right)\Bigg],
\end{align}
such that after setting $k_\perp^{'}= k_\perp$ one obtains 
\begin{align}
   &| \hat{S}_b\ket{(\kvp,k_\perp) \mu a}|^2=\bra{(\kvp,k_\perp) \mu a}\hat{S}^{\dagger}_b  \hat{S}^{\phantom{\dagger}}_b\ket{(\kvp,k_\perp) \mu a}\nnl 
   &=\frac{ |A|^2    \omega_{\kv}}{ k_\perp d^2} \sum_{ nmn'm'} \left( \rho \mathbb{I}- \rho P(\kvp,a)\right)^{-1}_{n,m} \left( \rho \mathbb{I}- \rho P(\kvp,a)\right)^{*,-1}_{n',m'}g_{\kv \mu, n,a }g_{\kv \mu, n',a }  2 i \Im \rho D(\kvp+ \kv_{L,ba},b)_{m, m'}.\label{abconveff}
\end{align} Details on how the imaginary part of $\rho D(\kvp+ \kv_{L,ba},b)$ was obtained can be found in \cref{App_Dyadic_Green}. Similarly, one obtains 

\begin{align}
    &| \hat{S}_a\ket{(\kvp,k_\perp) \mu a}|^2=1- i  \frac{  |A|^2    \omega_{\kv}}{ k_\perp d^2} \sum_{ nm} \left[\left( \rho \mathbb{I}- \rho P(\kvp,a)\right)^{-1}_{n,m} -\left( \rho \mathbb{I}- \rho P(\kvp,a)\right)^{*,-1}_{n,m} \right] g_{\kv \mu, n,a }g_{\kv \mu, n',a }  \nnl
   & +\frac{ |A|^2    \omega_{\kv}}{ k_\perp d^2} \sum_{ nmn'm'} \left( \rho \mathbb{I}- \rho P(\kvp,a)\right)^{-1}_{n,m} \left( \rho \mathbb{I}- \rho P(\kvp,a)\right)^{*,-1}_{n',m'}g_{\kv \mu, n,a }g_{\kv \mu, n',a }  2 i \Im \rho D(\kvp,a)_{m, m'} ,
\end{align} and one can easily verify that
\begin{align}
    \left| \hat{S}\ket{(\kvp,k_\perp) \mu a}\right|^2= \left| \hat{S}_a\ket{(\kvp,k_\perp) \mu a}\right|^2+ \left| \hat{S}_b\ket{(\kvp,k_\perp) \mu a}\right|^2 =1. 
\end{align}

To determine the probability to be scattered into specific $b$-modes as introduced in \cref{methods_efficiency} and presented in \cref{efficiency_near_resonance,angled_fig}, one can define a projection operator $\hat{P}_{\text{specific}}$ and compute the transduction efficiency into these specific modes as 
\begin{align}
    \left|\hat{P}_{\text{specific}} \hat{S}_b \ket{(\kvp,k_\perp) \mu a}\right|^2.
\end{align}
This projection may for example be a restriction to specific reciprocal lattice vectors $\pv \in X_C=\left\{ \pv \in \frac{2 \pi}{d} \mathbb{Z}_2| |\pv|= |C|\right\}$ or $\pv \in X_C=\left\{ \pv \in \frac{2 \pi}{d} \mathbb{Z}_2| |\pv|< |C|\right\}$, $C\in \mathbb{R}$, such that 
\begin{align}
   \hat{P}_{\text{specific}}= \sum_{\mu' \pv s} \Theta(\pv \in X_C)  \ket{\kv'(\pv, s,b) \mu' b} \bra{\kv'(\pv, s,b) \mu' b}. 
\end{align}
To compute $ \left|\hat{P}_{\text{specific}} \hat{S}_b \ket{(\kvp,k_\perp) \mu a}\right|^2$, one can then use \cref{abconveff}, while replacing $\Im \rho D(\kvp+ \kv_{L,ba},b)_{m, m'}$ with a modified function where in \cref{gparallelimaginary} only $\pv \in X_C$ are summed over.

\section{The dyadic Green's function }\label{App_Dyadic_Green}
In the following we relate the function $D(\kvp, \sigma)$ to the dyadic Green's function, the inverse of the differential operator $(- k^2\mathbb{I}+ \nabla \times \nabla\times)$ which acts on vectors with three entries. To begin, let us define the spatial function 
\begin{align}
    D(\rv,\sigma)_{m,n}&=  \frac{1}{V}\sum_{\kv'' \mu''} L_{\kv'' \mu'',m,\sigma} H_{\kv''\mu'',n,\sigma} e^{i \kv'' \rv}\\
    &=  \frac{|\sigma|^2}{\rho} \frac{1}{V}\sum_{\kv'' \mu''}  \frac{ g_{\kv'' \mu'',m,\sigma} g_{\kv''\mu'',n,\sigma}}{\rho - (\omega_{\kv''}- \omega_{e_1 }+ \omega_\sigma)} e^{i \kv'' \rv}\\
    &= \frac{ |\sigma|^2 \wp_\sigma^2}{\rho} \frac{1}{V}\sum_{\kv''}  \frac{1}{2} \frac{ \delta_{m,n}- \kv^{''}_m \kv^{''}_n / |\kv''|^2}{\rho - (\omega_{\kv''}- \omega_{e_1 }+ \omega_\sigma)} \omega_{\kv''} e^{i \kv'' \rv}\label{rhoD},
\end{align}

which is related to $D(\kvp, \sigma)$ (defined in \cref{D_def}) by an in-plane Fourier transform
\begin{align}
    \sum_i e^{-i \kvp \rv_i} D(\rv_i,\sigma)_{m,n}=\sum_i e^{-i \kvp \rv_i}  \frac{1}{V}\sum_{\kv'' \mu''} L_{\kv'' \mu'',m,\sigma} H_{\kv''\mu'',n,\sigma} e^{i \kv^{''}_{\parallel} \rv_i}= D(\kvp,\sigma)_{m,n}. \label{spatialsum_Dkvp}
\end{align}

Using the notation from Ref. \cite{J_rgensen_2022}, on can see that 
\begin{align}
D(\rv, \sigma)_{m,n}&= \frac{(\rho+ \omega_{e_1}- \omega_\sigma)^2 }{\rho }|\sigma|^2 \wp_\sigma^2  K^{+} (\rv,\rho+ \omega_{e_1}- \omega_\sigma)_{m,n}, 
\end{align}
which is related to the classical dyadic Green's function \cite{Novotny2012} by $G= K^+ - K^-$, where $K^-$ is obtained analogously by replacing $\omega_{\kv''}\to - \omega_{\kv''}$ in the denominator of \cref{rhoD}. 

These functions are given in Ref. \cite{J_rgensen_2022} and we reproduce them here for completeness 
\begin{align}
    G(\rv, k)_{m,n}&= - \frac{e^{i k r}}{4 \pi r}\left[\delta_{m,n} \left(1 + \frac{i}{k r}- \frac{1}{k^2 r^2}\right)+ \frac{\rv_m \rv_n}{r^2} \left(-1 - \frac{3i }{kr }+ \frac{3}{k^2 r^2}\right)\right],\label{Gdef}\\
    K^{+}(\rv, k)_{m,n}&= G(\rv, k)_{m,n}- \frac{1}{4 \pi^2 k r^2}\left[ I_2(kr)\left( \delta_{m,n}- \frac{\rv_m \rv_n}{r^2}\right)+  [I_1(kr)+ I_0(kr)] \left( \delta_{m,n}- 3 \frac{\rv_m \rv_n}{r^2}\right)\right], \label{RWAGdef}\\
    I_{j} (x)&= \int_{0}^{\infty} d u \frac{u^j e^{-u}}{u^2 + x^2}.
\end{align}
Notably, the rotating-wave approximation corresponds to replacing $G$ with $K^+$ \cite{J_rgensen_2022}. Hence, in \cref{app_explicitcalculation}, rather than obtaining the classical dyadic Green's function $G$, we obtain $K^+$, because the rotating-wave approximation was taken in deriving \cref{Ham}. Importantly, the imaginary parts of $G$ and $K^+$ coincide, $\Im G(\rv,k)= \Im K^+(\rv, k)$,  and the same goes for their in-plane Fourier transforms $\Im G(\kvp,k)=\Im K^+(\kvp,k)$ \cite{J_rgensen_2022}.

Formally, the real part of $D(\kvp, \sigma)$ diverges as its sum in \cref{spatialsum_Dkvp} contains a diverging term with $\rv_i= \zerov$, which leads to UV divergences (the same is true for $G(\kvp,k)$ and $K^+(\kvp,k)$). This term corresponds to the Lamb shift and can be absorbed into the atomic transition frequencies. Thus in computing $D(\kvp, \sigma)$, $G(\kvp,k)$ and $K^+(\kvp,k)$, we need to manually exclude this term, $\Re D(\rv_i=\zerov, \sigma)=\Re G(\rv_i=\zerov)= \Re K^+(\rv_i=\zerov)=0 $.  As a result, no finite, closed expressions exist for the real parts of $D(\kvp, \sigma)$, $G(\kvp,k)$ and $K^+(\kvp,k)$. However,  these three can easily be computed numerically for arbitrary lattices. 

For the imaginary part of $D(\kvp, \sigma)$  a closed, finite expression can be obtained as also noted in Ref. \cite{Shahmoon2017}
\begin{align}
& \Im \rho  P(\kvp,\sigma')_{m,n}= \Im \rho  \sum_{\sigma=a,b} \sum_\pv \frac{1}{L d^2} \sum_{k_\perp \mu} L_{\kvp + \kv_{L,\sigma\sigma'} + \pv, k_\perp\mu,m,\sigma}  H_{\kvp + \kv_{L,\sigma\sigma'} + \pv, k_\perp\mu,n,\sigma}\nnl
 &= -i\pi \sum_{\sigma=a,b} \frac{|\sigma|^2 d^2_{\sigma}}{2 \epsilon_0 L d^2}   \sum_{\pv k_\perp}\Bigg[  \left(\delta_{mn}-\frac{(\kvp + \kv_{L,\sigma\sigma'} + \pv, k_\perp)_m (\kvp+\kv_{L,\sigma\sigma'} + \pv, k_\perp)_n }{(\kvp+\kv_{L,\sigma\sigma'} + \pv)^2+ k^2_\perp} \right) \omega_{(\kvp+\kv_{L,\sigma\sigma'} + \pv, k_\perp)}\nnl
 &\quad \quad \quad \quad \quad \quad \quad \quad \quad  \times \delta\left(\rho- (\omega_{(\kvp+\kv_{L,\sigma\sigma'} + \pv, k_\perp)}-\omega_{e_1}+ \omega_\sigma)\right)\Bigg]\nnl
&= -i \sum_{\sigma=a,b} \sum_{s=+, -}  \sum_\pv \Bigg[ \frac{|\sigma|^2 d^2_{\sigma}}{4 \epsilon_0  d^2} \left(\delta_{mn}-\frac{\mathbf{X}(\kvp, \sigma, \sigma', \pv, \rho,s)_m \mathbf{X}(\kvp, \sigma, \sigma', \pv, \rho,s)_n }{(\rho+ \omega_{e_1}- \omega_\sigma)^2} \right) \nnl&
     \quad\quad\quad\quad\quad\quad\quad\quad\quad\quad \times\frac{(\rho+ \omega_{e_1}- \omega_\sigma)^2 \Theta\left((\rho+ \omega_{e_1}- \omega_\sigma)> |\kvp+\kv_{L,\sigma\sigma'} + \pv|\right)}{\sqrt{(\rho+ \omega_{e_1}- \omega_\sigma)^2-(\kvp+\kv_{L,\sigma\sigma'} + \pv)^2}} \Bigg], \label{gparallelimaginary}
\end{align}
where
\begin{align}
    \mathbf{X}(\kvp, \sigma, \sigma', \pv, \rho,s)_m&=\left(\kvp+\kv_{L,\sigma\sigma'} + \pv, s\sqrt{(\rho+ \omega_{e_1}- \omega_\sigma)^2-(\kvp+\kv_{L,\sigma\sigma'} + \pv)^2}\right)_m.
\end{align}
Notably, in \cref{gparallelimaginary} only modes with $\mathbf{X}(\kvp, \sigma, \sigma', \pv, \rho,s)_\perp \in \mathbb{R}$ may contribute to the imaginary part as enforced by the Heaviside function $\Theta$. Cooperativity limits the set of reciprocal lattice vectors $\pv$ which may contribute to the imaginary part.

For the real part, the analytical structure is more involved due to the presence of branch cuts in the square root function contained in $\omega_{(\kvp+\kv_{L,\sigma\sigma'} + \pv, k_\perp)}$
\begin{align}
    \rho P(\kvp,\sigma')_{m,n} &=  \sum_{\sigma=a,b} \frac{|\sigma|^2 d^2_{\sigma}}{2 \epsilon_0 L d^2} \sum_{\pv k_\perp}\Bigg[  \left(\delta_{mn}-\frac{(\kvp + \kv_{L,\sigma\sigma'} + \pv, k_\perp)_m (\kvp+\kv_{L,\sigma\sigma'} + \pv, k_\perp)_n }{(\kvp+\kv_{L,\sigma\sigma'} + \pv)^2+ k^2_\perp} \right) \nnl
&\times \frac{\sqrt{(\kvp+\kv_{L,\sigma\sigma'} + \pv)^2+  k_\perp^2}}{\rho+\omega_{e_1}- \omega_\sigma-\sqrt{(\kvp+\kv_{L,\sigma\sigma'} + \pv)^2+ k_\perp^2}}\Bigg],\label{real_part}
\end{align}
but also due to the uncured UV divergences present in the unregularized expressions. Within \cref{real_part}, there are different origins that result in UV divergences to arise. First, the integral over $k_\perp$ is not formally converging for some of the entries as not all of the entries decay faster than $1/k_\perp$. This divergence can be cured by evaluating the in-plane Fourier transform in \cref{spatialsum_Dkvp} slightly above or below the array plane at $z= \pm 0^+$ rather than at $z=0$ within the array plane. This will lead to factors of $e^{\pm i  k_\perp 0^+  }$, which generate convergence when closed in the upper/lower half. For example, on may take the average of transforms for $z=0^+$ and $z=0^-$.  Second, even if the integration over $k_\perp$ is converging, then the summation over lattice vectors may still yield a UV divergent expression. The latter divergence can be regularized by introducing an upper momentum cutoff and subtracting the Lamb shift \cite{ThesisDominikWild}.

While the UV divergences will be cured upon regularization, it instructive to consider the instabilities of the $P(\kvp,\sigma')$ matrix, which takes the role of a dipolar self-energy. While the eigenvectors and eigenvalues of $D(\kvp, \sigma)$ are related to the spectrum of dipolar surface modes, the instabilities of $P(\kvp,\sigma')= \sum_\sigma D(\kvp+ \kv_{L,\sigma \sigma'}, \sigma)$ occur when due to a choice of $(\rho + \omega_{e_1}- \omega_\sigma)$, $\kvp$ and $\kv_{L,\sigma \sigma'}$ the regularized dipolar self-energy diverges. That is, at least one of the eigenvalues of the $D(\kvp+ \kv_{L,\sigma \sigma'}, \sigma)$ diverges and this divergence is due to the choice of parameters and not due to regularization issues. For the classical dyadic Green's function these instabilities have been described \cite{Shahmoon2017,Asenjo-Garcia2017} and in the following we will show that under the rotating-wave approximation one obtains the same instability conditions.

To begin, we note that while the imaginary part in  \cref{gparallelimaginary} is finite for every combination of $(\rho+ \omega_{e_1}- \omega_\sigma)$, $\kvp$ and $\kv_{L,\sigma \sigma'}$, it is not bounded because of the possibility to choose parameter combinations for which reciprocal lattice vectors exist, such that $(\rho+ \omega_{e_1}- \omega_\sigma)^2-(\kvp+\kv_{L,\sigma\sigma'} + \pv)^2$ may become arbitrarily close to $0$. These, however, are only the free space mode contributions, which contribute to the imaginary part. The real part contains further information about contributions from evanescent modes, which may lead to further instabilities.

To analyze the structure of $P(\kvp, \sigma')$ and its instabilities, we first consider the expression of the in-plane Fourier transform that is obtained without the rotating-wave approximation. Upon reversing  the rotating-wave approximation by subtracting $\rho P(\kvp,\sigma')^{-}_{m,n}$ (obtained by replacing $\sqrt{(\kvp+\kv_{L,\sigma\sigma'} + \pv)^2+ ( k_\perp)^2}\to -\sqrt{(\kvp+\kv_{L,\sigma\sigma'} + \pv)^2+ ( k_\perp)^2}$ in the denominator of  \cref{real_part}, analogous to $K^-$) the branch cuts along the imaginary axis in $k_\perp$ disappear and only simple poles are left at $k_\perp= \pm  \sqrt{(\rho+\omega_{e_1}- \omega_\sigma)^2-(\kvp+\kv_{L,\sigma\sigma'} + \pv)^2}$. Thus, closing the contour either in the upper or lower half of the complex plane, one may  carry out a contour integration in $k_\perp$ to obtain
\begin{align}
\rho P(\kvp,\sigma')_{m,n}- \rho P(\kvp,\sigma')^{-}_{m,n} &=\sum_{\sigma=a,b} \frac{|\sigma|^2 d^2_{\sigma}}{ \epsilon_0 L d^2}   \sum_{\pv k_\perp}\Bigg[  \left(\delta_{mn}-\frac{(\kvp + \kv_{L,\sigma\sigma'} + \pv, k_\perp)_m (\kvp+\kv_{L,\sigma\sigma'} + \pv, k_\perp)_n }{(\kvp+\kv_{L,\sigma\sigma'} + \pv)^2+ k^2_\perp} \right) \nnl
&\times\frac{(\kvp+\kv_{L,\sigma\sigma'} + \pv)^2+  k_\perp^2}{(\rho+\omega_{e_1}- \omega_\sigma)^2-(\kvp+\kv_{L,\sigma\sigma'} + \pv)^2-  k_\perp^2} \Bigg]\nnl
&=-  i  \sum_{\sigma=a,b} \frac{|\sigma|^2 d^2_{\sigma}}{2 \epsilon_0  d^2}   \sum_{\pv }\Bigg[  \left(\delta_{mn}-\frac{  \mathbf{X}(\kvp, \sigma, \sigma', \pv, \rho,+)_m  \mathbf{X}(\kvp, \sigma, \sigma', \pv, \rho,+)_n }{(\rho+\omega_{e_1}- \omega_\sigma)^2} \right) \nnl
&\times\frac{(\rho+\omega_{e_1}- \omega_\sigma)^2}{\sqrt{(\rho+\omega_{e_1}- \omega_\sigma)^2-(\kvp+\kv_{L,\sigma\sigma'} + \pv)^2}} \Bigg]+ (\text{UV divergent terms}), \label{fullGexpression}
\end{align}
where due to the anti-symmetry of the integrand in $k_\perp$ for $m=z, n=x,y$ and $n=z, m=x,y$ we require the expression to vanish identically for these indices. As it can be seen, an expression similar to \cref{gparallelimaginary} is obtained which furthermore includes contributions from evanescent modes. The UV divergent terms in \cref{fullGexpression} stem from the $k_\perp$ integration over the infinite arc in the upper or lower complex half, which due to the evaluation at $z=0$ does not vanish. As mentioned before, performing the Fourier transform infinitesimally above or below the array plane (i.e. at $z=\pm 0^+$) instead, the convergence generating factor of $e^{\pm i k_\perp 0^+}$  causes the UV divergent terms in \cref{fullGexpression} to vanish. The remaining UV divergences in \cref{fullGexpression} stemming from the summation over reciprocal lattice vectors can then be cured introducing an upper-momentum cutoff and subtracting the Lamb shift \cite{ThesisDominikWild}. From \cref{fullGexpression} it can be seen that for some combinations of  $(\rho+ \omega_{e_1}- \omega_\sigma)$, $\kvp$ and $\kv_{L,\sigma \sigma'}$,  reciprocal lattice vectors exist, such that $(\rho+ \omega_{e_1}- \omega_\sigma)^2-(\kvp+\kv_{L,\sigma\sigma'} + \pv)^2\to 0$, causing $\rho P(\kvp,\sigma')- \rho P(\kvp,\sigma')^{-}$ to diverge. In other words, the poles at $k_\perp= \pm  \sqrt{(\rho+\omega_{e_1}- \omega_\sigma)^2-(\kvp+\kv_{L,\sigma\sigma'} + \pv)^2}$ approach the origin of the complex plane. These are the instability conditions of the Green's function when no rotating-wave approximation is undertaken.  

Finally, we consider $\rho P(\kvp,\sigma')^{-}$ to show that the instability conditions remain the same 
\begin{align}
    \rho P(\kvp,\sigma')^{-}_{m,n} &=\sum_{\sigma=a,b} \frac{|\sigma|^2 d^2_{\sigma}}{2 \epsilon_0 L d^2}   \sum_{\pv k_\perp}\Bigg[  \left(\delta_{mn}-\frac{(\kvp + \kv_{L,\sigma\sigma'} + \pv, k_\perp)_m (\kvp+\kv_{L,\sigma\sigma'} + \pv, k_\perp)_n }{(\kvp+\kv_{L,\sigma\sigma'} + \pv)^2+ k^2_\perp} \right) \nnl
&\times \frac{\sqrt{(\kvp+\kv_{L,\sigma\sigma'} + \pv)^2+  k_\perp^2}}{\rho+\omega_{e_1}- \omega_\sigma+\sqrt{(\kvp+\kv_{L,\sigma\sigma'} + \pv)^2+ k_\perp^2}}\Bigg]. \label{Pminus}
\end{align} 
Again, due to the anti-symmetry of the integrand in $k_\perp$ for $m=z, n=x,y$ and $n=z, m=x,y$ this expression vanishes identically. For the indices $(m,n)=(z,z)$ , $(x,y)$ and $(y,x)$ the integrand decays faster than $1/k_\perp$ and hence the $k_\perp$-integration may be carried out easily. Finally, for $(m,n)= (x,x)$ and $(y,y)$, the integration may be carried out including the convergence generating factor. While this leads to a UV divergent term (upon taking $z\to 0$ this expression  diverges logarithmically), importantly for $|z|>0 $ there is no combination $(\rho+ \omega_{e_1}- \omega_\sigma)$, $\kvp$ and $\kv_{L,\sigma \sigma'}$ which may lead to a divergence in \cref{Pminus}, beyond the curable UV divergences. This can also be seen by noting that the last term in \cref{Pminus} does not have a pole with respect to $k_\perp$. While there is a branch cut along the imaginary axis starting at $k_\perp= \pm i \sqrt{(\kvp+\kv_{L,\sigma\sigma'} + \pv)^2} $, when $(\kvp+\kv_{L,\sigma\sigma'} + \pv)^2=0$, the integration may be carried out using the convergence generating factor and the limit $z\to 0$ may be taken safely without any divergence.

Thus, we can conclude  that there are no parameter combinations for which $\rho P(\kvp, \sigma')^{-}$ becomes diverging beyond the uncured UV divergences. Since $\rho P(\kvp, \sigma')=\rho P(\kvp, \sigma')- \rho P(\kvp, \sigma')^{-}+ P(\kvp, \sigma')^{-}$, it thus  has the same instability conditions as the classical dyadic Green's function, namely combinations of   $(\rho+ \omega_{e_1}- \omega_\sigma)$, $\kvp$ and $\kv_{L,\sigma\sigma'}$ for which reciprocal lattice vectors exist, such that $\sqrt{(\rho+ \omega_{e_1}- \omega_\sigma)^2-(\kvp+\kv_{L,\sigma\sigma'} + \pv)^2} \to 0$.

\section{Solution of classical Maxwell equations in steady state}\label{App_Maxwell}
To illuminate how the quantum mechanical scattering problem corresponds to a semiclassical solution of classical Maxwell equations, let us consider how an array of emitters, each in their single-emitter steady state, scatters electric fields of different frequencies. This is an extension of the treatment performed in Ref. \cite{Shahmoon2017} to include four-wave mixing. 

To this end, let us consider the steady state polarizability of an atom at site $i$ for strong, resonant drives of the $g_{1,i}\leftrightarrow e_{2,i}$ and $g_{2,i}\leftrightarrow e_{2,i}$ transitions as described in \cref{app_explicitcalculation}. For weak fields $\Ev_a $ and $\Ev_b $ acting on their respective transitions,  $g_{1,i}\leftrightarrow  e_{1,i}$ and $g_{2,i}\leftrightarrow  e_{1,i}$, the single atom is effectively trapped in the dark state $\ket{\text{dark}_i}$ corresponding to the strong drive. As a result, the steady-state polarization on these two transitions is given by \cite{Lukin2000} 
\begin{align}
    \mathbf{P}_a (\rv_i)&=i \frac{\wp_{a} }{\frac{\gamma_a + \gamma_b}{2}- i \delta} \left( \wp_{a} |A|^2 \Ev_a(\rv_i) + \wp_{b} B^* A e^{i( \kv_{L_1} - \kv_{L_2})\rv_i} \Ev_b(\rv_i)\right)\nnl
    &\equiv \alpha_{aa} \Ev_a(\rv_i) + \alpha_{ab} e^{i( \kv_{L_1} - \kv_{L_2})\rv_i}\Ev_b(\rv_i)\\
    \mathbf{P}_b (\rv_i)&=i \frac{\wp_{b} }{\frac{\gamma_a + \gamma_b}{2}- i \delta} \left( \wp_{b} |B|^2 \Ev_b(\rv_i) + \wp_{a} A^* B  e^{i( \kv_{L_2} - \kv_{L_1})\rv_i}  \Ev_a(\rv_i)\right)\nnl
    &\equiv \alpha_{ba}  e^{i( \kv_{L_1} - \kv_{L_2})\rv_i} \Ev_a(\rv_i) + \alpha_{bb} \Ev_b(\rv_i), 
\end{align}
where $\alpha_{aa}\alpha_{bb}= \alpha_{ab}\alpha_{ba}$ and $\gamma_{a}, \gamma_b$ denote the vacuum decay rates on the $g_{1,i}\leftrightarrow  e_{1,i}$ and $g_{2,i}\leftrightarrow  e_{1,i}$ transitions, projected onto the dark state $\ket{\text{dark}_i}$ (see also \cref{polarizability}). The frequencies $k_a$ and $k_b$ of the electric fields $\Ev_a$ and $\Ev_b$ are both detuned from the atomic transition frequencies by the same detuning $\delta$.

For an incoming field $\Ev_a^0$, Maxwell's equations thus yield 
\begin{align}
    \Ev_a (\rv)&= \Ev_a^0 (\rv)- k_a^2 \sum_i G(\rv- \rv_i, k_a)\left[\alpha_{aa} \Ev_a(\rv_i) + \alpha_{ab}e^{i( \kv_{L_1} - \kv_{L_2})\rv_i} \Ev_b (\rv_i)\right],\\
    \Ev_b (\rv)&= \phantom{\Ev_a^0 (\rv)}- k_b^2 \sum_i G(\rv- \rv_i, k_b)\left[\alpha_{ba}e^{i( \kv_{L_2} - \kv_{L_1})\rv_i} \Ev_a(\rv_i) + \alpha_{bb} \Ev_b (\rv_i)\right],
\end{align}
and evaluated at the lattice sites one obtains
\begin{align}
    \Ev_a (\rv_j)&= \Ev_a^0 (\rv_j)- k_a^2 \sum_{i\neq j} G(\rv_j- \rv_i, k_a)\left[\alpha_{aa} \Ev_a(\rv_i) + \alpha_{ab}e^{i( \kv_{L_1} - \kv_{L_2})\rv_i} \Ev_b (\rv_i)\right],\label{Easite}\\
    \Ev_b (\rv_j)&= \phantom{\Ev_a^0 (\rv)}- k_b^2 \sum_{i\neq j} G(\rv_j- \rv_i, k_b)\left[\alpha_{ba}e^{i( \kv_{L_2} - \kv_{L_1})\rv_i} \Ev_a(\rv_i) + \alpha_{bb} \Ev_b (\rv_i)\right]. \label{Ebsite}
\end{align}
Note, that in \cref{Easite,Ebsite} the $i=j$ terms corresponding to $\rv_j-\rv_i=\zerov$ need to be omitted as we have absorbed the Lamb shift into the atomic transition frequencies $\omega_{e_1}- \omega_\sigma$ and the vacuum decay rates $\gamma_\sigma=-2 k^2_\sigma \wp_\sigma^2 |\sigma|^2  \Im G(\rv=0, k_\sigma) $  are already contained within the single-site steady state solutions which gave the polarizabilities $\alpha_{\sigma,\sigma'}$.

Upon considering an incoming plane wave of the form $\Ev^{0}_a(\rv)= \mathbf{E_0} e^{i \kvp \rv_\parallel} e^{ i k_\perp z }$, the Maxwell equations can be solved using an in-plane Fourier transform  (\cref{spatialsum_Dkvp}) to yield 
\begin{align}
      \Ev_{\sigma=a,b} (\rv)&=\delta_{\sigma, a} \Ev_\sigma^0 (\rv)- k_\sigma^2  \alpha_{\sigma a} G^{\text{sc}} ((\kv+ \kv_{L_\sigma}-\kv_{L_1})_{\parallel}, \rv, k_\sigma) \left[ \mathbb{I}+ k_a^2 \alpha_{aa}\left(G(\kvp, k_\sigma)+ i\frac{\gamma_a}{2 k_a^2 |A|^2 \wp_a^2}\right)\right.\nnl
      &\left.+ k_b^2 \alpha_{bb} \left(G((\kv+ \kv_{L_2}-\kv_{L_1})_{\parallel}, k_b) +i \frac{\gamma_b}{2 k_b^2 |B|^2\wp_b^2}\right)\right]^{-1}\mathbf{E_0}\label{Maxwellsolution1},
\end{align} where 
 \begin{align}
    G^{\text{sc}} (\kvp, \rv, k)= \sum_{i} e^{i \kvp \rv_i } G(\rv- \rv_i, k).
\end{align}

\subsection{Comparison of classical Maxwell solution to quantum mechanical scattering calculation}\label{Comparison_Maxwell_quantum}

To compare the solution of the classical Maxwell equations to the expressions obtained in the scattering calculation shown in \cref{app_explicitcalculation}, we note that taking the rotating-wave approximation in the Hamiltonian in \cref{Ham} corresponds to neglecting the $K^{-}$ function within $G$, effectively replacing $G\to K^{+}$.

Taking the rotating-wave approximation $G\leftrightarrow K^{+}$ and identifying the detuning as $\delta\leftrightarrow \rho$ and the energies $k_a \leftrightarrow \omega_\kv= \rho + \omega_{e_1} - \omega_{g_1}$,  $k_b\leftrightarrow \omega_{\kv'}= \rho+\omega_{e_1}- \omega_{g_2}$, one can see that the inverse matrix appearing in \cref{Maxwellsolution1} is proportional to the inverse matrix appearing in \cref{Soperatoraction} 
\begin{align}
  i\left( \frac{\gamma_a + \gamma_b}{2}- i \delta \right)  \left[ \mathbb{I}+\sum_{\sigma=a,b} k_\sigma^2 \alpha_{\sigma\sigma}\left(G(\kvp+ \kv_{L_\sigma} - \kv_{L_1}, k_\sigma)+ \frac{i\gamma_\sigma}{2 k_\sigma^2 |\sigma|^2\wp_\sigma^2}\right) \right]   
  \leftrightarrow \rho  \mathbb{I} - \rho P(\kvp, a). 
\end{align}

Let us now consider $G^{sc}$. Taking the rotating-wave approximation it can be related to $K^+$
 \begin{align}
    G^{\text{sc}} (\kvp, \rv, k)\leftrightarrow K^{+,\text{sc}} (\kvp, \rv, k) \equiv \sum_{i} e^{i \kvp \rv_i } K^+(\rv- \rv_i, k). 
\end{align}
Evaluating this for the case of the $a$-field one obtains
\begin{align}
    K^{+,\text{sc}} (\kvp, (\rv,z), k_a)&= \sum_{i} e^{i \kvp \rv_i }\frac{1}{\omega^2_{\kv} \wp_a^2} \frac{1}{V}\sum_{\kv'' \mu''}  \frac{ g_{\kv'' \mu'',m,a} g_{\kv''\mu'',n,a}}{\rho - (\omega_{\kv''}- \omega_{e_1 }+ \omega_{g_1})} e^{i \kv'' (\rv-\rv_i)}\\
    &=\frac{1}{\omega^2_{\kv} \wp_a^2}\sum_{\pv \mu''} \frac{e^{i (\kvp+\pv) \rv}}{ d^2}\int_0^{\infty}dk^{''}_{\perp}  \left(\frac{ g_{(\kvp+\pv,k^{''}_{\perp}) \mu'',m,a} g_{(\kvp+\pv,k^{''}_{\perp}) \mu'',n,a}}{\omega_{\kv}+ i 0^+ - \sqrt{(\kvp + \pv)^2 + (k^{''}_\perp)^2}}  e^{i k^{''}_{\perp}z }\right.\nnl
&\left.\quad\quad\quad\quad\quad\quad\quad\quad\quad\quad\quad\quad+\frac{ g_{(\kvp+\pv,-k^{''}_{\perp}) \mu'',m,a} g_{(\kvp+\pv,-k^{''}_{\perp}) \mu'',n,a}}{\omega_{\kv}+ i 0^+ - \sqrt{(\kvp + \pv)^2 + (k^{''}_\perp)^2}}  e^{-i k^{''}_{\perp}z }\right). \label{hourglass}
\end{align}Upon  considering $k^{''}_\perp \in \mathbb{C}$ as a complex variable, for $\omega_\kv- (\kvp+ \pv)^2>0$ the  denominator in \cref{hourglass} has  poles slightly above (below) the real axis near $k^{''}_\perp= \pm \left(\sqrt{\omega_\kv- (\kvp+ \pv)^2}+ i 0^+ \right)$. Similarly, for $\omega_\kv- (\kvp+ \pv)^2<0$ these poles are near the complex axis, where the integrand may have additional branch cuts and poles stemming from the square roots and the coupling constants $g_{\kv'' \mu'',m,\sigma}$. 

For $z>0$ we now integrate the term with $e^{i k^{''}_{\perp}z }$ in \cref{hourglass} along a contour that contains the positive real and positive imaginary axis along with the infinite arc between them. Similarly, the $e^{-i k^{''}_{\perp}z }$ term is integrated along the positive real and negative imaginary axis along with the connecting infinite arc. The integrands vanish along the infinite arcs and for $z<0$ one proceeds analogously. The integrations along the imaginary axis yield evanescent field contributions which decay exponentially in $z$. As these cannot propagate away from the array, they do not correspond to in- or out-states and thus cannot appear in the quantum scattering calculation described in \cref{app_explicitcalculation}. Disregarding these evanescent fields, from the pole residues one thus finds 
\begin{align}
    G^{+,\text{sc}} (\kvp, (\rv,z), k_a)_{m,n}
    &\leftrightarrow\! - i\!\!\sum_{\pv \mu''}\!\!\Theta(\omega_{\kv}>\!|\kvp + \pv|)  \frac{ g_{(\kvp+\pv,k^{''}_\perp) \mu'',m,a} g_{(\kvp+\pv,k^{''}_\perp) \mu'',n,a}}{ \omega_{\kv} \wp_a^2 d^2\sqrt{\omega^{2}_{\kv}-(\kvp + \pv)^2}} e^{i (\kvp+\pv) \rv}  e^{i k^{''}_{\perp}z }\Bigg|_{k^{''}_\perp= \sgn(z)\sqrt{\omega^{2}_{\kv}-(\kvp + \pv)^2}}.
\end{align}

As the incident field is transversal, the polarization vector $\Ev_0 $ can be decomposed as $\Ev_0 = \sum_\mu \Ev_0^\mu \vec{\epsilon}_{\kv \mu} $
and because the relation between the scattered fields $\Ev_a, \Ev_b$ and the incident field $\Ev_0$ is linear, without loss of generality we can consider $\Ev_0=\vec{\epsilon}_{\kv \mu}$ to obtain 
\begin{align}
    \Ev_a (\rv)\leftrightarrow \Ev_a^0 (\rv)-&\frac{i}{a^2}  \sum_{\pv \mu''n,m} \left[ \frac{  |A|^2 g_{\kv \mu,m,a} g_{(\kvp+\pv,k^{''}_\perp) \mu'',n,a}\omega_\kv}{\sqrt{\omega^{2}_{\kv}-(\kvp + \pv)^2}}  \left(\rho  \mathbb{I} - \rho P(\kvp, a)\right)^{-1}_{n,m} \right.  \\
   & \left.\times \Theta(\omega_{\kv}>|\kvp + \pv|)  e^{i (\kvp+\pv) \rv} \vec{\epsilon}_{(\kvp+\pv,k^{''}_\perp) \mu''}  e^{i k^{''}_{\perp}z }\Bigg|_{k^{''}_\perp= \sgn(z)\sqrt{\omega^{2}_{\kv}-(\kvp + \pv)^2}}\right]. 
\end{align}
Finally, identifying  $\Ev_a^0 \leftrightarrow \ket{(\kvp,k_\perp) \mu a}$ and 
\begin{align}
e^{i (\kvp+\pv) \rv} \vec{\epsilon}_{(\kvp+\pv,k^{''}_\perp) \mu''}  e^{i k^{''}_{\perp}z }\Bigg|_{k^{''}_\perp= \sgn(z)\sqrt{\omega^{2}_{\kv}-(\kvp + \pv)^2}} \leftrightarrow  \sum_{s=+,-}\ket{\kv'(\pv, s,a) \mu' a}, 
\end{align}one can see that enforcing the rotating-wave approximation and identifying allowed out-states, these two approaches are equivalent. Performing an analogous calculation for the $b$-field $\Ev_b$ yields a similar equivalence.

\section{Spectral analysis of finite emitter array scattering}
\label{Fourier_scattering}

\begin{figure}[t!]
   \centering
     \begin{tikzpicture}
    \node[anchor=south west,inner sep=0] (image) at (0,0) {\includegraphics[width=.48\linewidth]{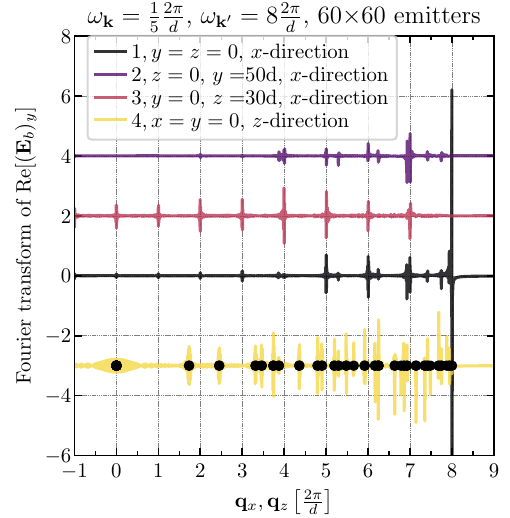}};
    \begin{scope}[x={(image.south east)},y={(image.north west)}]
\end{scope}
\end{tikzpicture}
    \caption{\textbf{Spectral components at normal incidence for the data shown in \cref{maxwell_normal_incidence}(\textbf{d}).} The numbered, red-dashed lines in \cref{maxwell_normal_incidence}(\textbf{d}) correspond to the axis of the  Fourier transforms shown here.  Fourier transform of the real part of the $y$-component  $\Re[(\Ev_b)_y]$ in $x$-direction for $y=z=0$ (1, black), $z=0,y=50d  $ (2, purple), and $y=0, z=30 d$ (3, red) as a function of the wavevector component $\qv_x$ in units of $(2 \pi /d)$.  Furthermore, a  Fourier transform of the $y$-component  $\Re[(\Ev_b)_y]$  in $z$-direction for $x=y=0$ (4, yellow) is shown as a function of the wavevector component $\qv_z$ in units of $(2 \pi /d)$. The Fourier transform is given in arbitrary units and the different transforms are shifted by $4$ (2, purple), $2$ (3, red) and $-5$ (4, yellow), additionally, the purple line is enhanced by a factor of 5 to improve visibility. As the scattering at normal incidence is both symmetric wrt. $x\to-x$ and $z\to -z$, the Fourier transforms are even functions of frequency and we only show them for positive wavevector components. The black dots shown on the transform in $z$-direction (yellow line) denote the location of components $\qv_z= \pm \sqrt{(\omega_{\kv'})^2- \pv^2}\in \mathbb{R}$ with $\pv \in \mathbb{Z}_2 (2\pi/d)$. }
    \label{fig:Fourier_Maxwell}
\end{figure}

To analyze the spectral makeup of the scattered signal, in \cref{fig:Fourier_Maxwell} we show Fourier transforms of the $y$-component of the scattered field shown in \cref{maxwell_normal_incidence}(\textbf{d}). The $y$-component is chosen here because for a normally-incident $y$-polarized electric field and an infinite emitter array, the outgoing signal must have a non-vanishing $y$-polarization component. The Fourier transforms are performed both in $x$-direction as well as in $z$-direction to obtain spectral components at momenta $\qv_x$ and $\qv_z$, respectively. The red dashed lines in \cref{maxwell_normal_incidence}(\textbf{d}) correspond to the axes along which the Fourier transforms in \cref{fig:Fourier_Maxwell} are performed. The Fourier transform performed in $x$-direction for $y$=$z$=$0$ (i.e. within the array plane, crossing the array), shows strong peaks for $\qv_x= \pm 8 (2\pi/d)$ along with smaller peaks near integer multiples of $(2\pi/d)$. The strong peaks at $\qv_x=\pm 8 (2\pi/d)$ correspond to the scattering into the critical modes ($\pv=[\pm 8,0] (2\pi/d)$) with vanishing perpendicular wavevector component, while the smaller peaks near integer multiples of $(2\pi/d)$ correspond to scattering into free space modes with $k_\perp\neq 0 $.
A Fourier transform performed in $x$-direction for $y=50d, z=0$ (in the array plane,  outside the array and outside the main scattering lobe) is much smaller in magnitude and shows weak peaks at integer multiples of $(2\pi/d)$, suggesting scattering into free space modes (note that the plot  of this transform in \cref{fig:Fourier_Maxwell} is enhanced by a factor of 5 to improve visibility). The strong scattering peak from scattering into the critical modes [$\pv=(\pm 8,0) (2\pi/d)$] is absent here, due to the location at which this Fourier transform is carried out. The strongest peaks visible here are near $\qv_x=\pm 7 (2\pi/d)$ and $\qv_x=\pm 4 (2\pi/d)$ corresponding to the off-resonant directions mentioned before. Transforming in $x$-direction for $y=0$, $z= 30d $ instead (i.e. behind the array, parallel to the array) where one expects to see scattering into free space modes, on can see that there are peaks at integer multiples of $(2\pi/d)$, corresponding to scattering into modes with $|k_\perp|>0$. Fourier transforming in $z$-direction for $x=y=0$ (i.e. perpendicular to the array, through the array) to obtain $\qv_z$, one can see that the scattered electric field spectrum is peaked for components $\qv_z = \pm \sqrt{(\omega_{\kv'})^2- \pv^2}$ with $\pv \in \mathbb{Z}_2 (2\pi/d)$ and $|k_\perp|>0$, while there is also a contribution at $|\qv_z|=0$ corresponding to the scattering into the critical mode with $\pv= [\pm 8,0](2 \pi /d)$.

\begin{figure}[ht!]
    \centering
\begin{tikzpicture}
\node[anchor=south west,inner sep=0] (image) at (0,0) {\includegraphics[width=.48\linewidth]{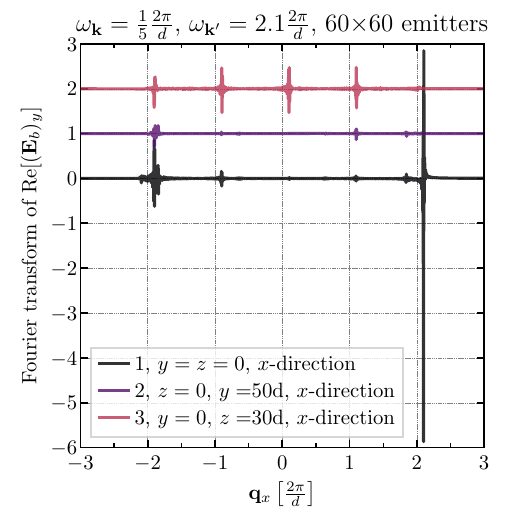}};
\begin{scope}[x={(image.south east)},y={(image.north west)}]
\end{scope}
\end{tikzpicture}
\caption{\textbf{Spectral components for  $30^\circ$-degree incidence for the data shown in \cref{fig:non_normal_maxwell}(\textbf{b}).} The numbered, red-dashed lines in \cref{fig:non_normal_maxwell}(\textbf{b}) correspond to the axis of the  Fourier transforms shown here. Fourier transforms of the real part of the $y$-component  $\Re[(\Ev_b)_y]$ in x-direction for $y=z=0$ (1,black), $z=0,y=50d  $ (2, purple), and $y=0, z=30 d$ (3, green) are shown as a function of the wavevector component $\qv_x$ in units of $(2 \pi /d)$. The Fourier transform is given in arbitrary units and the different transforms are shifted by $1$ (2, purple) and $2$ (3, red) and, additionally, the purple line is enhanced by a factor of 2 to improve visibility. The Fourier transform performed within the array (1, $y=z=0$), shows a strong peak at $\qv_x= 2.1 (2\pi/d)$ as expected from the criticality of the $\pv+ \kvp$ mode with $\pv=[2,0] (2\pi /d)$. Weaker peaks are visible at integer multiples of $(2\pi/d)$, offset by $(\kvp)_x= 0.1 (2 \pi/d)$. The transform performed in the array plane, outside the array (2, $z=0$, $y=50d$) shows much weaker peaks, the strongest being at $\qv_x= 1.9 (2\pi/d)$, in line with the off-resonant modes seen in \cref{fig:non_normal_maxwell}(\textbf{b}, left plot). Behind the array (3, $y=0$, $z=30d$) peaks at $\qv_x= -1.9$, $-0.9$, $0.1$ and $1.1 (2\pi/d)$ can be seen, which correspond to photonic free space modes $\left(\kvp + \pv,k_\perp= \pm \sqrt{(\omega_{\kv'})^2 - |\kvp+ \pv|^2 } \right)$ with $|k_\perp|>0 $ and $\pv \in (2\pi/d) \mathbb{Z}_2$.}
    \label{fig:Fourier_Maxwell_nonnormal}
\end{figure}

To further illuminate the  directionality of the scattering at non-normal incidences, in \cref{fig:Fourier_Maxwell_nonnormal} we show $x$-direction Fourier transforms of the $y$-component of the scattered $b$-field $(\Ev_b)_y$ for the data shown in \cref{fig:non_normal_maxwell}(\textbf{b}). The numbered, dashed lines in \cref{fig:non_normal_maxwell}(\textbf{b}) correspond to the axes of the Fourier transforms.  It can be seen that for $y=z=0$, which is within the array plane and crosses through the array, there is a strong peak at $\qv_x = 2.1 (2\pi /d)$ along with weaker peaks at $-1.9$,  $-0.9$, $0.1$ and $1.1 (2\pi /d)$. The strong peak at $\qv_x = 2.1 (2\pi /d)$ originates from the scattering into the critical mode $\kvp+ \pv$ with $\pv=[2,0] (2 \pi / d)$, while the weaker peaks originate from scattering into free space modes $(\kvp+\pv, k_\perp)$ with $|k_\perp|>0$ and $\pv \in (2\pi/d) \mathbb{Z}_2$; that is the peaks are shifted by $(\kvp)_x $ from integer multiples of $2\pi/d$. Performing instead a Fourier transform at $y=50d, z=0$ in $x$-direction (within the array plane but outside the array), the strong peak is absent and only the weaker peaks are visible, as expected geometrically. Here, the strongest peak is near $\qv_x= - 1.9 (2\pi /d)$, which aligns with the scattering into off-resonant modes near $\kvp +\pv \approx(-2+0.1, \pm 1) (2\pi/d)$ seen in \cref{fig:non_normal_maxwell}(\textbf{b}). Finally, performing a Fourier transform in $x$-direction behind the array at $y=0, z=30d$, there are peaks visible at $\qv_x =1.9$,  $-0.9$, $0.1$ and $1.1 (2\pi /d)$, which correspond to the scattering into free space modes observed behind the array in \cref{fig:non_normal_maxwell}(\textbf{b}).

\section{Effect of the rotating-wave-approximation}\label{RWA_no_RWA_comp}

In the following, we compare the effect of the rotating-wave-approximation (RWA) to Green's functions for which the RWA was not undertaken. To this end, we carry out several of the semiclassical calculations shown in the main text, but replacing the full dyadic Green's function $G= K^+- K^-$ shown in \cref{Gdef} with the RWA Green's function $K^+$ shown in \cref{RWAGdef}.

\begin{figure}[ht!]
    \centering
    \begin{tikzpicture}
    \node[anchor=south west,inner sep=0] (image) at (0,0) {\includegraphics[width=\linewidth]{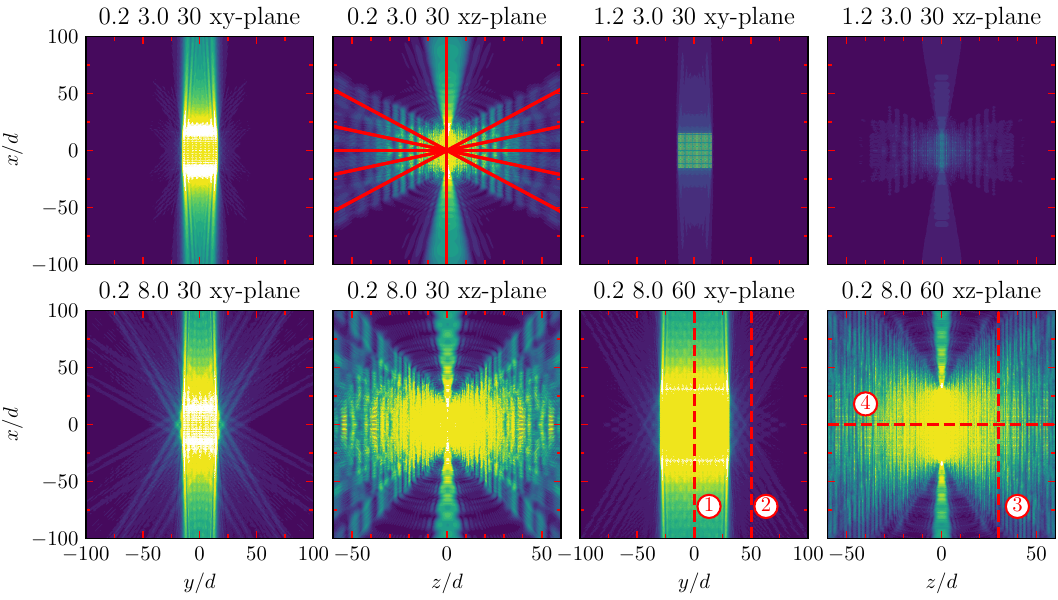}};
    \begin{scope}[x={(image.south east)},y={(image.north west)}]
        \node[align=right] at (0.02,0.98) {$\textbf{(a)}$};
    \end{scope}
    \end{tikzpicture}
        \begin{tikzpicture}
    \node[anchor=south west,inner sep=0] (image) at (0,0) {\includegraphics[width=\linewidth]{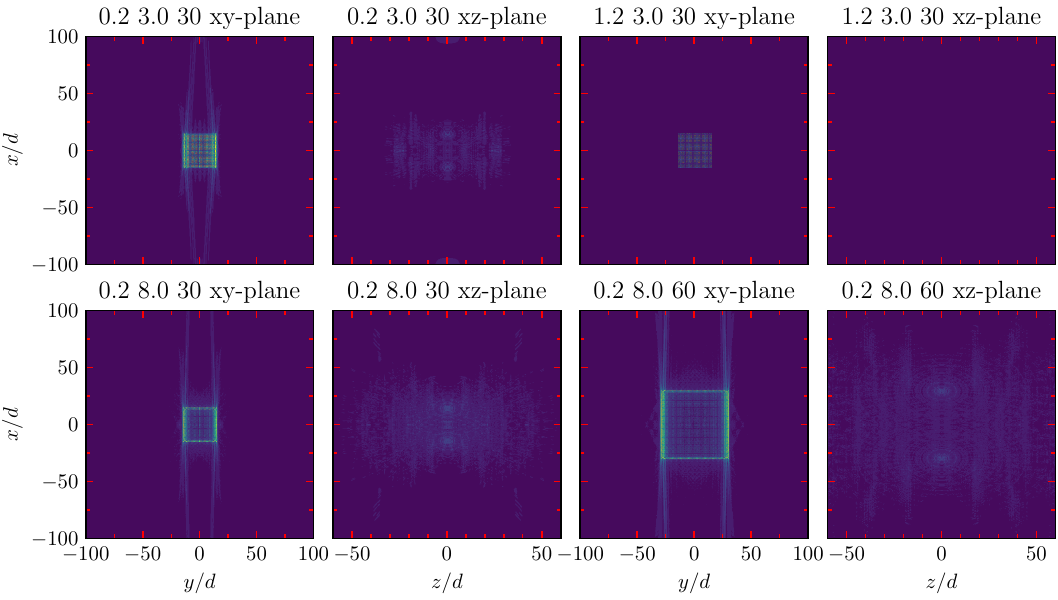}};
    \begin{scope}[x={(image.south east)},y={(image.north west)}]
        \node[align=right] at (0.02,0.98) {$\textbf{(b)}$};
    \end{scope}
    \end{tikzpicture}
    \caption{(\textbf{a}) Same as \cref{maxwell_normal_incidence}, but replacing $G=K^+  - K^-$ (\cref{Gdef}) with $K^+$  (\cref{RWAGdef}) to undertake the RWA. (\textbf{b}) Absolute difference between \cref{maxwell_normal_incidence} and plots shown in (\textbf{a}). The color scale is $[0,1]$ and the same qualitative  behavior can be seen. Only at short distances within the array, are qualitative differences noticeable.}
    \label{Fig6comp.}
\end{figure}

In \cref{Fig6comp.}, the results of \cref{maxwell_normal_incidence} are shown under the RWA, which is taken by replacing $G=K^+  - K^-$ (\cref{Gdef}) with $K^+$  (\cref{RWAGdef}). As it can be seen, no qualitative differences can be observed and significant qualitative differences are only visible at short distances within the array. Similarly, in \cref{Fig789comp.}, the same results as in \cref{fig:non_normal_maxwell,fig:Fourier_Maxwell,fig:Fourier_Maxwell_nonnormal} are shown but taking the RWA. Again, no qualitative differences are visible and only minimal quantitative differences can be seen.

\begin{figure}
    \centering
    \begin{tikzpicture}
    \node[anchor=south west,inner sep=0] (image) at (0,0) {\includegraphics[width=\linewidth]{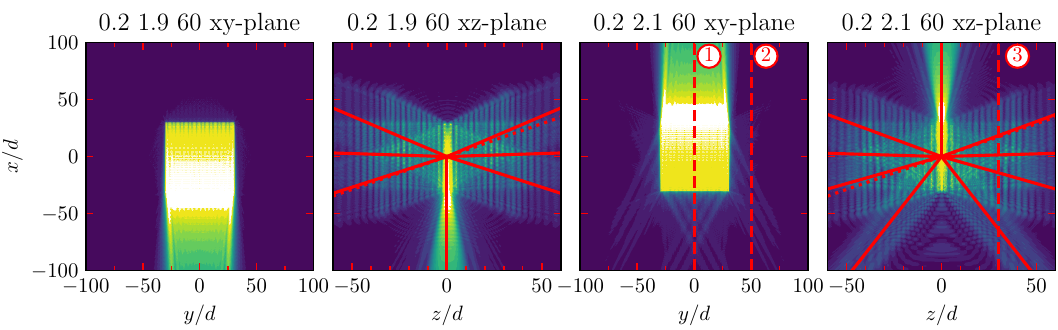}};
    \begin{scope}[x={(image.south east)},y={(image.north west)}]
        \node[align=right] at (0.02,0.98) {$\textbf{(a)}$};
    \end{scope}
    \end{tikzpicture}
     \begin{tikzpicture}
    \node[anchor=south west,inner sep=0] (image) at (0,0) {\includegraphics[width=0.5\linewidth]{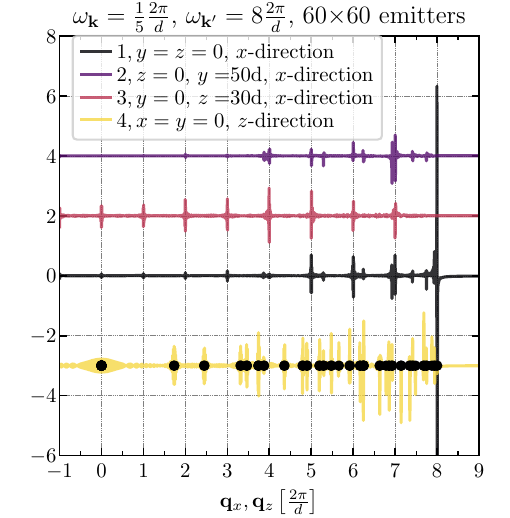}\includegraphics[width=0.5\linewidth]{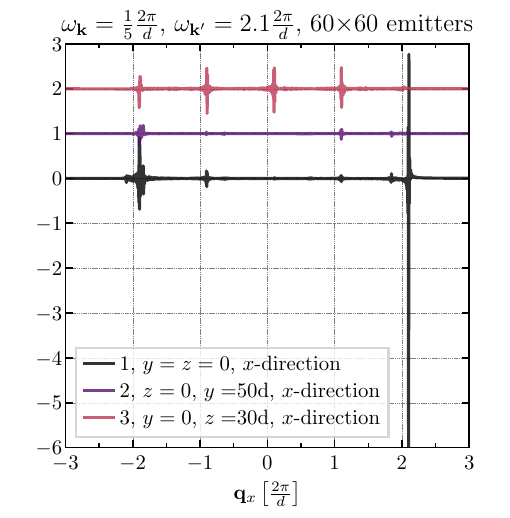}};
    \begin{scope}[x={(image.south east)},y={(image.north west)}]
        \node[align=right] at (0.02,0.98) {$\textbf{(b)}$};
        \node[align=right] at (0.51,0.98) {$\textbf{(c)}$};
    \end{scope}
    \end{tikzpicture}
    \caption{(\textbf{a}) Same as \cref{fig:non_normal_maxwell} but with rotating-wave-approximation. (\textbf{b}) Same as \cref{fig:Fourier_Maxwell} but with rotating-wave-approximation. (\textbf{c}) Same as \cref{fig:Fourier_Maxwell_nonnormal} but with rotating-wave-approximation.}
    \label{Fig789comp.}
\end{figure}

 \end{widetext}

\end{document}